%% file: Planck_Early_Paper_1_v3.1.tex
\documentclass[twocolumn,traditabstract,longauth]{aa}
\usepackage{graphicx}
\usepackage[breaklinks, colorlinks, citecolor=blue]{hyperref}
\usepackage{natbib}
\usepackage{subfigure}
\usepackage{epsfig}
\usepackage{multirow}
\usepackage{color}
\usepackage{txfonts}
\newcommand{\ie}{{i.e.}}
\newcommand{\eg}{{e.g.}}
\usepackage{url}

\input{Planck.tex}

\begin{document}

\input{Proj_Ref_7_7b_authors_and_institutes.tex}

\title{ \textit{Planck} Early Results: The Galactic Cold Core Population revealed by the first all-sky survey}

\abstract
{We present the statistical properties of the first version of the Cold Core Catalogue of 
Planck Objects (C3PO), in terms of their spatial distribution, 
temperature, distance, mass, and morphology. We also describe the statistics of the 
Early Cold Core Catalogue (ECC) that is a subset of the complete catalogue, and
that contains only the 915 most reliable detections. ECC is delivered as a part of the Early Release Compact Source Catalogue (ERCSC).
We have used the CoCoCoDeT algorithm to extract about 10 thousand cold sources. 
The method uses the IRAS $100 \mu m$ data as a warm template 
that is extrapolated to the {\Planck}  bands and subtracted from the signal, 
leading to a detection of the cold residual emission. We have used cross-correlation
with ancillary data to increase the reliability of our sample, and to derive other key properties such
as  distance and mass. 

Temperature and dust emission spectral index values are derived using the fluxes in the IRAS 100 $\mu m$ 
band and the three highest frequency {\Planck} bands. 
The range of temperatures explored by the catalogue spans from 7\,K to 17\,K, and peaks around 13\,K. 
Data are not consistent with a constant value of the associated spectral index $\beta$ over the all temperature range. 
$\beta$ ranges from 1.4 to 2.8 with a mean value around 2.1, and
several possible scenarios are possible, including $\beta(T)$
and the effect of multiple temperature components folded into the measurements.

For one third of the objects the distances are obtained using various methods 
such as  the extinction signature, or the association with known molecular complexes or Infra-Red Dark Clouds.
 Most of the detections are within 2\,kpc in the Solar neighbourhood, but a few are 
 at distances greater than 4\,kpc. The cores are distributed over the whole range of longitude and latitude,
 from the deep Galactic plane, despite the confusion, to high latitudes ($>30^{\circ}$).
The associated mass estimates derived from dust emission range from 
1 to $10^5$ solar masses. Using their physical properties such as temperature, mass, luminosity, density and size, 
these cold sources are shown to be cold clumps, defined as the intermediate cold sub-structures between clouds and cores.
These cold clumps are not isolated but mostly organized in filaments associated with molecular clouds.
The Cold Core Catalogue of Planck Objects (C3PO) is the first unbiased all-sky catalogue 
of cold compact objects and contains 10783 objects. It gives an 
unprecedented statistical view to the properties of these potential pre-stellar clumps 
and offers a unique possibility for their classification 
in terms of their intrinsic properties and environment.
}


\authorrunning{Planck Collaboration}
\titlerunning{The Galactic Cold Core Population revealed by the first {\Planck} all-sky survey}
\maketitle

\section{Introduction}
The main difficulty in understanding star formation lies in the vast
range of scales involved in the process. If star formation itself is
the outcome of gravitational instability occurring in cold and dense
structures at sub-parsec scales, the characteristics of these
structures (usually called pre-stellar cores) depend on their
large-scale environment, up to Galactic scales because their formation
and evolution is driven by a complex coupling of self-gravity with
cooling processes, turbulence and magnetic fields, to name a few.  To
progress in the understanding of star formation pre-stellar cores need
to be observed, in a variety of environments. More importantly, broad
surveys are required to address statistical issues, and probe
theoretical predictions regarding the initial mass function (IMF)
largely determined at the stage of fragmentation of pre-stellar cores.
    
Unfortunately, the properties of the pre-stellar cores
are still poorly known mostly because of  observational difficulties.  The
total number of Galactic pre-stellar cores is estimated to be
around 3$\times$10$^5$ \citep{Clemens1991} but most of them have so far
escaped detection, simply because they are cold and immersed in warmer 
(therefore brighter) environments.

 The thermal dust emission of 
nearby molecular clouds has been mapped from the ground in the millimeter and submillimeter
ranges with
instruments such as  SCUBA, MAMBO, SIMBA, and Laboca. Because of
limited sensitivity, but also the presence of the atmospheric fluctuations that 
call for beam-throw of at most a few arcmin, 
the studies have concentrated on the brightest and most compact
regions that  are already in an active phase of star formation. Thanks
to  sub-arcminute resolution, these observations (together with
dedicated molecular line studies) have been the main source of
information also on the structure of the pre-stellar cores 
\citep{Motte1998,Curtis2010, Hatchell2005, Enoch2006, Kauffmann2008}.

Many {\em compact} clouds were detected as absorption features on
photographic plates. A new population of thousands of cold dark clouds
was discovered by observations of mid-infrared absorption towards the
bright Galactic background \citep[MSX and ISOGAL surveys;
see][]{Egan1998, Perault1996}. The absorption studies are, however,
strongly biased towards the low latitudes and do not directly provide
information on the temperature of the detected sources. For a definitive
study of the cold cloud cores, one must turn to high resolution
observations in the submillimetre or millimetre range
\citep{Andre2000}. 
The Bolocam Galactic Plane Survey (BGPS) is producing mm data for the
central part of the Galactic plane \citep{Aguirre2010}. The first
results suggest that at kpc distances, even with a half arcmin
resolution, one is detecting mainly cluster forming {\em clumps}
rather than cores that would produce,  at most,  a small multiple system
\citep{Dunham2010}.

Balloon borne experiments have provided larger blind surveys of higher
latitudes. PRONAOS discovered cold condensations also in cirrus-type
clouds \citep{Bernard1999, Dupac2003} 
Similarly, Archeops \citep{Desert2008} detected  hundreds of sources
with temperatures down to 7\,K. 
The latest addition to the
balloon borne surveys is the BLAST experiment which has located
several hundred submillimetre sources in Vulpecula \citep{Chapin2008}
and Vela \citep{Netterfield2009, Olmi2009},  including a number of cold
and probably pre-stellar cores.

Since its launch in May 2009, the Herschel satellite has already
provided hundreds of new detections of both starless and protostellar
cores \citep{Andre2010, Bontemps2010, Konyves2010, Molinari2010,
WardThompson2010}.  There is an intriguing similarity between the
core mass function (CMF) derived from these data, 
 and the IMF that need to be investigated in different environments, towards
the inner Galaxy in particular.The Herschel studies will
eventually cover a significant fraction of the Galactic mid-plane and
the central parts of the nearby star-forming clouds
but cannot cover high Galactic latitudes where star formation is known to occur. 
In this endeavor the main
challenge is how to locate the cores because, even with Herschel,
detailed studies must be limited to a small fraction of the whole sky.

The {\Planck} \footnote{\Planck\ (http://www.esa.int/\Planck ) is a project of the European Space 
Agency (ESA) with instruments provided by two scientific consortia funded by ESA member 
states (in particular the lead countries France and Italy), with contributions from NASA 
(USA) and telescope reflectors provided by a collaboration between ESA and a scientific 
consortium led and funded by Denmark.}  satellite \citep{tauber2010a} improves over the previous
studies by providing an {\em all-sky} submillimetre/millimetre survey
that has both the sensitivity and resolution needed for the detection
of compact sources. The shortest wavelength channels of {\Planck} cover
the wavelengths around and longwards of the intensity maximum of the
cold dust emission: $\nu^2B_{\nu}(T=10{\rm K})$ peaks close to
300\,$\mu$m while, with a temperature of $T\sim 6$\,K, the coldest
dust inside the cores has its maximum close to 500\,$\mu$m. Combined
with far-infrared data such as the IRAS survey, the data enable accurate
determination of both the dust temperature and the spectral index. 
We use the {\Planck} observations to search for Galactic cold cores,  \ie\
compact cloud cores with colour temperatures below 14\,K.  Because of
the limited resolution, we are likely to detect mainly larger clumps
inside which the cores are located. The cores will be pre-stellar
objects before (or at the very initial stages) of the protostellar
collapse,  or possibly more evolved sources that still contain
significant amounts of cold dust. The Cold Core Catalogue of Planck
Objects (C3PO) which will be made public at the end of the {\Planck}
proprietary period, will be the first all-sky catalogue of cold cloud cores and clumps.
It will reveal the locations where the next generations of stars will
be born and will provide an opportunity to address a number of key
questions related to  Galactic star formation: What are the characteristics of
this source population? How does the distribution of the cores/clumps
correlate with the current star formation activity and the location of
the molecular cloud rings and the spiral arms? How are the sources
related to large-scale structures like the FIR loops, bubbles, shells,
and filaments? Are there pre-stellar cores at high latitudes? How much
do the core properties depend on their environment? Investigations such as these
will help us understand the origin of the pre-stellar cores, the
instabilities that initiate the collapse, and the roles of turbulence
and magnetic fields. The catalogue will prove invaluable for 
follow-up studies to investigate in detail the internal properties of the
individual sources.

In this paper we describe the general properties of the current cold
cores catalogue that is based on  data that the {\Planck} satellite has
gathered during its first two scans of the full sky. In particular, we
will describe the statistics of the Early Cold Cores Catalogue (ECC)
that is part of the recently published Planck Early Release Compact
Source Catalogue \citep[ERCSC][]{planck2011-1.10}. ECC forms a subset of the full C3PO and
contains only the most secure detections of all the sources with
colour temperatures below 14\,K. The final version of C3PO will be
published in 2013. For historical reasons, we use "Cold Cores" to designate the entries in
the C3PO and in the ECC, and similarly in much of this paper.  However,
as this paper and the companion paper \citep[][hereafter Paper II]{planck2011-7.7a} demonstrate,
most of these are more correctly described as "cold clumps", intermediate
in their structure and physical scale between a true pre-stellar core
and a molecular cloud.

{\Planck} \citep{tauber2010a, planck2011-1.1} is the third generation space 
mission to measure the anisotropy of the cosmic microwave background (CMB).  
It observes the sky in nine frequency bands covering 30--857\,GHz with high 
sensitivity and angular resolution from 31\arcmin\ to 5\arcmin.  The Low Frequency 
Instrument LFI; \citep{Mandolesi2010, Bersanelli2010, planck2011-1.4} covers the 30, 44, 
and 70\,GHz bands with amplifiers cooled to 20\,\hbox{K}.  The High Frequency 
Instrument (HFI; \citealt{Lamarre2010, planck2011-1.5}) covers the 100, 143, 217, 353, 
545, and 857\,GHz bands with bolometers cooled to 0.1\,\hbox{K}.  Polarization is
 measured in all but the highest two bands \citep{Leahy2010, Rosset2010}.  
 A combination of radiative cooling and three mechanical coolers produces the 
 temperatures needed for the detectors and optics \citep{planck2011-1.3}.  Two 
 Data Processing Centers (DPCs) check and calibrate the data and make maps 
 of the sky \citep{planck2011-1.7, planck2011-1.6}.  {\Planck}'s sensitivity, angular 
 resolution, and frequency coverage make it a powerful instrument for galactic and 
 extragalactic astrophysics as well as cosmology.  Early astrophysics results are given 
 in {\Planck} Collaboration, 2011h--z.

\section{Source Extraction} 
\label{sec_source_extraction}

\subsection{Data Set}
\label{data}

As cold cores are traced by their cold dust emission in the submillimetric bands,
we use {\Planck} channel maps of the HFI at 3 frequencies : 
353, 545 and 857~GHz as described in
detail in~\citet{planck2011-1.7}. The temperature maps at these frequencies 
are based on the  first two sky surveys of {\Planck}, provided in Healpix format \citep{Gorski2005} at nside=2048. 
We give here a very brief
summary of the data reduction, cf~\citet{planck2011-1.7} for further details.
Raw data are first processed to produce cleaned timelines (TOI) and
associated flags identifying various systematic effects. The data analysis
includes application of a  low-pass filter, removal and correction of  glitches,  
conversion to absorbed power and  decorrelation of thermal stage fluctuations. For
the cold core detection, and more generally for source detection, 
Solar System objects (SSO) are identified in the TOI data using the publicly 
available Horizon ephemerides and an SSO flag is created to ensure 
that they are not projected onto the sky. 

Focal plane reconstruction and beam-shape estimates are obtained using
observations of Mars. Beams are described by an elliptical Gaussian
parameterisation leading to FWHM $\theta_{\mathrm{S}}$ given in Table~2 of \citet{planck2011-1.7}.
 The attitude of the satellite as a function
of time is provided by the two star trackers installed on the {\Planck}
spacecraft. The pointing for each bolometer is computed by combining
the attitude with the location of the bolometer in the focal plane reconstructed 
from Mars observations.

From the cleaned TOI and the pointing, channel maps have been made
using bolometers at a given frequency. The path from TOI to maps in
the HFI DPC is schematically divided into three steps, ring-making,
destriping and map-making. The first step averages circles within a
pointing period to make rings with higher signal-to-noise ratio taking
advantage of the redundancy of observations provided by the {\Planck}
scanning strategy. The low amplitude $1/f$ component is accounted for in
a second step using a destriping technique. Finally,
cleaned maps are produced using a simple co-addition of the rings.

The noise in the channel maps is essentially white with a mean
standard deviation of $1.4\times10^{-3}$, $4.1\times10^{-3}$,
$1.4\times10^{-3}~\rm{MJy/sr}$ at 353, 545 and 857~GHz respectively
\citep{planck2011-1.7}. The photometric calibration is performed
either at the ring level using the CMB dipole, for the lower frequency
channels, or at the map level using FIRAS data, for the higher
frequency channels at 545 and 857\,GHz. The absolute gain calibration
of HFI {\Planck} maps is known to better than 2\% at 353\,GHz and 7\%
at 545 and 857\,GHz \citep[see Table~2 in][]{planck2011-1.7}.

The detection algorithm requires the use of ancillary data to trace
the warm component of the gas.  Thus we combine {\Planck} data with
the IRIS all-sky data \citep{Miville2005}.  The choice of the IRIS
$100~\mu \rm{m}$ as the {\it warm template} is  motivated by the
following:  (i) $100~\mu \rm{m}$ is very close to the peak frequency of
a black body at 20 K, and traces  the warm component of
the Galaxy; (ii) the fraction of small grains at this wavelength
remains very small and does not significantly the estimate of the
emission from large  grains that is  extrapolated to longer
wavelengths; (iii) the IRAS survey covers almost the entire sky (only 2
bands of $\sim$2\% of the whole sky are missing); (iv) the resolution of
the IRIS maps is similar to  the resolution of {\Planck} in the high
frequency bands, \ie\ around 4.5\arcmin.  Using the
map at $100~\mu \rm{m}$ as the {\it warm template} is,  of course,  not
perfect, because a non-negligible fraction of the cold emission is
still present at this frequency. This lowers the  intensity in
the {\Planck} bands after removal of the extrapolated background.  We
will describe in detail,  especially in Sect.~\ref{sec_photometry}, 
 how we deal with
this issue for the photometry of the detected cores.

All {\Planck} and IRIS maps have been smoothed at the same resolution
4.5\arcmin\, before source extraction and photometry processing.

\subsection{Source Extraction Method}

We have applied the detection method described in Montier et al. 2010,
known as {\it CoCoCoDeT} (standing for Cold Core Colour Detection
Tool), on the combined IRIS plus {\Planck} data set described in
Sect.~\ref{data}.  This algorithm uses the colour properties of the
objects to be detected  to separate them from the
background. In the case of cold cores, the method selects compact
sources colder than the surrounding envelope and the diffuse Galactic
background, that is at about 17\,K \citep{Boulanger1996} 
but can largely vary from one place to the other across the Galactic plane or at higher latitudes.  
This {\it Warm Background Subtraction} method is applied on each one of the three {\Planck}
maps, and consists of  6 steps:
\begin{description}
\item [1.] for each pixel, the background colour is estimated as the 
 median value of the {\Planck}  map divided by the 100$\mu m$ map within
 a disc of radius 15\arcmin\ around the central pixel;
\item [2.] the {\it warm component} in a pixel at the {\Planck} frequency is
obtained by multiplying the estimate of the background colour with the value of the pixel in the $100\mu m$ map;
\item [3.] the {\it cold residual} map is computed by subtracting the {\it warm component} from the {\Planck} map;
\item [4.] the local standard deviation around each pixel in the {\it cold residual} map is estimated in a radius of 30\arcmin\ using the so-called 
Median Absolute Deviation that ensures robustness against a high confusion level of the background
and presence of other point sources within the same area;
\item [5.] a thresholding detection method is  applied in the {\it cold residual} map to detect sources at  a signal-to-noise ratio SNR$>$4;
\item [6.] final detections are defined as local maxima of the SNR constrained so that there is a minimum distance of 5\arcmin\ between them.
\end{description}

This process is performed at each {\Planck} band yielding individual catalogues
at $857~\mathrm{GHz}$, $545~\mathrm{GHz}$ and $353~\mathrm{GHz}$.
The last step of the source extraction consists in merging these three independent catalogues
requiring a detection in all three bands at SNR$>$4.  
This step rejects spurious detections that are due to map artifacts 
associated with a single frequency (\eg\  stripes or under-sampled features).
It increases the robustness of the final catalogue, which contains 10783 objects.

We stress that no any other a-priori constraints are imposed on the
size of the expected sources, other than the limited area on which the
background colour is estimated. Thus the maximum scale of the C3PO
objects is about 12\arcmin.  Note also that this {\it Warm
  Background Subtraction} method uses local estimates of the colour,
identifying a relative rather than an absolute colour excess.  Thus
cold condensations embedded in cold regions can be missed, while in
hot regions condensations may be detected that are not actually cold.  A
more detailed analysis in temperature is required to assess the nature
of the objects.

\subsection{Photometry}
\label{sec_photometry}

\begin{table*}[t!]

\begin{center}
\newdimen\digitwidth 
\setbox0=\hbox{\rm 0} 
\digitwidth=\wd0 
\catcode`*=\active 
\def*{\kern\digitwidth} 

\newdimen\signwidth 
\setbox0=\hbox{+} 
\signwidth=\wd0 
\catcode`!=\active 
\def!{\kern\signwidth} 

\newdimen\signwidth 
\setbox0=\hbox{.} 
\signwidth=\wd0 
\catcode`?=\active 
\def?{\kern\signwidth} 

\begin{tabular}{l|cc|cc|cc|cc}
\hline
\hline
  & \multicolumn{2}{c}{Normal} & \multicolumn{2}{|c}{{\it Bad Sfit $100~\mu$m}} &   \multicolumn{2}{|c}{{\it Aper Forced}} &  \multicolumn{2}{|c}{{\it Upper 100$\mu$m}}   \\
\multicolumn{1}{c|}{Quantity} &      Bias (\%) & 1$\sigma$(\%)    &        Bias(\%) & 1$\sigma$(\%)          &         Bias(\%) & 1$\sigma$(\%)      &        Bias (\%) & 1$\sigma$(\%)       \\
\hline
Flux at $100~\mu\rm{m}$ &       !1.4   &    31.7 &    *1.0 &   *4.7  &   -58.1  &    14.1  & !117.1  & 190.0   \\
Flux at $857~\rm{GHz}$ &   -5.0   &   *6.2 &    *3.9 &     *3.2 &      -56.3  &   13.8  &   *-11.0   &        **6.0\\
Flux at $545~\rm{GHz}$ &    -3.6   &   *6.4   &    *3.7  &   *3.7   & -55.8   &  14.4   &  **-9.0  & **6.0 \\
Flux at $353~\rm{GHz}$ &  -5.0    &  *7.3    &   *2.4   &    *4.7  &  -58.9 &   14.8  &   *-10.0  &  **6.7  \\
FWHM &   -0.6    &  16.2 &  30.9 &    27.7  & -25.2    &  16.3   & **-6.7   & *15.3 \\
Ellipticity & !0.0 & *8.2 & *0.0 & *9.5 & - & - & **!0.0 & **9.0\\ 
T &   -4.2    &   *5.2 &      -4.1  &   *1.6  &   *-6.5  &     *3.8 &  !**0.4 &  *16.0 \\
$\beta$ & !9.8   &  *7.3   &    10.5   &   *2.4  & !11.2   &  *6.7  &  !**2.7  & *18.7\\
\hline
\end{tabular}
\caption{Statistics of the Monte-Carlo analysis performed to estimate the robustness of the photometry algorithm. 
The bias (expressed in \%) is defined as the relative error between the median of the output distribution of the photometry algorithm and the injected input.
The 1$\sigma$ (expressed in \%) represents the discrepancy around the most probable value of the output distribution.
Those quantities are given in the various cases corresponding to the output flags provided by the algorithm. 
Statistics of the temperature and spectral index is also given here to show the impact of the observed error on fluxes.}
\label{tab:mc_photometry_1}
\end{center}
\end{table*}

\begin{table*}[t!]
\begin{center}

\newdimen\digitwidth 
\setbox0=\hbox{\rm 0} 
\digitwidth=\wd0 
\catcode`*=\active 
\def*{\kern\digitwidth} 

\newdimen\signwidth 
\setbox0=\hbox{+} 
\signwidth=\wd0 
\catcode`!=\active 
\def!{\kern\signwidth} 

\newdimen\signwidth 
\setbox0=\hbox{.} 
\signwidth=\wd0 
\catcode`?=\active 
\def?{\kern\signwidth} 

\begin{tabular}{l|cc|cc|cc|cc}
\hline
\hline
 & \multicolumn{2}{c}{Normal} & \multicolumn{2}{|c}{{\it Bad Sfit $100~\mu$m}} &   \multicolumn{2}{|c}{{\it Aper Forced}} &  \multicolumn{2}{|c}{{\it Upper 100$\mu$m}}   \\
\multicolumn{1}{c|}{Quantity} &      Bias (\%) & 1$\sigma$(\%)    &        Bias(\%) & 1$\sigma$(\%)          &         Bias(\%) & 1$\sigma$(\%)      &        Bias (\%) & 1$\sigma$(\%)       \\
\hline
Flux at $100~\mu\rm{m}$ &    !11.5   	&  44.3    	&    *!0.8    &   *8.4 	& -51.6 	&   21.1 	&     !204.5 	& 278.2 \\
Flux at $857~\rm{GHz}$   &  *-4.0    	&  *8.1  	&  *!2.1   	& *4.7  	&  -58.3  	&    20.1 	& *-10.4 		&  **7.1 \\ 
Flux at $545~\rm{GHz}$   &   *-2.5 	& *8.0     	&   *!2.4  	&  *4.9 	&   -57.4   	&    21.3 	&    **-7.8  	& **7.0 \\
Flux at $353~\rm{GHz}$   & *-3.4    	&   *8.7   	&  *!1.9  	& *5.5    	&   -59.3  	&   21.3    	&  **-8.7   		& **7.4  \\
FWHM 				&  *!0.0  	&   *18.1  	&  !31.0  	& 31.1   	&  -24.4   	& 16.9  	&  **-5.2 		& *17.6 \\
Ellipticity 				&   *!0.0  	&   *9.3   	& *-0.5  	&   *9.2 	& - 		&   -    	&  **!0.1  		&  *10.4 \\
T 					& *-2.1  	&  *6.3  	&    *-3.2 	&   *1.8  	&  *-4.4   	&     *6.2  	& **!6.8  		&  *20.6 \\
$\beta$ 				& !*7.1 	&    *8.2 	&   *!9.3  	&  *2.6  	&   !*5.6   	& 12.3 	&   **-4.9 		&  *20.4 \\
\hline
\end{tabular}
\caption{Same as Table \ref{tab:mc_photometry_1}  in the Galactic plane ($|\rm{b}|<25^{\circ}$).}
\label{tab:mc_photometry_2}
\end{center}
\end{table*}

We have developed a dedicated algorithm to derive the photometry of the clump itself. 
The fluxes are estimated 
from the {\it cold residual} maps, instead of working on the initial maps where the clumps are embedded in their warm
surrounding envelope. As already stressed above,
the main issue is to perform the photometry on the IRIS $100~\mu \rm{m}$ maps that also include a fraction of
the cold emission. The flux of the source at $100~\mu \rm{m}$ has to be well determined for two reasons: 
(1) an accurate estimate of the flux at this frequency is required because it is constrains significantly
the rest of the analysis (in terms of spectral density distribution (SED) and temperature); 
(2) an incorrect estimate of the flux at $100~\mu \rm{m}$
 will propagate through the {\Planck} bands after removal of the extrapolated {\it warm component}. 
 The main steps of the photometry processing are described in the following subsections. An illustration of this process
 is provided in Fig.~B.5 of the associated Planck Early Paper on Cold Clumps describing 
 in detail a sample of 10 sources \citep{planck2011-7.7a}.

\subsubsection{Step1: Elliptical Gaussian fit}
\label{sec_step1}

An elliptical Gaussian fit is performed on the $1^{\circ} \times 1^{\circ}$ colour map 
 $857~\mathrm{GHz}$ divided by $100~\mu\mathrm{m}$ centered on each C3PO object. This
results in estimates of three parameters: major axis  extension $\sigma_{\rm Maj}$, 
minor axis extension $\sigma_{\rm Min}$ and position angle $\psi$.
The relation between the extension $\sigma$ and the FWHM $\theta$ of a Gaussian is given by : 
\begin{equation}
\sigma = \theta  / \sqrt{8 \, {\rm ln}(2)}
\end{equation} 
If the  elliptical Gaussian fit is indeterminate,  a symmetrical Gaussian is assumed with a 
FWHM fixed to $\theta=4.5\arcmin$, and
the flag {\it Aper Forced} is set to on.
In these cases, the  source fluxes are severely underestimated at all frequencies.
This flagged population contains 978 sources which are rejected from the physical analysis of
 Sect.~\ref{sec_physical_properties}, but not from the entire catalogue, which  is used
to assess the association with ancillary data (cf Sect.~\ref{sec_spatial_distribution}) and to 
 study morphology at large scale (cf Sect.~\ref{sec_large_scale_morphology}).

\subsubsection{Step2: 100$\mu$m photometry}
The photometry on the $100~\mu\rm{m}$ map is obtained by  surface fitting,  performed on local 
maps of $1^{\circ} \times 1^{\circ}$ centered on each candidate. All components of the map are fitted as a whole:
 a polynomial surface of an order between three and six for the background;
a set of elliptical Gaussians when other point sources are detected inside the local map; 
and a central elliptical Gaussian corresponding to the cold core candidate for which the elliptical
shape is set by the parameters obtained during step 1.
When the fit of the background is poor, \ie\ a clear degeneracy is observed between the polynomial fit and 
the central Gaussian, 
we switch to a simple aperture photometry on the local map. Note that the aperture 
photometry is performed taking into account the elliptical shape of the cold core provided by step 1. 
In such cases (140 sources), the flag {\it Bad Sfit 100$\mu$m} is set to on. 
Occasionally  no counterpart at all is observed at $100~\mu\rm{m}$, when 
the cold core candidate is too faint or very cold, or the confusion of the Galactic background is too high. 
In such case, we are not able to derive any reliable estimate of the $100~\mu\rm{m}$ flux of the core, 
so only an upper-limit  can be provided. This upper limit is defined as three times the standard deviation 
of the {\it cold residual} map
within a 25\arcmin\ radius circle, and the flag {\it Upper 100$\mu$m} is set to on. 
There are  2356 objects for which only an upper limit of the temperature is derived. 
This population represents a very interesting sub-sample of the whole catalogue, probably the coldest objects, but we do not
have  confidence in the physical properties derived from the \Planck\ data and so it is excluded 
 from the physical analysis.

\subsubsection{Step 3: 100$\mu$m correction}
Once an  estimate of the flux at $100~\mu\rm{m}$ has been provided by steps 1 and 2, 
the {\it warm template} at  $100~\mu\rm{m}$ is corrected by removing an elliptical 
Gaussian corresponding to the flux of the central clump.
This new {\it warm} template is then extrapolated and subtracted from the {\Planck} maps to
build the {\it cold residual} maps.
When only an upper limit has been obtained at $100~\mu\rm{m}$, the {\it warm template} is not changed.

\subsubsection{Step 4: {\Planck} bands photometry}
Aperture photometry is performed on local {\it cold residual} maps centered on each candidate in the {\Planck} bands, at 
 $857~\rm{GHz}$,  $545~\rm{GHz}$ and  $353~\rm{GHz}$. This aperture photometry takes into account the real extension
 of each object by integrating the signal inside the elliptical Gaussian constrained by the parameters obtained at step 1.
 The background is estimated by taking the median value on an annulus around the source. 
 Nevertheless, in 229 cases, no positive estimate of the flux has been obtained, 
 because of  the presence of  cold point sources that are too close or because the background is highly confused.
 These sources (for which the flag {\it PS Neg} is set to on) are simply removed from the physical analysis described in
this paper.

\subsection{Monte-Carlo Quality Assessment}
\label{sec_mcqa}

To assess the quality of our photometry algorithm, we have performed a Monte-Carlo analysis.
A total of 10000 simulated sources are randomly distributed over the whole sky in the IRIS and {\Planck} maps.
The sources are assumed to follow the emission of a modified black body with
a temperature randomly, $T$,  distributed between $6~\rm{K}$ and $20~\rm{K}$, and an associated spectral index
given by $\beta = 11.5 \times T^{-0.66}$ within a 20\% error bar, based on the work done on Archeops data by \citet{Desert2008}.
The FHWM of the simulated sources spans from 4.5\arcmin\ to 7\arcmin\, with an ellipticity ranging from 0 to 0.87. The flux at $857~\rm{GHz}$
is taken from 10 to 500~Jy following a logarithmic random distribution.
The derived fluxes in all IRAS and {\Planck} bands take into account the colour correction. 
We apply our complete process of photometry on this set of simulated data, and retrieve an estimate of all quantities 
(fluxes, FWHM, ellipticity) in the various cases described by the flags listed before (cf Fig.~\ref{fig:mc_analysis}). 
Statistical bias and 1$\sigma$ errors are derived for all quantities and cases, and are listed 
 in Table~\ref{tab:mc_photometry_1} and \ref{tab:mc_photometry_2} for all-sky and $|b|<25^{\circ}$ respectively.
 We  the temperature and spectral index estimates 
 recovered at the end of the processing are also listed to illustrate 
 the impact of the errors on the fluxes.

This Monte-Carlo analysis confirms,  firstly,  why sources with  {\it Aper forced} set to on  should be rejected from the physical study,
since for these sources 
fluxes are systematically under-estimated by about 60\%. Sources with  {\it Upper $100~\mu$m} set to on, 
for which only an upper limit at $100~\mu\rm{m}$ has been provided by the algorithm, the flux at $100~\mu\rm{m}$
is over-estimated by a factor of two,  with an associated discrepancy that can reach a factor of three times the input
value in regions close to  the Galactic plane.
Moreover the fluxes in the {\Planck} bands are significantly biased to lower values, 
with  a bias greater than the 1$\sigma$ discrepancy. The resulting temperature estimate is, as expected, greater than the injected
value and the uncertainties in the temperature and spectral index are around 20\%. 
This illustrates  the limitations on any physical conclusions that could be drawn from  this population of sources. When a bad fit of the 
$100~\mu\rm{m}$ background has been obtained, {\it Bad Sfit $100~\mu$m} flag set to on, 
the main error comes from the highly biased estimate of the FWHM 
($\sim$31\%), leading to an over-estimate of the fluxes in all bands. 
This happens when a strong source is embedded in a faint background (\eg\ at high latitude), introducing a degeneracy between the 
fit of the central elliptical Gaussian and the polynomial fit of the background surface at $100~\mu$m. Although bias and 1-$\sigma$
values are smaller than in the {\it normal} case due to the strong signal of these sources, 
we reject this population from the physical analysis, because they could introduce wrong estimates 
of the physical properties based on a highly biased extension.

If we focus now on the {\it normal} case, when the 
photometry algorithm has performed  well, we first observe a slight bias of all fluxes estimates. 
The bias at $100~\mu\rm{m}$ becomes larger when looking into the Galactic plane
(11.5\% for $|\rm{b}|<25$ compared to  1.4\% over the whole sky). The 
fluxes {\Planck} bands, however,  are less under-estimated  when looking inside the Galactic plane, with biases
spanning from 2.5\% to 5\%. The associated 1$\sigma$ errors are about 6 to 7\% on all-sky and 8-9\% in the 
Galactic plane. 
The impact of such a biased estimate of the fluxes will be discussed together with the study on the calibration
uncertainty in Sect.~\ref{sec_physical_properties_temperature}.
On the other hand, the FWHM estimate are typically biased by less than 1\%  and have an accuracy of $\sim$18\%,
when the ellipticity presents no bias and an accuracy of $\sim$9\%.
Finally the temperature and spectral index are derived using the method described in Sect.~\ref{sec_physical_properties_temperature}.
Whereas the temperature is slightly under-estimated ($\sim$2\% in the Galactic plane), 
the associated spectral index is over-estimated by $\sim$7\%.  
The statistical 1-$\sigma$ uncertainties are about 6\% and 8\% for $T$ and $\beta$ respectively. 
These results will be taken into account in detail
when discussing the physical properties of these cold sources in Sect.~\ref{sec_physical_properties_temperature}.

The Monte-Carlo simulations described here demonstrate  the robustness of our photometry algorithm, and justify
the rejection of entire categories of objects using the photometry flags, such as the {\it Aper Forced}, {\it PS Neg} 
and {\it Bad Sfit 100$\mu$m}. The remaining sample consists of  9465 objects, divided into two categories:
1840 objects have only an upper limit estimate of the flux at $100~\mu\rm{m}$ and 7625 have
well defined photometry in IRAS and {\Planck} bands. We will focus on this last category of 7625 sources for the rest of 
the analysis on the physical properties. Based on this Monte-Carlo analysis, we will adopt the following estimate of the 1$\sigma$ 
uncertainty on fluxes: 40\% on IRAS $100~\mu$m, and 8\% on {\Planck} bands. 
This error is much larger than the intrinsic pixel noise and so instrumental errors are  neglected.

\subsection{Cross-Correlation with existing catalogues}
\label{sec_xcheck_simbad}

\begin{figure}[t]
  \center
  \includegraphics[width=8cm]{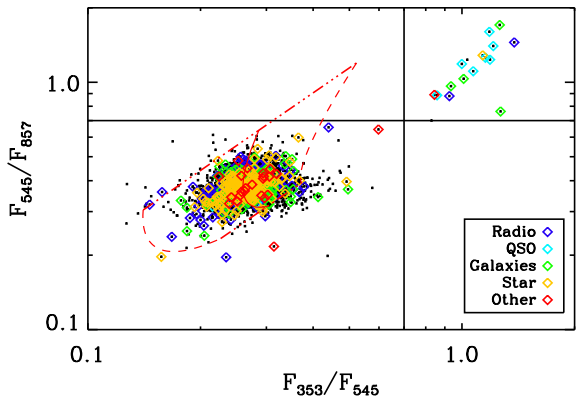} \\
  \caption{Colour-Colour diagram of the catalogue. The over-plotted symbols stand for the positive cross-matches with non ISM objects. 
  The red contours give the domain of the diagram filled by Archeops cold cores assumed to follow a grey-body law, 
  with a temperature ranging from $6\,\rm{K} < T < 25\,\rm{K}$, and a spectral index $\beta$ given by \citet{Desert2008}.}
  \label{fig:simbad_colour_colour}
\end{figure}

As one step of the validation of our detections, we have performed an astrometric
 search on the Simbad database\footnote{http://simbad.u-strasbg.fr/simbad/}
  for all known sources within a 5\arcmin\ radius of our sources. There 
are a large number of objects in the Simbad database which raises the 
question of chance alignments. This is especially true for extragalactic objects which have 
a reasonably isotropic sky distribution. To judge the number of chance 
alignments that can be expected by performing this kind of search, we have also conducted 
a Simbad cross check on the positions of a set of 100 Monte-Carlo simulated catalogues presented in Sect.~\ref{sec_groups}. 
These Monte-Carlo realizations reproduce the object density
of the {\Planck} catalogue per bin of longitude and latitude. 
The results presented in Table~\ref{tab:simbad} show that the number of coincidences in the {\it ISM} 
category is greater in the C3PO catalogue than the probability of
chance alignment estimated from  the Monte-Carlo simulations. On the contrary, the fraction of contaminants (\ie\ {\it Galaxies}, {\it QSO}, 
{\it Radio Sources}, {\it stars}) is  always lower in C3PO than in the Monte-Carlo realizations. 
Thus extragalactic objects and Galactic non-dusty objects are mostly rejected by the detection algorithm,
whereas actual ISM structures are preferentially detected. A more detailed comparison between C3PO and IRDCs catalogues 
 is presented in Sect.~\ref{sec_nature_clumps}.

Nevertheless the association with probable contaminants in C3PO is quite high ($\sim$10\%)
and not all are necessarily the result of chance alignments.
To disentangle between chance alignment and real matches, we use colour-colour
information as shown in Fig. \ref{fig:simbad_colour_colour}. Mostly  objects are distributed in the bottom-left corner of the diagram, 
typical of dust-dominated emitters. The red contours of this figure show the domain filled by dusty objects assuming a grey-body emission law, 
with $6\,\rm{K} < T < 25\,\rm{K}$, and a spectral index $\beta$ given by \citet{Desert2008}. The match between {\Planck} detections and this colour-colour domain is strong.
Only a few objects ($17$) show the  colour-colour properties of radio emitters, located in the top-right corner, 
indicating real matches with extragalactic objects. For the rest of the sample, the probability of chance alignment is high. 
Concerning the association with {\it stars}, except for a few X-ray emitters, mostly all Simbad matches seem associated with dusty emission, 
and thus represent chance alignment.

We finally reject only the obvious extragalactic matches, located in the top-right corner of the colour-colour diagram, leading to 7608 objects.

\begin{figure}
  \center
  \includegraphics[width=8cm]{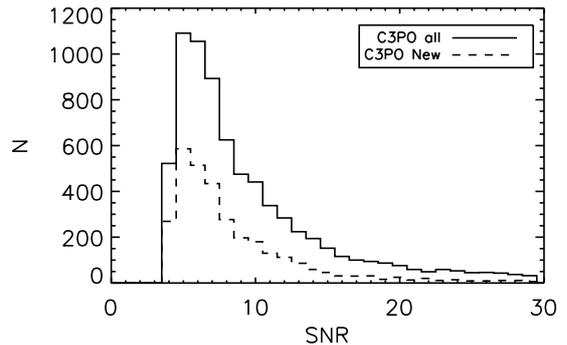} \\
  \caption{Signal-to-noise ratio (SNR) of new sources (dash line) overlaid on the SNR of all sources (solid line).}
  \label{fig:snr_new_sources}
\end{figure}

Out of the 7608 sources in the photometric reliable catalogue, 40~\% have no counterpart
 in the Simbad database. In addition, these {\it new} detections 
have a similar SNR distribution as the entire catalogue as shown 
in Fig.~\ref{fig:snr_new_sources}, and can be considered as reliable as the
entire catalogue.

\begin{table}
\begin{center}

\newdimen\digitwidth 
\setbox0=\hbox{\rm 0} 
\digitwidth=\wd0 
\catcode`*=\active 
\def*{\kern\digitwidth} 

\newdimen\signwidth 
\setbox0=\hbox{+} 
\signwidth=\wd0 
\catcode`!=\active 
\def!{\kern\signwidth} 

\newdimen\signwidth 
\setbox0=\hbox{.} 
\signwidth=\wd0 
\catcode`?=\active 
\def?{\kern\signwidth} 

\begin{tabular}{l|c|c}
\hline
\hline
Simbad type & C3PO &  $<$ MC $>$ \\
 & [\%] & [\%] \\
\hline
ISM     	&  49.0   	&    21.7 \\
Star   	& *2.3   	&   *4.9   \\
Gal    	&  *2.1    	&  *7.4    \\
Radio  	& *5.3   	&   *7.7  \\
QSO  	& *0.1   	&   *0.3   \\
Others  	&     *0.3   	&  *0.2   \\
New detections  & 40.9  & 57.8 \\
\hline
\end{tabular}
\caption{Cross match with Simbad database for C3PO and simulated catalogues,  for each category of Simbad type. 
The $<\rm{MC}>$ column gives an estimate of the probability of chance alignment for each Simbad type.}
\label{tab:simbad}
\end{center}
\end{table}

\section{Spatial Distribution}
\label{sec_spatial_distribution}

\subsection{Association with Galactic structures}

\begin{figure*}[!]
  \center
  \begin{tabular}{c}
  \includegraphics[width=16cm, viewport=70 40 510 240]{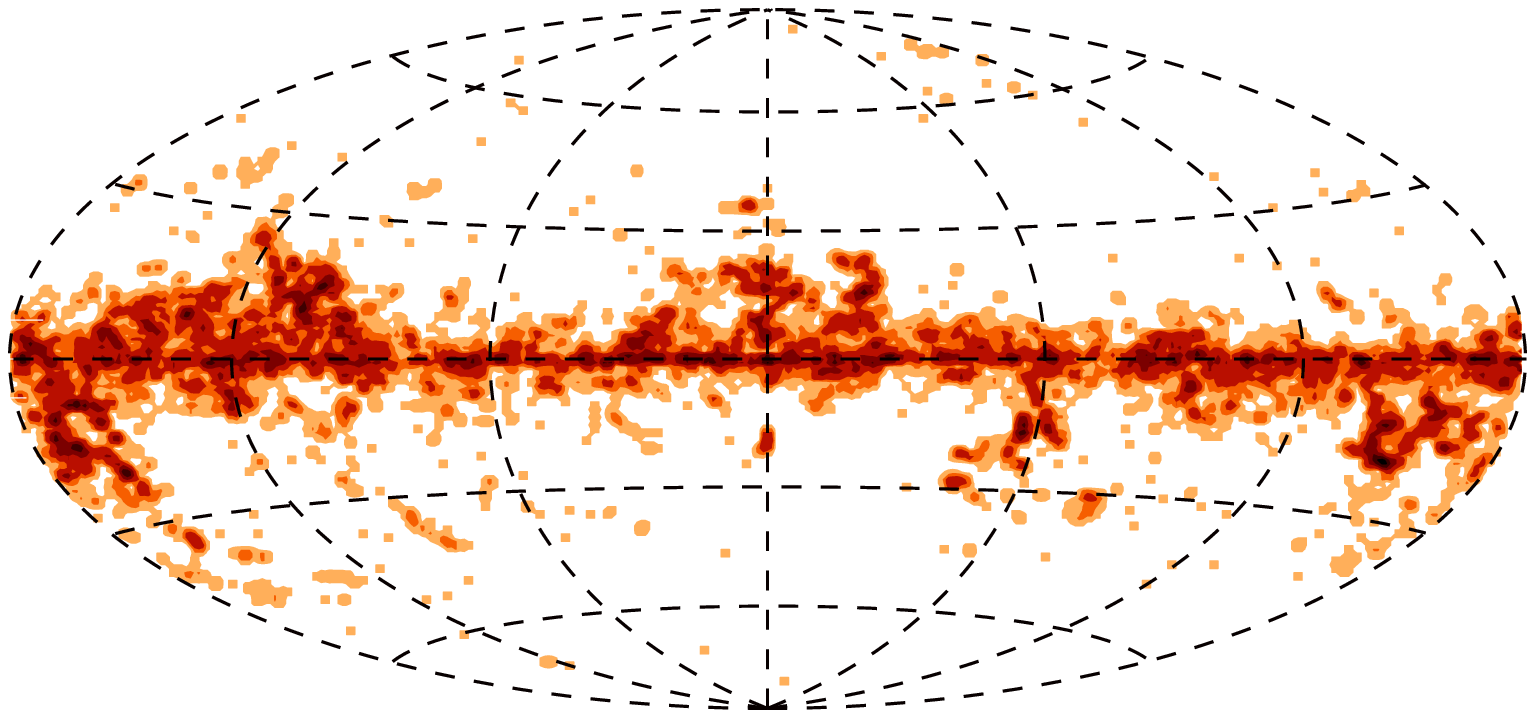} \\
{\bf Cold Core Density Map} \\
\\
 \includegraphics[width=16cm,viewport=70 40 510 240]{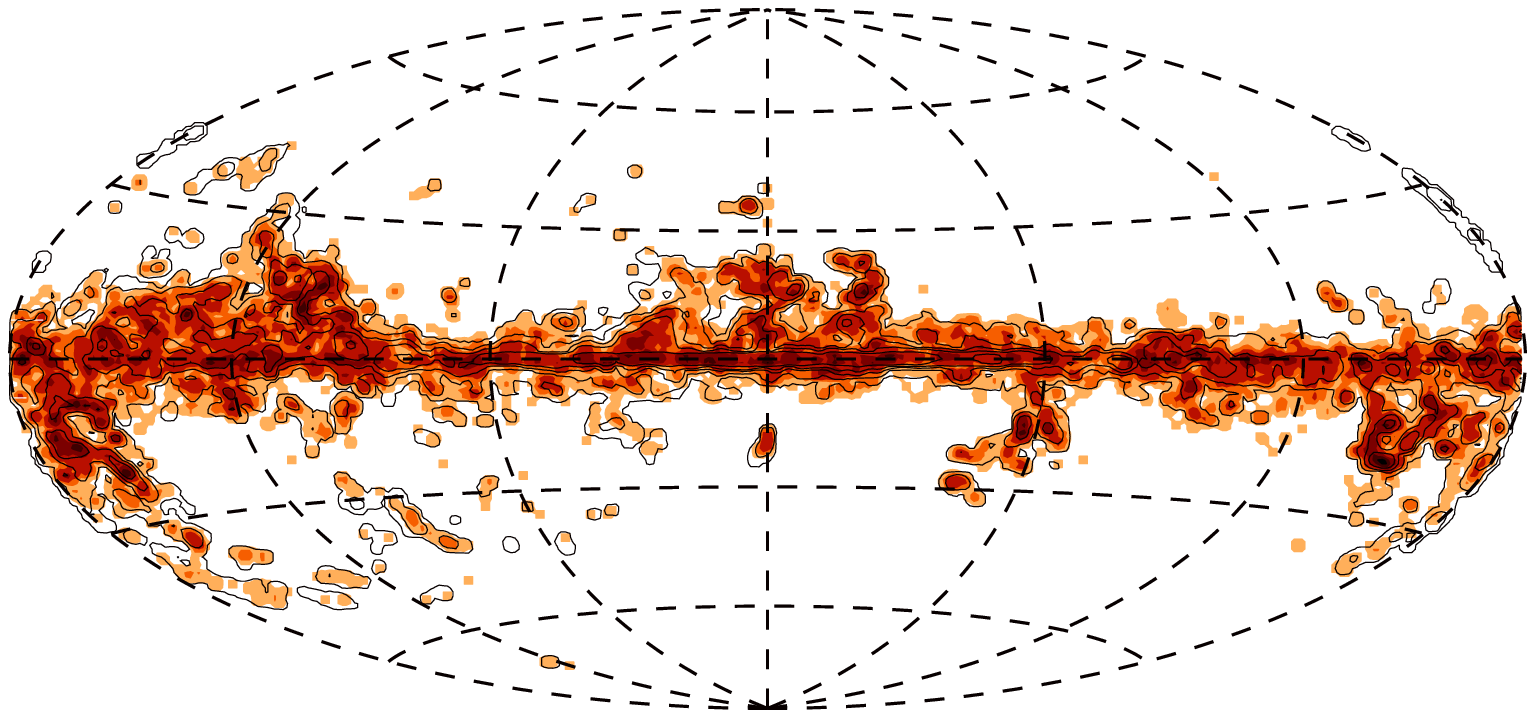} \\
{\bf CO contours on Cold Core Density Map} \\
\\
 \includegraphics[width=16cm,viewport=70 40 510 240]{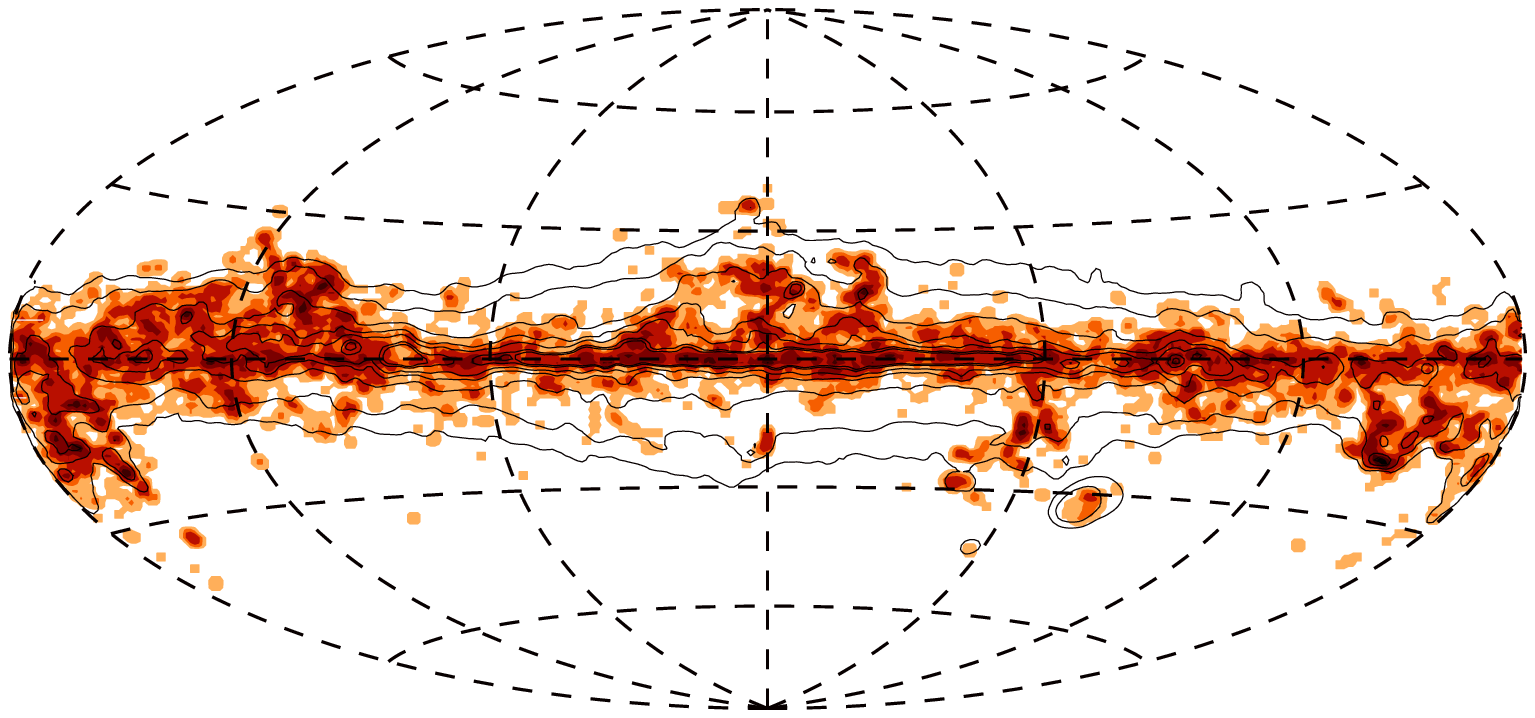} \\
{\bf Av contours on Cold Core Density Map} \\
\\
 \end{tabular}
  \caption{Upper panel: All-sky map of the number of C3PO {\Planck} cold clumps per sky area, smoothed at $3^{\circ}$. 
  Middle panel:  CO contours are over-plotted on the C3PO density map which is set to 0 where CO map is not defined.
  Lower panel: Av contours are over-plotted on the C3PO density map which is set to 0 where Av map is lower than 0.1 Av. }
  \label{fig:cc_spatial_distribution}
\end{figure*}

The all-sky distribution of the 10783 C3PO sources is presented in the
upper panel of Fig.~\ref{fig:cc_spatial_distribution}.  Mostly
concentrated in the Galactic plane, the distribution clearly follows
Galactic structures between latitudes of $-20^{\circ}$ and
$+20^{\circ}$.  A few detections are observed at high Galactic
latitude ($|\rm{b}|>30^{\circ}$) and after cross-correlation with
external catalogues have been confirmed not to be extra-galactic
objects (see Sect.~\ref{sec_xcheck_simbad}).

In the middle panel of Fig.~\ref{fig:cc_spatial_distribution},  contours of the integrated intensity map of the CO J1-0 line
are overlaid on the {\Planck} cold clumps density all-sky map. This CO map is a combination of 
CO data from \citet{Dame2001} and NANTEN data \citep{Fukui1999, Matsunaga2001, Mizuno2004}, 
as defined in \citet{planck2011-7.0}.
The correlation between CO and C3PO Cold Clumps is quite impressive 
and demonstrates once again the robustness of the detection process and the consistency of the physical nature of these {\Planck} cold objects.
A detailed analysis shows that more than 95\% of the clumps are associated with CO structures.

The lower panel shows the same kind of spatial correlation with the all-sky Av map \citep{Dobashi2011}. 
The Av map traces more diffuse regions of the Galaxy and extents to higher latitude, where cold clumps are also present. 
About 75\% of the C3PO objects are associated with an Av signature greater than 1.

\subsection{Distance Estimation}
\label{sec_distance_estimate}

Distance estimates are essential  to properly analyse the population of detected cold clumps. 
We have used four different methods: 
association with IRDCs, association with known molecular complexes, three dimensional extinction method using 2MASS data, 
and extinction method using SDSS data.

\subsubsection{Distances to IRDCs}

\citet{Simon2006b} and \citet{Jackson2008} provide kinematic distance
estimates for a total of 497 IRDCs extracted from the MSX catalogue
\citep{Simon2006a} that consists of 10931 objects.  Kinematic
distances are obtained via the observed radial velocity of gas tracers
in the plane of the Galaxy. By assuming that the Galactic gas follows
circular orbits and a Galactic rotation curve, an observed radial
velocity at a given longitude corresponds to a unique Galactocentric
radius. Of course,  this means that in the inner Galaxy, two
heliocentric distances are possible. This technique is only applicable
in the plane and requires the availability of appropriate molecular
data.  We find 127 {\Planck} cold clumps, over the complete catalogue,
associated with IRDCs that already have a kinematic distance
estimate. This number decreases to 32 associations over the 7608
objects of the {\it photometric reliable} C3PO catalogue.

A more recent work by \citet{Marshall2009} uses
an extinction method, detailed in Sect.~\ref{sec_sed_3D_ext}, on the same MSX catalogue of IRDCs to derive the distance of 1259 objects.
This yields 188 associations with C3PO clumps over the entire catalogue, and 47 over the {\it photometric reliable} C3PO catalogue.

\subsubsection{Distances to known molecular complexes}

The all-sky distribution of cold clumps follows known molecular
complexes. Many of these have distances estimates in the
litterature. To assign the distance of a complex to a particular
cold clump we use the CO map of \citet{Dame2001} to trace the
structure of the molecular cloud above a given threshold, and test
for the presence of cold clumps inside this region. The association has
been performed on 14 molecular complexes (see
Table~\ref{tab:molecular_complexes}), leading to 1152 distance
estimates over the entire catalogue  and 947 on the photometrically
reliable catalogue.
catalogue.

\begin{table}
\begin{center}
\begin{tabular}{lrrrrr}
\hline
\hline
Name & Lon & Lat & Area & Distance & Nb  \\
 & [deg] & [deg] & [$\rm{deg}^2$] & [pc] & \\
\hline
Aquila Serpens  	&    	3     &28     &   30   &    260   &     59 \\
Polaris Flare     		&	24   & 123    &   134  &     150    &    55 \\
Camelopardalis 	&   	 20  &  148     &  159    &   240     &   11 \\
Ursa Major    		& 	35   & 148      & 44     &  240     &   13 \\
    Taurus  			& 	 -15 &   170      & 883    &   140   &      393 \\
Taurus Perseus  	&  	-15  &  170     &  883     &  350    &   227 \\
   $\lambda$ Ori 	&   	-13  &  196     &  113    &   400     &   66 \\
     Orion    		& 	-9    &212       &443     &  450   &    353 \\
Chamaeleon 		&   	-16  &  300     &   27     &  150   &    114 \\
 Ophiuchus    		& 	17   & 355      & 422     &  150   &    311 \\
  Hercules     		& 	9    &45      &  35     &  300     &   16 \\
  \hline
\end{tabular}
\caption{Molecular complexes used to associate C3PO cold clumps to Galactic well-known structures, for which an estimate of the distance is available.}
\label{tab:molecular_complexes}
\end{center}
\end{table}

\subsubsection{Distances from extinction signature}
\label{sec_sed_3D_ext}

Genetic forward modelling \citep[using the PIKAIA code][]{Charbonneau1995} is used along with the 
Two Micron All Sky Survey \citep{Skrutskie2006} 
and the Besan\c{c}on Galactic model \citep{Robin2003} to deduce the three dimensional 
distribution of interstellar extinction towards the cold clump detections.
The derived dust distribution can then be used to determine the distance and mass of 
the sources, independently of kinematic models of the Milky Way.
Along a line of sight that crosses a cold clump, the extinction is seen to rise sharply at the distance of the cloud. 
The method is fully explained in \citet{Marshall2006} and \citet{Marshall2009}.

The distance,  as determined by this technique, 
provides line of sight information on the dust distribution. However, 
it does not have sufficient angular resolution to perform morphological   matches on the cold clumps. 
To ensure that the extinction rise detected along the line of sight 
is indeed related to the inner structure we perform a consistency 
check on the column density derived from the extinction and from the source flux, corrected for its temperature. 
Only detections where the two column densities are in agreement within a factor of two are retained. 
This leads to  distance estimates for 978 objects of the entire and {\it photometric reliable} catalogue.

\subsubsection{Distances from SDSS}

Distances to cold clumps within 1 kpc are obtained by analysis of
distance-reddening relations for late spectral type stars within the line
of sight to each source \citep{McGehee2011}. Specifically, we use Sloan
Digital Sky Survey photometry of M1 to M5 dwarfs colour-selected by the
reddening-invariant index 
\begin{equation}
Q_{gri} = (g-r) - \frac{E(g-r)}{E{r-i}} (r-i).
\end{equation}
The updated $ugriz$ reddening coefficients of \citet{Schlafly2010} are used. The
median stellar locus of \citet{Covey2007} forms the basis of a calibration
between $Q_{gri}$ and the intrinsic $g-i$ colour. After dereddening, the
distance to each star is determined including corrections for Galactic
metallicity variation following \cite{Bochanski2010}.

The distance-reddening profile is constructed by computing the median
reddening for stars within a circular patch centered on the core location
for 25 pc wide distance bins spanning 0 to 2000 pc. We fit the observed
reddening profile to the model defined by
convolution of the near-field plus single cloud profile with a
Gaussian (in distance modulus), this function is:
\begin{equation}
E(B-V)_{obs} = a +  c \int^{x-x_0}_{-\infty} \frac{1}{\sqrt{2 \pi \sigma^2}}
    {\tt exp}\biggl(\frac{-t^2}{2 \sigma^2}\biggr) dt
\end{equation}
where
$x$ is the independent variable (distance modulus),
$x_0$ is the location of the single cloud,
$a$ is the near-field reddening,
$c$ is the reddening associated with the cloud, and
$\sigma$ is the width of the Gaussian.
The fitted $\sigma$ values are typically 0.4 to 0.5 magnitudes in $m-M$, as
expected from the standard deviation of the $(r-z, M_r)$ used to assign
absolute magnitudes.

Analysis of calibration fields containing well-studied molecular cloud
complexes, \eg\ the Orion B Cloud, reveal that the recovered distance
moduli are underestimated by 0.2 to 0.3 magnitudes, consistent with the bias expected
from the M dwarf multiplicity fraction.

This processing leads to 1452 distance estimates over the entire catalogue and 
1004 over the {\it photometric reliable} one.

\subsubsection{Combined results}
\label{sec_distance_combined_results}

\begin{figure}
  \center
  \includegraphics[width=8cm, viewport=50 0 450 450]{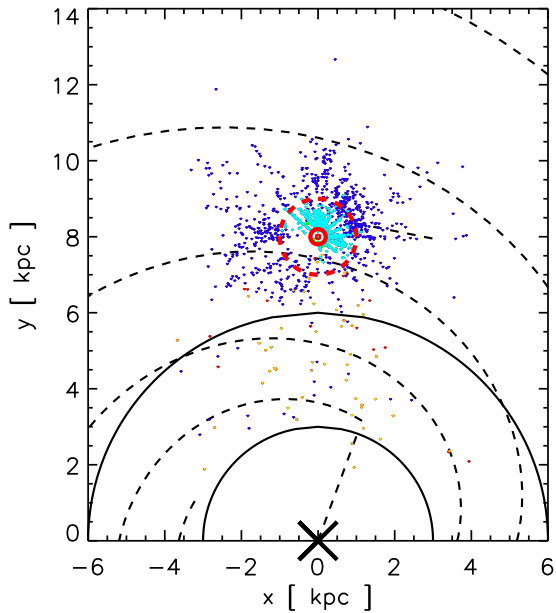} \\
  \caption{Distribution of C3PO cold clumps as seen from the North Galactic Pole.
  Colours stand for methods used to estimate distance: Molecular Complex association (green), 
  SDSS extinction (light blue), 2MASS extinction (dark blue), IRDCs extinction (orange) and IRDCs kinematic (red).
  The red dashed circle shows the 1\,kpc radius around the sun. Black dashed lines represent the spiral arms and local bar.
  The black circles give the limits of the molecular ring.}
  \label{fig:galactic_plane}
\end{figure}

The number of sources for which distances could be recovered depends on the method used (cf Table~\ref{tab:distances}).
There is some overlap but each method has its distinct advantages according to the distance range being considered.  
The 2MASS extinction method is not very sensitive nearby (D$<$1\,kpc),  as there are not enough stars to determine accurately  the line 
of sight information. In contrast, the extinction method using SDSS is especially designed for nearby objects. 
For objects with 1\,kpc, we have used SDSS distances when availble or  molecular complex distances.

\begin{figure}
\center
\includegraphics[width=8cm]{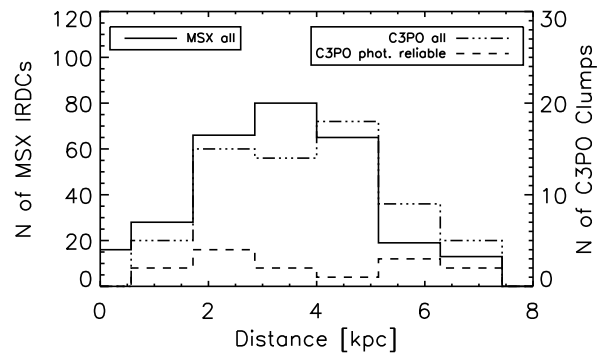}
\caption{Distance distribution of the MSX IRDCs \citep{Simon2006b} (solid line) and of the subset associated to 
the cold clumps of the entire C3PO catalogue (dot-dash-dash line) and the {\it photometric reliable} subset of C3PO (dotted line).}
\label{fig:c3po_msx_distance}
\end{figure}

The  number of objects for which we have a distance estimate is
2619 out of a total of 7608 objects in our {\it photometric reliable}
subset, \ie\ $\sim$34\%.  The distances of the cold clumps span from
0.1 to 7\,kpc, but they mainly concentrated in the nearby Solar
neighbourhood as shown on Fig.~\ref{fig:galactic_plane}.  This type of distribution 
has been already demonstrated  using simulations, see Fig.~10 of
\citet{Montier2010}.  The lack of  detections at large distances is mainly caused 
by the effects of confusion within the Galactic plane, from which suffers the
detection method.  Nevertheless, when comparing the distance
distribution of the C3PO cold clumps associated to MSX IRDCs with the
total sample of \citet{Simon2006b} in
Fig.~\ref{fig:c3po_msx_distance}, we notice that the fraction of C3PO
- IRDCs matches does not depend on distance and extends to  8\,kpc.

Because the subset of C3PO cold clumps with a distance estimate has been 
obtained using different methods,  exploring various regions and distances over the sky, 
this sample appears heterogeneous. The completeness of the catalogue with distances is quite difficult to assess. 
Thus we define two subsets for further analysis, especially when looking at number counts, 
for which we know that the sample is more homogeneous: 
the first subset (1790 objects) deals with the local objects ($D<1\,\rm{kpc}$) 
and uses only estimates from molecular complexes association and SDSS extinction; 
the second subset (674 objects) focuses on distant objects ($D>1\,\rm{kpc}$) and
 uses only 2MASS extinction estimates and IRDCs associations.

\begin{table}

\begin{center}
\newdimen\digitwidth 
\setbox0=\hbox{\rm 0} 
\digitwidth=\wd0 
\catcode`*=\active 
\def*{\kern\digitwidth} 

\newdimen\signwidth 
\setbox0=\hbox{+} 
\signwidth=\wd0 
\catcode`!=\active 
\def!{\kern\signwidth} 

\newdimen\signwidth 
\setbox0=\hbox{.} 
\signwidth=\wd0 
\catcode`?=\active 
\def?{\kern\signwidth} 

\begin{tabular}{ccc}
\hline
\hline
Method & Entire C3PO & Reduced C3PO \\
 & (10783) & (7608)\\
 \hline
IRDCs (Kinematic) & *127 & **32 \\
IRDCs (Extinction) & *188 & **47 \\
2MASS Extinction & *978 & *978 \\
SDSS Extinction & 1452 & 1004\\
Molecular Complexes & 1152 & *947\\
Total & 3411 & 2619 \\
\hline
\end{tabular}
\caption{Number of distance estimates available of the C3PO sources for each method. 
Notice that the total numbers are not equal to the sum of all methods, due to overlap between them.}
\label{tab:distances}
\end{center}
\end{table}

\section{Physical Properties}
\label{sec_physical_properties}

\subsection{Temperature}
\label{sec_physical_properties_temperature}

\begin{figure}
\begin{tabular}{c}
\includegraphics[width=8cm]{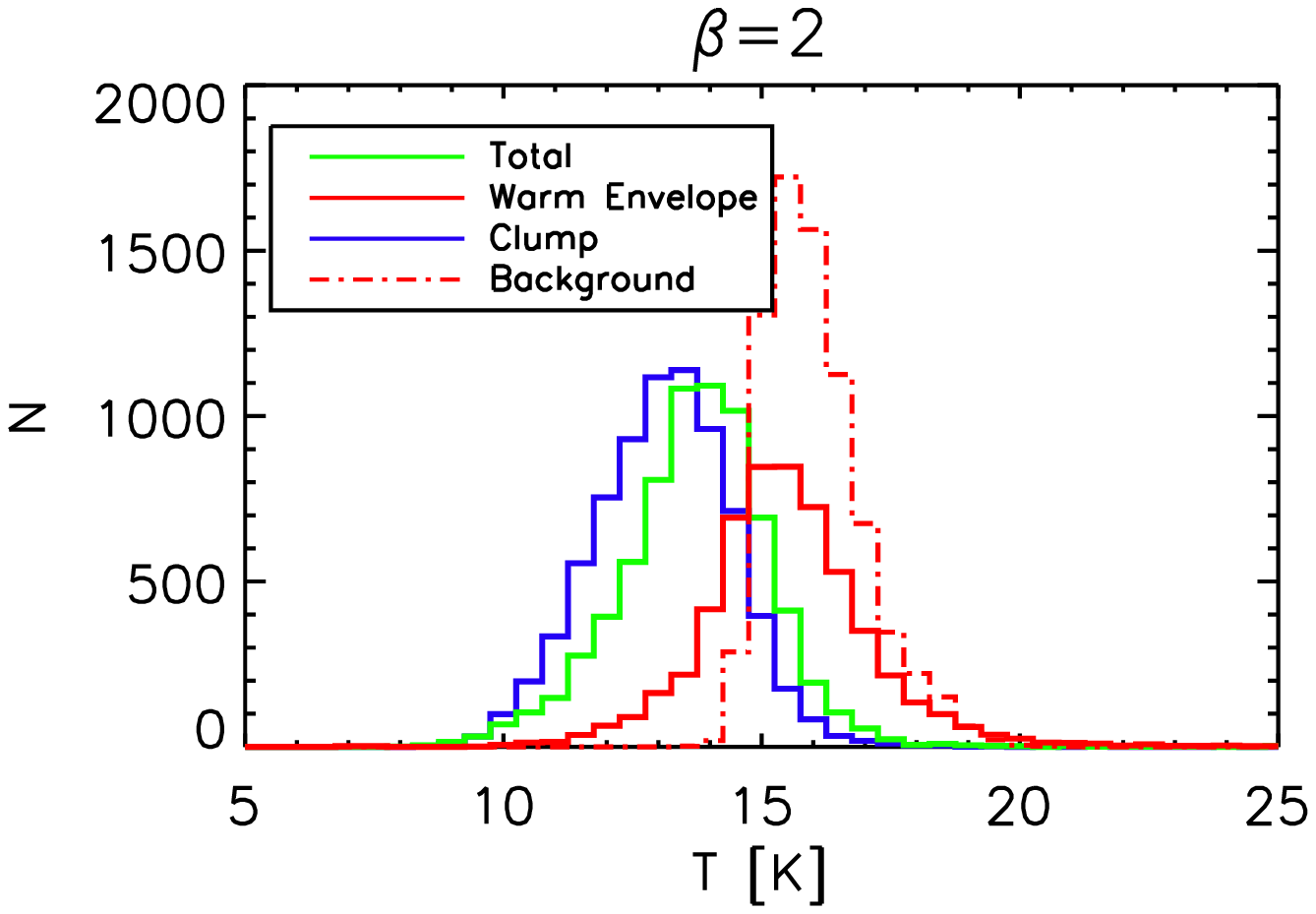} \\
\includegraphics[width=8cm]{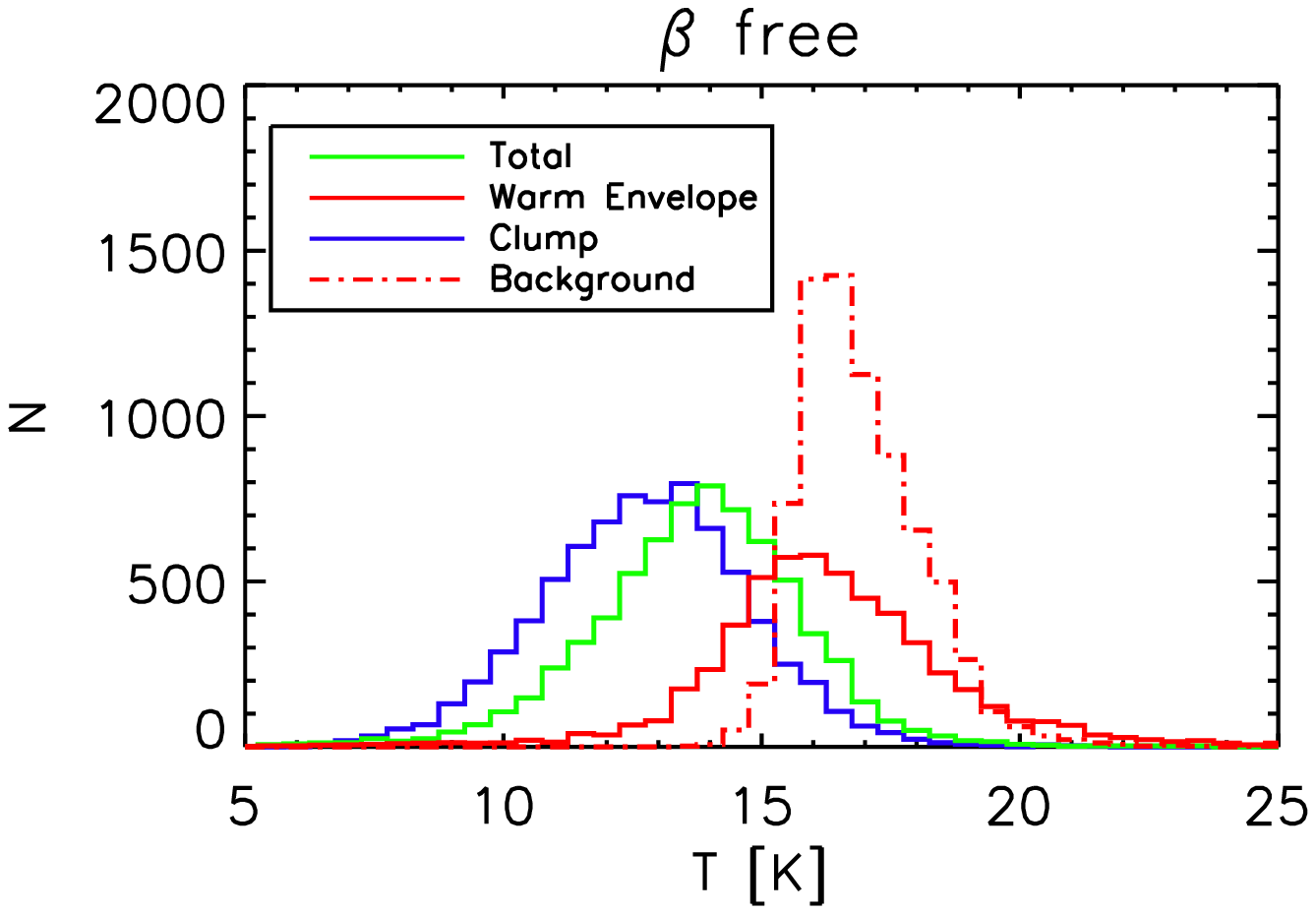}
\end{tabular}
\caption{Distribution of the temperature of the cold clumps (blue), of the warm envelope (red) and of the total (green)
estimated inside the elliptical Gaussian of the clump itself. The averaged temperature of the local background
is plotted in red dot-dash line.}
\label{fig:distribution_t}
\end{figure}

The temperature of the sources is estimated from SEDs using 4 bands: the IRAS $100~\mu\rm{m}$ and
the three highest frequency {\Planck} bands $857~\rm{GHz}$, $545~\rm{GHz}$ and $353~\rm{GHz}$. 
The assumed emission model is a modified black-body law, defined as:
\begin{equation}
S_{\nu} = A  B_{\nu}\left( T \right)  \nu^{\beta}, 
\end{equation}
where $S_{\nu}$ is the flux integrated over the solid angle $\Omega_{\rm C}~=~ \pi \sigma_{\rm Maj}\sigma_{\rm Min}$, 
$A$ is the amplitude,   $T$ is the temperature, 
$\beta$ is the spectral index and $B_{\nu}$ is the Planck function.

 For each source, a set of four temperatures is measured: (1)  the temperature of the clump $T_{C}$ is defined 
 as the temperature based on the SEDs of the {\it cold residual} as described in Sect.~\ref{sec_photometry};
(2) the temperature of the warm envelope $T_{env}$ is obtained from
 aperture photometry over the same region but performed on the {\it warm component}; (3) the total temperature
$T_{tot}$ is defined as the temperature of the source in the initial map, \ie\ without removing any warm component; 
(4) the temperature of the local background  $T_{bkg}$ is defined as the temperature of the average surface
brightness around the source.

We have first fixed the spectral index to $\beta=2$ \citep{Boulanger1996}. 
A $\chi^2$ fit is performed on the SEDs to derive all estimates of temperatures and associated 1-$\sigma$ errors. 
The distribution of these temperatures is shown on the upper panel of Fig.~\ref{fig:distribution_t}. 
The temperature of the cores $T_{C}$  (blue line) peaks at 13.4\,K and spans from 9\,K to 16\,K. 
The temperature of the total  $T_{tot}$ (green line), of the warm envelope $T_{env}$  (red line),   
and of the local background  $T_{bkg}$ (red dot dash line) distributions peak respectively at 13.9\,K, 15.1\,K and 16.1\,K.
The uncertainty on the temperature estimates is about 7\%.
These results are in good agreement with the expected values of cold cores  \citep[\eg\ ][]{Bergin2007} and consistent
with the results of our Monte-Carlo simulations demonstrating that our source extraction method
accurately recovers the cold source parameters in the presence of a warmer background.

\begin{figure}
\includegraphics[width=8cm]{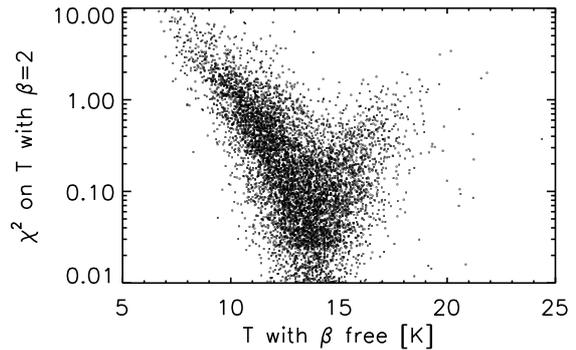} 
\caption{Reduced $\chi^2$ obtained in the case $\beta=2$ as a function of the temperature
obtained with $\beta$ free. When T becomes lower, the $\chi^2$ becomes larger.}
\label{fig:distribution_chi2_t}
\end{figure}

 In a second analysis, we performed a three parameter ($A$, $T$ and $\beta$) $\chi^2$ fit leading to the temperature distributions 
  shown in the lower panel of Fig.~\ref{fig:distribution_t}. The $\chi^2$ fit is performed on a grid taking into account the colour correction
 as defined in \citet{planck2011-1.7} and gives the exact minimum of the $\chi^2$  in the (A,T, $\beta$)
  space and providing the associated 1-$\sigma$ uncertainty. Evaluating the $\chi^2$ obtained with $\beta=2$ as a function 
  of the best fit temperature obtained from the full three parameter fits, we see that a model $\beta=2$ is
  reasonable for temperatures in the range $10\,\rm{K} < T < 18\,\rm{K}$ (for which the $\chi^2 < 1$),
  but does not provide a good fit at  lower temperature  $T<10\,\rm{K}$ (see Fig.~\ref{fig:distribution_chi2_t}). 
  In fact, the lower  the temperature, the worse  the fit.  
  Using  $\beta$ as a free parameter, the  temperature distributions peak
 at 13\,K, 13.9\,K, 15.5\,K and 17\,K for $T_{C}$, $T_{tot}$, 
    $T_{env}$ and  $T_{bkg}$  respectively, with an error of about 7\%. 
    The associated spectral index $\beta$ varies from 1.5 to 3, 
with an uncertainty of 21\% and a mean value of 2.1 for cold clumps and 1.8 for the total emission, consistent with
other studies based on {\Planck} data \citep{planck2011-7.0, planck2011-7.12, planck2011-7.13}. 
The temperature of the cold clumps span the range  7\,K to 17\,K. 
	
The bias and the uncertainty of the temperature and spectral index have to be adjusted,  taking into account
the Monte-Carlo analysis of the photometry algorithm (see Sect.~\ref{sec_mcqa}), 
and the impact of the calibration uncertainty detailed in Sect.~\ref{sec_calib_uncertainty}.
We recall that a bias of $\sim$ -2\% on T and $\sim$7\% on $\beta$ is induced by the photometry itself.
On the other hand, the calibration uncertainty of fluxes does not introduced any bias on T or $\beta$, but 
generates an error of $\sim$8\% on $\beta$ and from 3\% to 5\% on T, that should be added quadratically to the 
uncertainty due to statistical errors. All these considerations lead to a final range of temperature  spanning the range
7\,K to 17\,K with an uncertainty of about 9\%, and a spectral index $\beta$ varying from 1.4 to 2.8 with an uncertainty of about 23\%.

\subsection{Extension and ellipticity}
\label{sec_physical_properties_extension}

\begin{figure}
\center
\includegraphics[width=8cm,viewport=-50 -50 400 600]{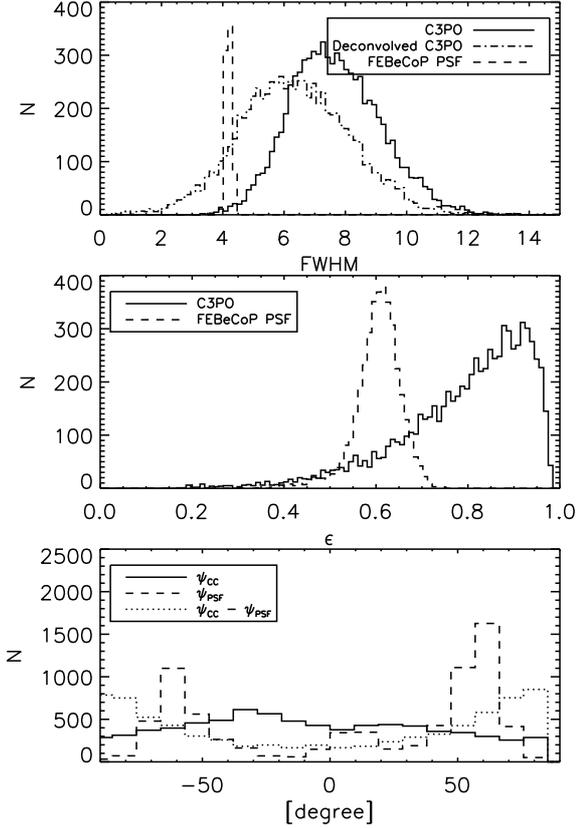}
\caption{Upper panel: distribution of the FWHM of the {\Planck} detections compared to the 
distribution of the local PSF at 857\,GHz (dashed line). Middle panel: distribution of the ellipticity of the cold clumps (solid line)
and of the local PSF (dashed line). Lower panel: distribution of the position angle of the elliptical Gaussian of the clumps
(solid line), of the local PSF (dashed line), and difference between both (dotted line).}
\label{fig:extension_ellipticity}
\end{figure}

The extension and ellipticity of the sources derived during step 1 of the photometry algorithm
described in Sect.~\ref{sec_photometry} have been compared with the local Point Spread Function
(hereafter PSF) provided by the FEBeCoP tool \citep{Mitra2010} at 857\,GHz. 
This PSF takes into account  the scanning strategy and the pixelization of the maps
at each location of the sky.
Thus for each source, an elliptical Gaussian fit is applied on the PSF smoothed at 4.5\arcmin\  to get the local FWHM $\theta_{\rm PSF}$, 
ellipticity $\epsilon_{\rm PSF}$ and position angle $\psi_{\rm PSF}$ of the effective beam. 
The FWHM $\theta$ is defined as the geometric mean of the major and minor axis widths:
\begin{equation}
\theta = \sqrt{ \left (\theta_{\rm Maj} \cdot \theta_{\rm Min} \right) } , 
\end{equation}
and the ellipticity is given by: 
\begin{equation}
\epsilon = \sqrt{1 - \left(\frac{\theta_{\rm Min}} {\theta_{\rm Maj}} \right)^2}.
\end{equation}
A few examples of FEBeCoP beams for {\Planck} HFI detectors are given in Fig.~B.5 of \citet{planck2011-7.7a}.
Fig.~\ref{fig:extension_ellipticity} compares  the statistical distributions of the FWHM (upper panel), ellipticity (middle) and 
position angle (lower panel) between C3PO sources (solid line) and the local PSF at 857\,GHz (dashed line).

Cold clumps are clearly extended, with 
an average value of  $\theta_{C}$ of 7.7\arcmin\, compared to the 4.3\arcmin\ of the average PSF over the sky.
Assuming that these compact sources are resolved by the {\Planck} beam, we can deconvolve them 
to derive the inferred intrinsic source size $\theta_i$ (dot-dash line in Fig.~\ref{fig:extension_ellipticity}):
\begin{equation}
\theta_i = \sqrt{ \theta_{C}^2 -  \theta_{PSF}^2 },  
\end{equation}
where $\theta_{C}$ is the extension of the source and $\theta_{PSF}$ is the PSF extension.
We find that $\theta_i / \theta_{PSF} \approx 1.4$, and  could conclude that we have resolved the sources.
Nevertheless, as pointed out by \citet{Enoch2007} and \citet{Netterfield2009}, 
the fact that the cold sources are mostly extended compared to the PSF 
is an indicator of the hierarchical structure of these objects. \citet{Netterfield2009} 
show that the BLAST sources present the same behavior with a ratio between
the inferred source size and the BLAST beam equal to 1.1. \citet{Enoch2007} obtained a value of 1.5 
for cold cores in Serpens, Perseus and Ophiucus observed with Bolocam.
Indeed these compact sources are associated to larger envelopes presenting radial density profiles in power law with an exponent equal to
-2 to -1 \citep{Young2003}.

Cold clumps are also mostly elongated, with a distribution of axial ratios extending to values as large as $5$ and  peaking at around 1.5,
compared to the  mean value of $1.3$ for the local PSF. Note that the C3PO 
cold clumps are not preferentially aligned with the major axis of the PSF,
the position angles of the elliptical clumps and of the PSF are uncorrelated. 
 As also stressed by \citet{planck2011-7.7a}, cold cores are often
associated with filaments and parts of larger elongated cold structures where star formation occurs. 
This was noted a long time ago  by \citet{Barnard1907} for Taurus, 
and it has more recently  been investigated by
Herschel observations in Polaris \citep{Menshchikov2010} and Aquila \citep{Konyves2010}. 
This characteristic of the cold core population can now addressed more generally using  the {\Planck} all-sky data and 
is discussed in  detail in Sect.~\ref{sec_large_scale_morphology}.

\begin{figure}
\includegraphics[width=8cm]{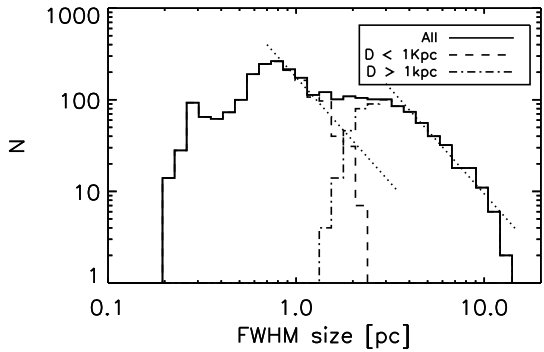}
\caption{Distribution of the physical size of the cold clumps in pc. The distinction is done between 
the local sample ($D<1\rm{kpc}$, dashed line ) and the distant sample ($D>1\rm{kpc}$, dot-dash line). 
A power law with $\alpha=-2.3$ is overlaid in dotted line over the 2 subsets.}
\label{fig:size_cc}
\end{figure}

When distances are available (see Sect.~\ref{sec_distance_estimate}), we can derive the physical size
of the sources,  defined as the FWHM in pc.
Fig.~\ref{fig:size_cc} presents the statistical distribution of the size obtained for 2619 sources. 
A distinction is made between local ($D<1\rm{kpc}$, dashed line ) and distant  ($D>1\rm{kpc}$, dot-dash line) sources
as defined in Sect.~\ref{sec_distance_combined_results}. 
The relation of  \citet{Elmegreen1996} and \citet{Heyer2001},  a size spectral index of -2.3 typical of dust clouds, 
is over-plotted on the distributions of the two subsets.

\subsection{Column Densities}
\label{sec_nh}

\begin{figure}
\begin{tabular}{cc}
\includegraphics[width=8cm]{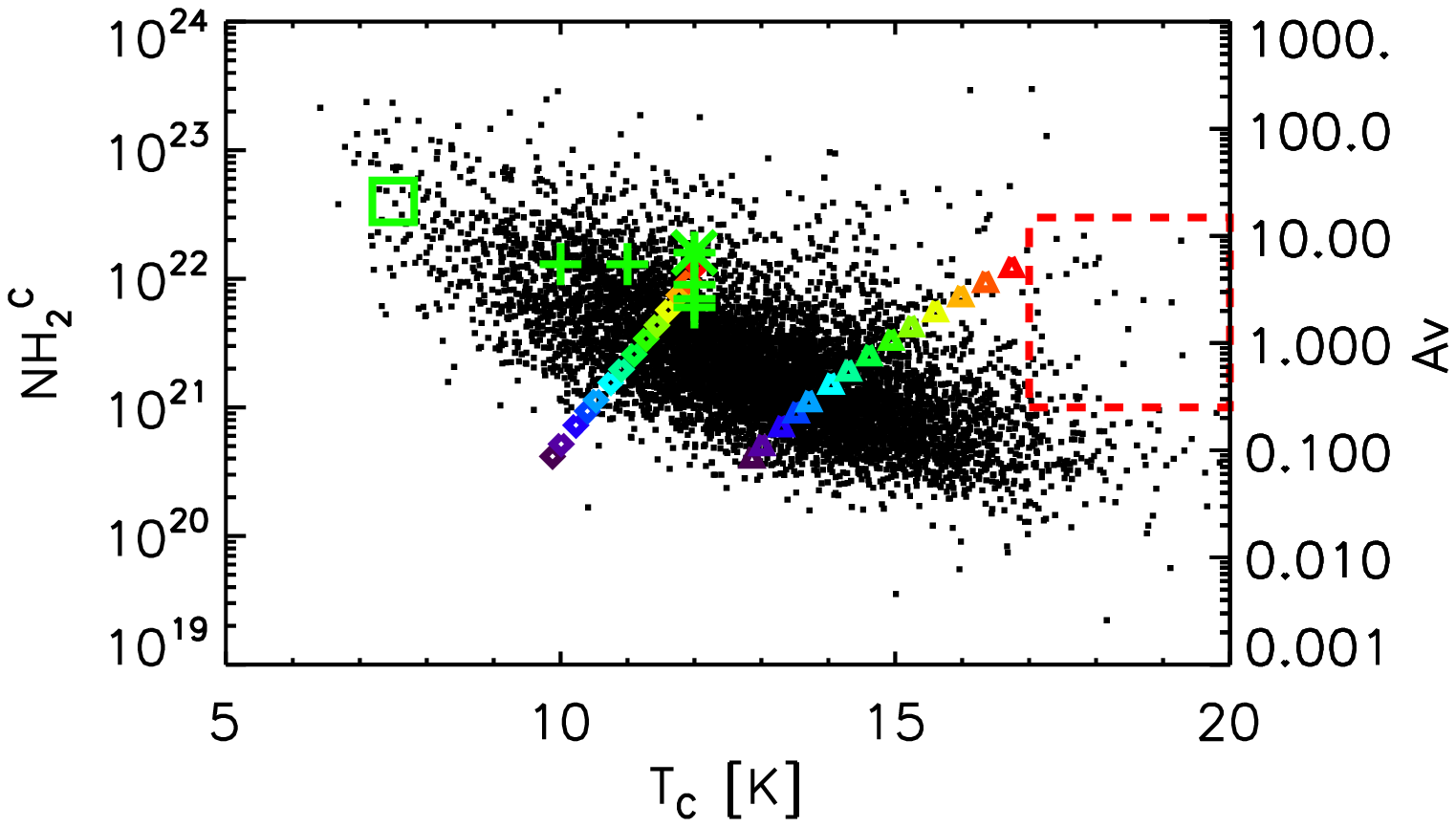} \\
\includegraphics[width=8cm]{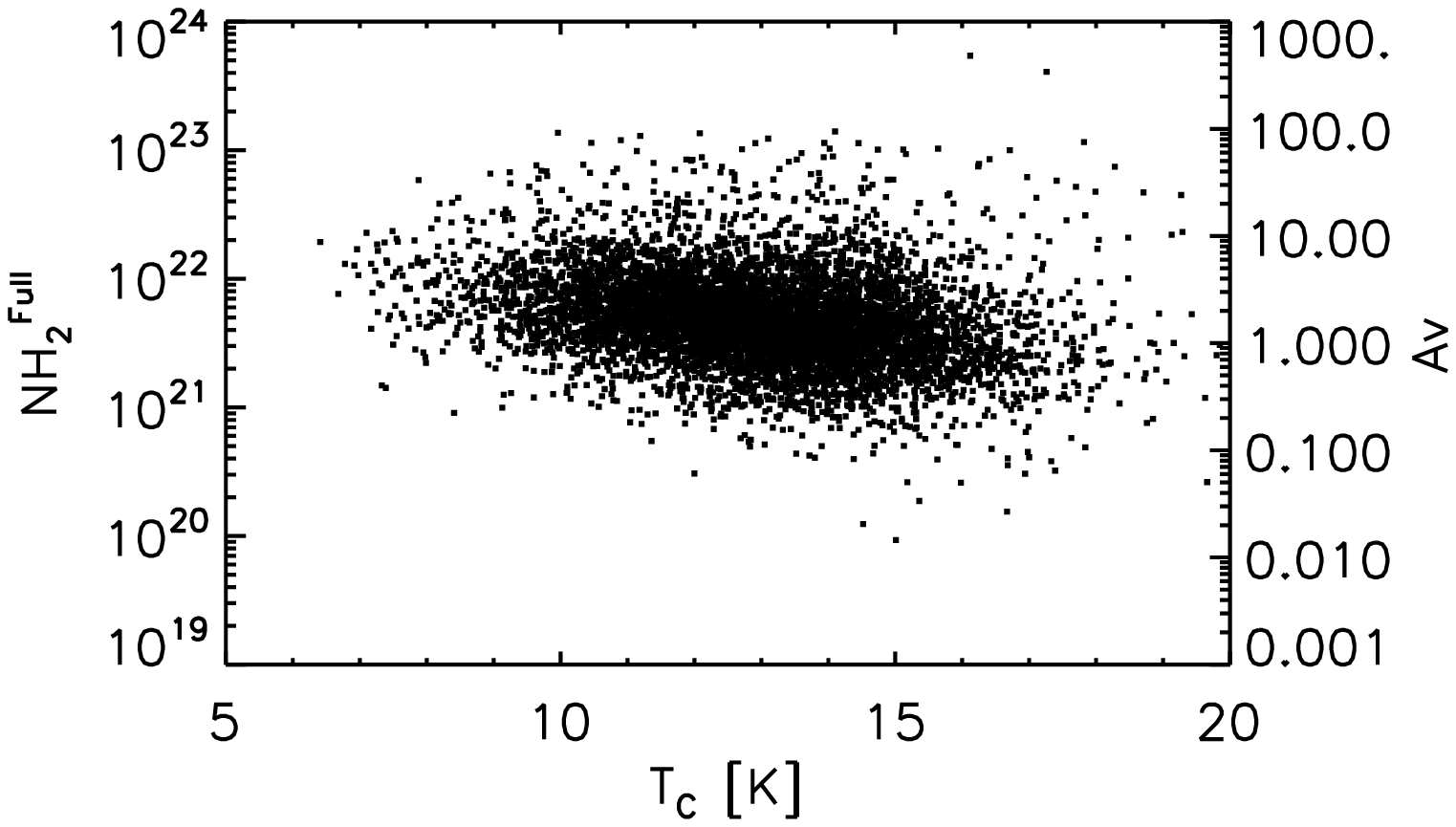}
\end{tabular}
\caption{Molecular column density of the clump itself $N_{\rm{H}_2}^{C}$ (upper panel) and molecular 
column density of the total line of sight  $N_{\rm{H}_2}^{full}$ (lower panel) as a function of the temperature of the cold clump $T_{C}$.
Modeling of Bonnor-Ebert spheres provides the temperature and column densities over-plotted
in coloured symbols (triangle and diamond) for mass spanning from 0.4 (blue) to 12 (red) solar masses. 
The triangles correspond to a normal radiation field \citep{Mathis1983} around the cold core, 
when the diamond correspond to the case of a radiation field already attenuated by external dust with Av=2.
The square, cross and plus green symbols are respectively the very cold core in L134N \citep{Pagani2004}, 
the Pronaos core in Taurus \citep{Stepnik2003}, and the starless cores of Herschel in Polaris Flare \citep{WardThompson2010}. 
The dashed red box gives the limits of the domain occupied  by the IRDCs of \citet{Rathborne2010}.}
\label{fig:nh2_t}
\end{figure}

The column density values averaged over the clump solid angle can be derived
from the integrated flux using :
\begin{equation}
N_{\rm{H}_{2}} =  \frac{S_{\nu_0}}{  \Omega_{\rm C} \mu m_{\rm{H}} \kappa_{\nu_0} \times B_{\nu_0}(T)}, 
\end{equation}
where $\Omega_{\rm C} = \pi \sigma_{\rm Maj}\sigma_{\rm Min}$ is the solid angle, $\mu=2.33$
is the mean molecular weight, $m_{\rm H}$ is the mass of atomic hydrogen, $\kappa_{\nu_0}$ is
the dust opacity (or mass absorption coefficient), and $B_{\nu_0}(T)$ is the Planck function for dust
temperature at $T$. We compute two different column densities, one for the core $N_{\rm{H}_2}^{\rm C}$
(with $T_{\rm C}$ and $S_{\nu0}^{\rm C}$) and the second for the total integrated flux along the line of site
$N_{\rm{H}_2}^{\rm full}$ (with $T_{\rm full}$ and $S_{\nu_0}^{\rm full}$) to give an  indication of the density of  the
surrounding environment.
The main source of uncertainty  here comes  from the value adopted for $\kappa_{\nu}$.
Large variations arise from one dust model to another, depending on the
dust properties considered : composition (with or without ice mantles), structure
(compact or fluffy aggregates), size... \citep[see reviews from ][]{Beckwith1990, Henning1995}. 
Dust models and observations show that $\kappa_{\nu}$ values can
vary  by a factor of 3-4 (or higher) 
from diffuse to dense and cold regions \citep{Ossenkopf1994, Kruegel1994, Stepnik2003, Juvela2010}.

For this study, we have adopted the dust opacity from \citet{Beckwith1990} in agreement with
the recommendation for dense clouds at intermediate densities ($n_{\rm{H}_2}\le 10^{5}$)
\citep{Preibisch1993, Henning1995, Motte1998}:  
\begin{equation}
\kappa_{\nu} =  0.1 (\nu/1000\rm{GHz})^{\beta} \ \rm{cm}^{2}\rm{g}^{-1}, 
\end{equation}
where we take a standard emissivity spectral index $\beta=2$. As $\nu_0$ is set to 857\,GHz
 that is close to the 1000\,GHz of the formula, the impact of variability of the spectral index $\beta$ 
 remains small compared to the uncertainty of $\kappa_\nu$.  For $\beta$ varying from 1 to 3, 
 $\kappa_{\nu_0}$ varies of a maximum of 15\% around the value obtained with $\beta=2$.

For the clumps for which the distance could be estimated, we have also determined an approximate
averaged volume density value with :
\begin{equation}
n_{\rm{H}_2}^{C} = N_{\rm{H}_{2}}^{C}/\sigma_{\rm Min}, 
\end{equation}
where the third size dimension of the object is taken as equal to the minimum value of the
clump 2D size $\sigma_{\rm Min}$. 

Fig.~\ref{fig:nh2_t} shows the column densities $N_{\rm{H}_2}^{\rm full}$ (lower panel) and $N_{\rm{H}_2}^{\rm C}$ (upper panel) .
The column density associated with the inner clump are systematically  higher than the full column density, 
because it is tracing the colder and denser phase of the medium. We also compare the observed column density of the clump
with  Bonnor-Ebert models of cold cores \citep{Bonnor1956, Ebert1955, Fischera2008} placed at 200\,pc and for masses spanning from $0.2\,M_{\sun}$ (blue) to $12\,M_{\sun}$ (red): 
The triangles correspond to a normal radiation field \citep{Mathis1983} around 
the cold core, while the diamond correspond to the case of a 
radiation field already attenuated by external dust with Av=2. 
This modeling does not match well with the observations. One explanation is first that the Bonnor Ebert sphere modeling is only valid until $M=20\,M_{\sun}$
as stressed in \citet{Montier2010}, whereas the mass range of the C3PO catalogue is much larger as detailed in Sect.~\ref{sec_mass_estimate}. 
Moreover it does not take into account the dilution inside the beam. This comparison shows also that the {\Planck} cold 
detections cannot be modeled in such a simple way and are probably more complex and extended objects.

We have also over-plotted a few other reference points of starless cores (see caption of Fig.~\ref{fig:nh2_t}). 
These few objects identified as cold cores are located in the upper distribution of the C3PO catalogue, 
in the coldest and densest part of the diagram. This underlines again the statistical property of the {\Planck} objects that have  a 
mean column density around a few $10^{21}$ hydrogen atoms per square $\rm{cm}$. This can be explained by the {\Planck} resolution 
that preferentially  selects quite extended objects, diluting objects smaller than  the 5\arcmin\  beam.
This will be discussed in detail in  Sect.~\ref{sec_nature_clumps}.
Nevertheless, we observe a few objects with column density greater than $10^{23}$, even at the {\Planck} resolution. 
These few objects could be precursors of massive stars,  or high mass formation regions.
Moreover, the locus of the IRDCs studied by \citet{Rathborne2010} is shown as a red dashed box in Fig.~\ref{fig:nh2_t}. 
This underlies the fact that {\Planck} detects clumps having the same column density but significantly colder temperatures than the IRDCs.

Finally, we observe that even the densest clumps (with high column densities) cannot reach temperature lower than 7\,K. 
This is in excellent agreement with recent observations of cold cores with Herschel (private communication).

\subsection{Mass Distribution}
\label{sec_mass_estimate}

\begin{figure}
\center
\includegraphics[width=8cm]{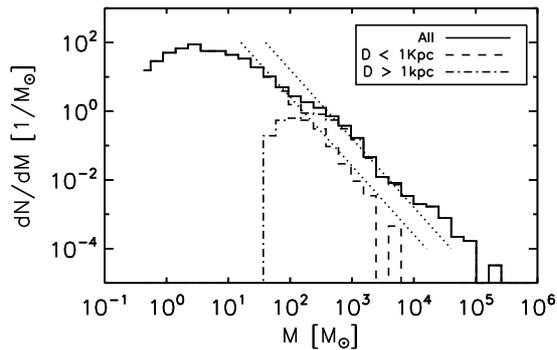}
\caption{Mass spectrum for total sample (solid line), close sample ($D<1\,\rm{kpc}$, dashed line) 
and far sample ($D>1\,\rm{kpc}$, dot-dash line). A power law $M^{-2}$ (dotted line) is overlaid for both subsets.}
\label{fig:mass_spec}
\end{figure}

The integrated mass over the clump is defined by:
\begin{equation}
M =  \frac{S_{\nu_0} D^2}{  \kappa_{\nu_0}  B_{\nu_0}(T)}, 
\end{equation}
where $S_{\nu_0}$ is the integrated flux at the frequency $\nu_0=857\rm{GHz}$, $D$ is the distance, 
$\kappa_{\nu_0}$ is
the dust opacity (or mass absorption coefficient) as defined in Sect.~\ref{sec_nh}, 
and $B_{\nu_0}(T)$ is the Planck function for dust temperature at $T$.
The range of masses of the detected cold clumps 
spans from $0.3\  {\rm M}_{\odot}$ up to $2.5\times 10^4 \ {\rm M}_{\odot}$, with a median mass of  $88 \ {\rm M}_{\odot}$.
The mass spectrum of the cold clumps is estimated by binning the mass 
distribution into logarithmically spaced bins in mass. 
The mass spectrum is then  calculated from  
\begin{equation}
  f(M) = \frac{dN}{dM} \approx \frac{N_i}{\Delta M_i} , 
\end{equation}
where $N_i$ is the number of clouds in bin $i$  and $\Delta M_i$  is the width of the $i^{\rm th}$ mass bin.

As already stressed in Sect.~\ref{sec_distance_combined_results}, it is very difficult to characterize the completeness of the catalogue
over the all-sky. The bias induced by the detection method inside the Galactic plane, due to confusion, and induced by
the various methods of distance estimate prevents any robust knowledge of the completeness of the sample. 
Thus the mass spectrum built here is not the mass spectrum of the cold core population of the entire Milky Way.
Fig.~\ref{fig:mass_spec} shows the mass spectrum of the total sample (solid line) and of the two subsets,
$D<1\,\rm{kpc}$ (dashed line) and $D>1\,\rm{kpc}$ (dot-dash line), as defined in Sect.~\ref{sec_distance_combined_results}. 
A power law,  $dN/dM \propto M^{-\alpha}$ with $\alpha=2$ is overlaid (dotted line) on each  mass function. 
We observe that the mass function for local objects is compatible with $\alpha \sim 2$ over the range 
$30\,M_{\sun} < M < 2000\,M_{\sun}$, and over the range $300\,M_{\sun} < M < 10^{4}\,M_{\sun}$  for the distant objects.
This slope  $\alpha=2$ is representative of the standard value $\alpha=2.1\pm0.4$ 
derived  for MSX IRDCs with $M>100\,M_{\sun}$ by  \citet{Rathborne2006} . 
Similar mass function have been obtained on the Pipe Nebula \citep{Alves2007}
\citep{Rathborne2008}, Perseus \citep{Enoch2006}, Ophiuchus
\citep{Young2006}, and Serpens \citep{Enoch2007}. They all derive mass function for cold cores in a range of mass
$0.5\,M_{\sun} < M < 20\,M_{\sun}$ with slopes spanning the range $\alpha=1.6$ to $ 2.77$ and peaking at around $\alpha=2.1$.
An excess of high mass objects is observed for $M>10^4\,M_{\sun}$, but the heterogeneity of our 
sample prevents any further interpretation at this stage of the analysis. A better knowledge of the completeness of the C3PO catalogue 
and a better consistency between the distance estimates are  needed.

\subsection{Luminosity}
\label{sec_luminosity}

\begin{figure}
\includegraphics[width=8cm]{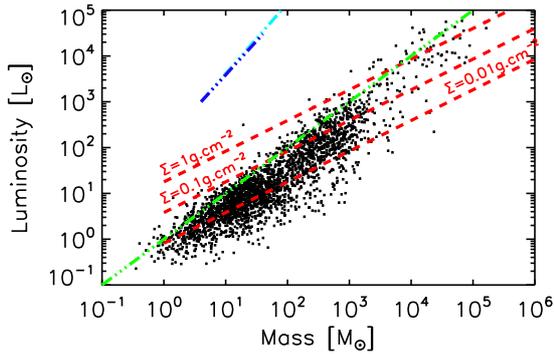}
\caption{Bolometric luminosity as a function of mass. The $L=M$ limit is over-plotted in dot-dash green line.
The loci of accretion-powered and nuclear burning phases of \citet{Molinari2008} are shown in light and dark blue lines.
The theoretical surface densities $\Sigma$ of \citet{Krumholz2006} are given in dashed red lines for 3 values:
$0.01\,\rm{g}\cdot \rm{cm}^{-2}$, $0.1\,\rm{g}\cdot \rm{cm}^{-2}$ and $1\,\rm{g}\cdot \rm{cm}^{-2}$.}
\label{fig:mass_luminosity}
\end{figure}

The bolometric luminosity is defined by:
\begin{equation}
L  =  4\pi D^2 \int_{\nu}{S_{\nu}d\nu}, 
\end{equation}
where $D$ is the distance, and $S_{\nu}$ is the integrated flux over the clump. 
The bolometric luminosity,  $L$, is integrated over the frequency range $1\,\rm{Hz}<\nu < 1\,\rm{THz}$, using the modeled SEDs derived from 
temperature and spectral index fitting (see Sect.~\ref{sec_physical_properties_temperature}).
The $L - M$ diagram is shown in Fig.~\ref{fig:mass_luminosity}. 
A large majority of the objects is located below the $L=M$ line (green dot-dash) over the whole range of mass.
The loci empirically derived by  \citet{Molinari2008} for sources in the accretion stage (light blue) and in the nuclear burning stage (dark blue)
are at least two order of magnitude above the domain covered by the C3PO clumps, indicating that 
accretion and nuclear burning  are not dominant in these sources. 

\begin{figure}
\includegraphics[width=8cm]{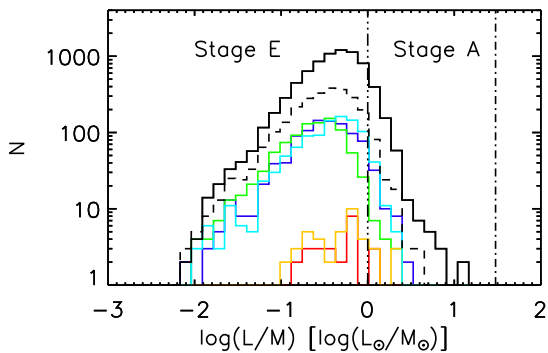}
\caption{Histogram of the $L/M$ ratio for the total C3PO catalogue (solid line) 
and the sub-sample of 2619 objects for which a distance estimate has been obtained (dashed line). The distinction is also done between
the various methods used to estimate the distance: 
Molecular Complex association 
(green), SDSS extinction (light blue), 2MASS extinction (dark 
blue), IRDCs extinction (orange) and IRDCs kinematic (red). The vertical lines indicate the theoretical frontiers 
between the three stages of the star evolution: the early stage 'E', the accretion-powered stage 'A' and the nuclear burning dominant stage.}
\label{fig:histo_mass_luminosity}
\end{figure}

The  quantity $L/M$ is very powerful in assessing the evolutionary stage of the sources, and has the advantage that it is
independent of distance:
\begin{equation}
\frac{L}{M}  =  \frac{4\pi \int_{\nu}{S_{\nu}d\nu}}{\mu m_{\rm{H}} \Omega_{\rm C} N_{\rm{H_2}}},
\end{equation}
 Fig.~\ref{fig:histo_mass_luminosity} shows the histogram of 
$L/M$ for the high-reliability C3PO catalogue (solid line), for the subset with distances (dashed line), and for all sub-samples 
corresponding to different  methods of distance estimate. Three domains are defined following the formalism of \citet{Roy2010}: 
{\it Stage E} ($L/M < 1 L_{\sun}/M_{\sun}$)  corresponding to the 'Early' stage in  which  external heating is dominant ; 
{\it Stage A} ($1 L_{\sun}/M_{\sun} < L/M< 30 L_{\sun}/M_{\sun}$) corresponding to the 'Accretion'-powered stage ; and 
 the nuclear burning dominant phase of the star formation  ($L/M> 30 L_{\sun}/M_{\sun}$) .
The C3PO clumps are mainly located in the $L/M$ domain of Stage E for all sub-samples of the catalogue, 
indicating that the nature of the objects in the catalogue is  homogeneous with  distance. 
Nevertheless about 15\% of the C3PO sources have a ratio $L/M>1L_{\sun}/M_{\sun} $ and represent a candidate population 
of evolved objects in which the accretion process has already started and so could already contain stars. 
The mean temperature of this population of clumps is around 16\,K and spans from 13\,K to 18\,K, 
indicating an internal heating due to star formation. 
For the other 85\% of sources that fall into the early stage domain, it is  difficult to assess the presence of stars inside the clumps.
The lack of angular resolution prevents us from seeing internal sub-structures, so the presence of low-mass YSOs cannot be rejected 
(as discussed in  \citet{Roy2010} for the BLAST population in CygX).

Moreover we have overlaid on the L-M diagram of Fig.~\ref{fig:mass_luminosity} 
the curves of constant surface density $\Sigma$ using the theoretical formula of \citet{Krumholz2006}:
\begin{equation}
L = 390 \left( \frac{\Sigma}{1\ \rm{g\ cm^{-2}}} \frac{M}{100\ M_{\sun}} \right)^{0.67} L_{\sun} , 
\end{equation}
for 3 values of $\Sigma$=0.01, 0.1 and 1 $\rm{g\ cm^{-2}}$ (=45, 450 and 4500 \ $M_{\sun}\ \rm{pc}^{-2}$).
Following recent theoretical work by \citet{Krumholz2008} suggesting
 that high-mass star form from clouds with $\Sigma > 1 \rm{g\ cm^{-2}}$, it appears that
 only a few {\Planck} cold clumps could be considered as precursors of high-mass stars,

\section{Large and medium scale distribution}
\label{sec_large_scale_morphology}

The spatial distribution of C3PO clumps is highly nonuniform; they seem to form arcs, 
groups and filaments (see Fig.~\ref{contour}). These large and 
small scale distribution anomalies were analysed. We performed an all-sky analysis ($|b|>5^{\circ}$) 
and we show results for both all-sky and Tau-Aur-Per-Ori region (hereafter TAPO), where C3PO 
surface density shows remarkable excess on known large scale loops and shells.

\subsection{Medium Scale Structures}

\subsubsection{Groups}
\label{sec_groups}

\begin{table}[t!]
\begin{center}

\newdimen\digitwidth 
\setbox0=\hbox{\rm 0} 
\digitwidth=\wd0 
\catcode`*=\active 
\def*{\kern\digitwidth} 

\newdimen\signwidth 
\setbox0=\hbox{+} 
\signwidth=\wd0 
\catcode`!=\active 
\def!{\kern\signwidth} 

\newdimen\signwidth 
\setbox0=\hbox{.} 
\signwidth=\wd0 
\catcode`?=\active 
\def?{\kern\signwidth} 

\begin{tabular}{llcc}
\hline
\hline
Region & & C3PO & MC \\
\hline
 & $N_G$ & 260 & $ ?161 \pm  9***?$ \\
TAPO  & $N_{G4}/N_G$ & 0.17 & $0.04 \pm 0.016$ \\
 & $\overline{\epsilon}$ & 2.67 & $2.55 \pm 0.920$ \\
 \hline
 & $N_G$ & 1833 & $?988 \pm  25**?$ \\
All-sky  & $N_{G4}/N_G$ & 0.11 & $0.06 \pm 0.007$ \\
 & $\overline{\epsilon}$ & 2.64 & $2.54 \pm 0.130$ \\
 \hline
\end{tabular}
\caption{The number and properties of identified groups in the C3PO data and the Monte Carlo simulations for the TAPO and for the all-sky. 
$\overline{\epsilon}$ means the average elongation of the groups.}
\label{tab:groups}
\end{center}
\end{table}

\begin{figure}[t]
\center
\includegraphics[width=4.3cm]{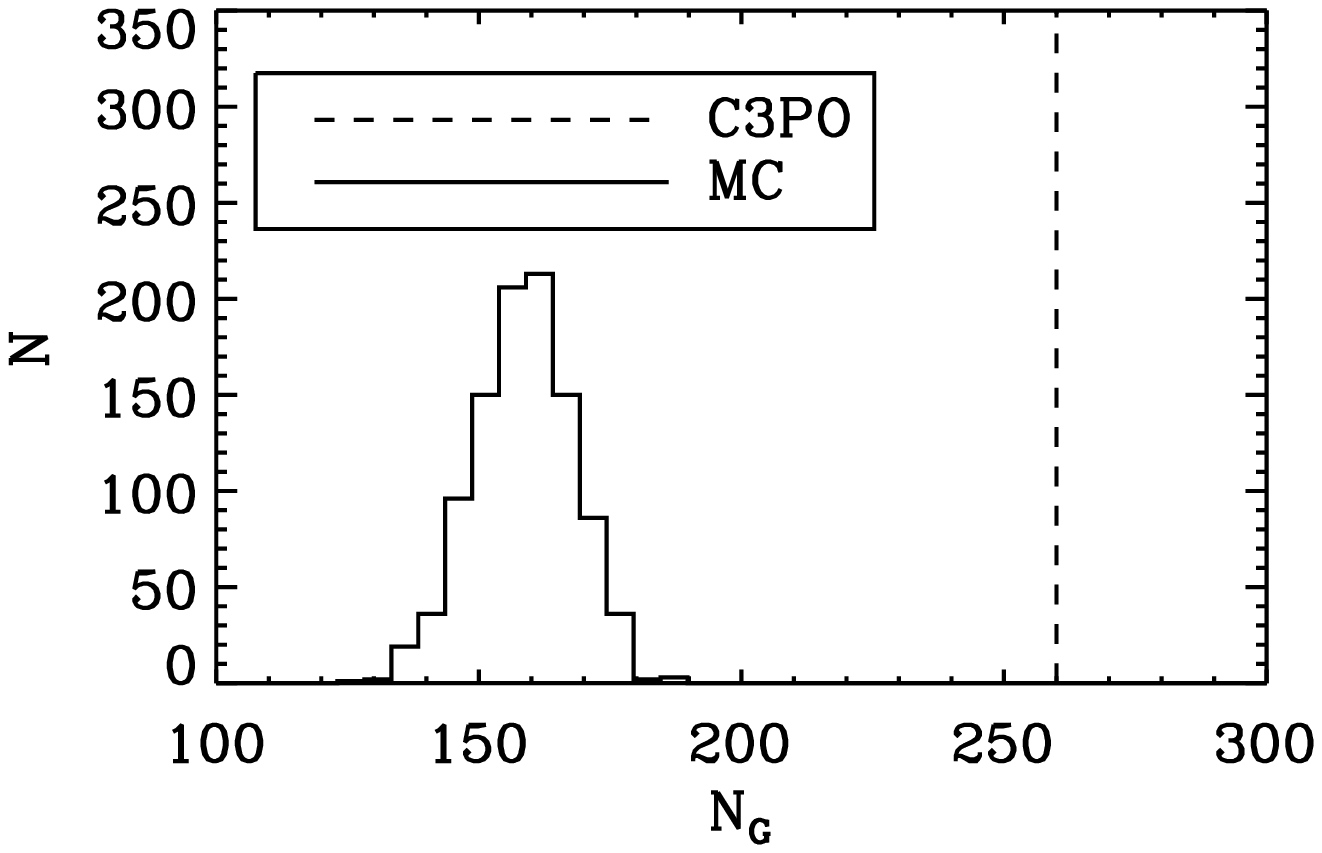}
\includegraphics[width=4.3cm]{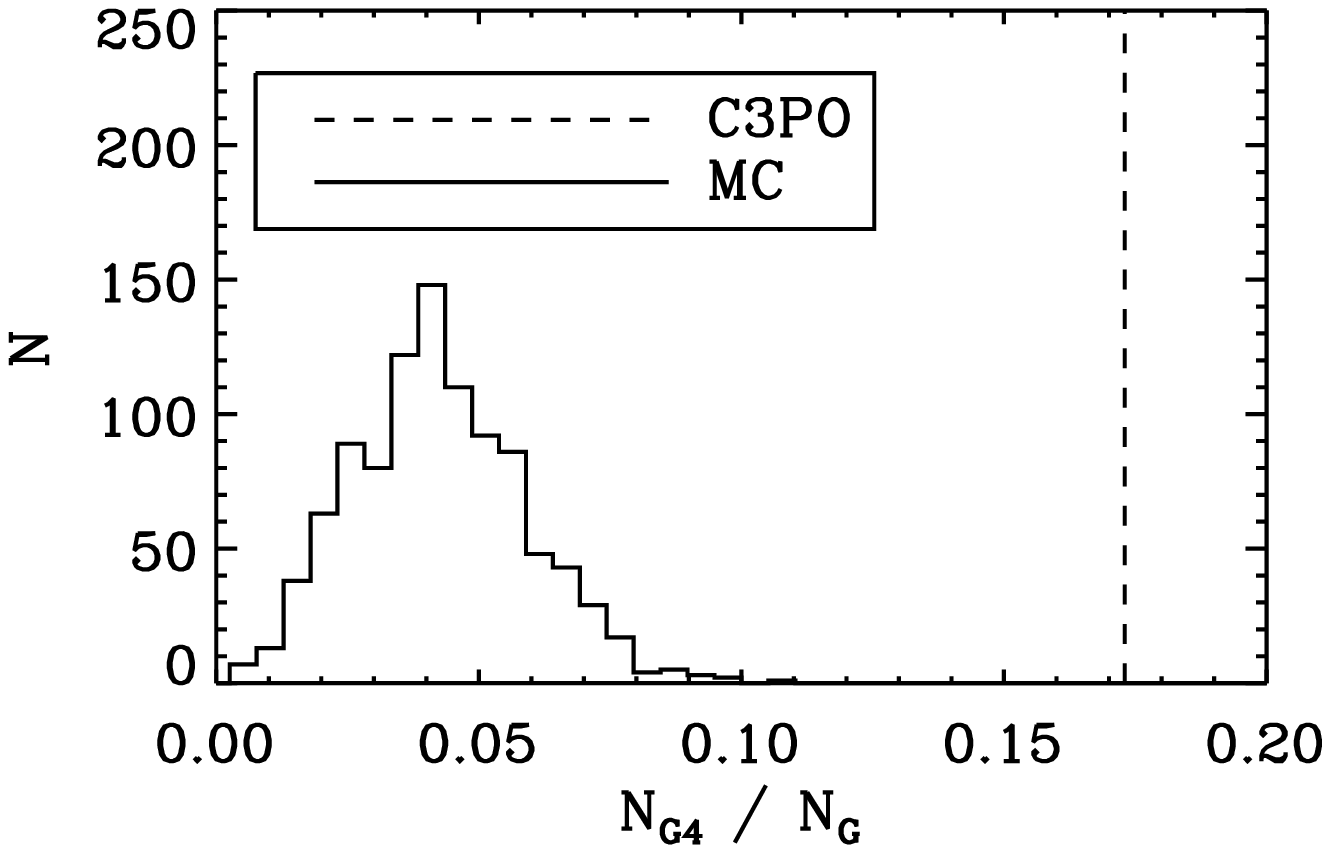}
\caption{Left: histogram of the number of groups, $N_G$ in the TAPO region in the Monte-Carlo simulations. 
Vertical dashed line shows the number of groups identified in the C3PO by using the same method in the same region. 
Right: histogram of the relative number of groups with 4 or more elements $N_{G4} / N_G$ for 1000 MC simulations 
in the TAPO region. Vertical dashed line shows the same value for the C3PO in this region.}
\label{fig:groups_fraction}
\end{figure}

We identified groups in the TAPO region using the Minimum Spanning Tree (MST)
 method of \citet{Cartwright2004} as described in \citet{Gutermuth2009} and \citet{Beerer2010}. 
A cut-off length (\ie\ maximum allowed distance between a group member core and a given subgraph) 
of 16 arcmins was used. It corresponds to the average distance between the nearest neighbours in the 
C3PO all-sky data. The number of C3PO groups, $N_G$ in the region is 260. The fraction of groups with 
more than 3 elements, $N_{G4}/N_G$ is 17\% (see Fig.~\ref{fig:groups_fraction}).
We identified groups in the all-sky data with the same method. The number of C3PO groups is 1833 and the
 value of $N_{G4} / N_G$ is 11\%.

In order to assess the reliability of the statistical estimate, we performed a Monte-Carlo analysis
 consisting in 1000 random realizations. The same number of sources as present in the C3PO 
 catalogue were randomly placed onto the sky, following the (l,b) marginal distributions of the 
 C3PO positions, averaged in latitude and longitude bins of $5^{\circ}$.

In the TAPO region in the simulated samples the average number of groups is 161, 
that is $\sim$40\% less than in the C3PO, see Table~\ref{tab:groups}. 
 There were at maximum 189 groups in the simulated samples, still 30\% less then 
 the value for C3PO. The histogram of the number of groups is shown in Fig.~\ref{fig:groups_fraction}, 
 where the value for C3PO is marked with a vertical dashed line. 
The all sky distribution shows similar grouping tendencies. The random samples show in 
average 988 groups, $\sim$47\% less then in the C3PO. We found at maximum 1083 groups in 
the simulated samples, still 40\% less then the value for C3PO.
Our result suggests that the point pattern unlikely has these many groups in random distributions. 
In the 1000 Monte-Carlo simulated samples the fraction of groups with more than 3 elements is 
in average 4\% in the TAPO region and 6\% in the all-sky and is never higher than 11\%
 in the TAPO region and 9\% in the all-sky. 
Their histogram is shown in Fig.~\ref{fig:groups_fraction}, where the vertical dashed line marks again the 
C3PO value. We note that the larger groups are more common in the C3PO than in an average random sample.

We also investigated the variation of $N_G$ and $N_{G4} / N_G $ as a function 
of the cut-off length, that we varied from 10\arcmin\  to 30\arcmin. We found that $N_G$ 
and $N_{G4} / N_G $ increase with the cut-off length. For every cut-off length the 
average $N_G$ and $N_{G4} / N_G $ in the simulated data are $\sim$25-45\% less than in the C3PO data.

\subsubsection{Filaments}

\begin{figure}[!t]
\center
\includegraphics[width=7cm]{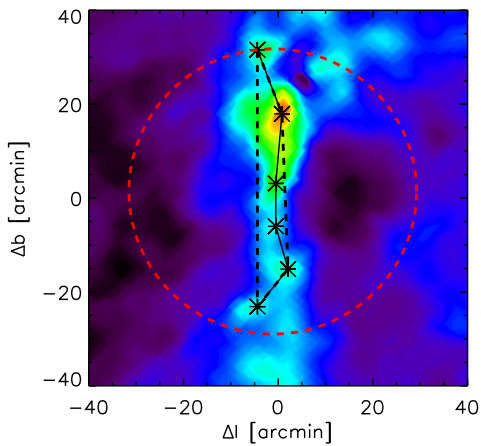}
\caption{A sample group from the C3PO in the TAPO region. Black asterisks show the clumps, black lines indicate the MST, dark dashed and dotted line indicate the convex hull and the radius of the red dashed circle is ($R^{c}$).}
\label{sample_group}
\end{figure}

The elongation of groups was analysed as described in \citet{Schmeja2006}. 
We used the \citet{Cartwright2004} definition of cluster radius: $R^{c}$ as the distance
 between the mean position of all cluster members and the most distant sources. The area $A$ 
 of the cluster was estimated using the convex hull 
 (the minimal convex set containing the set of points X in a real vector space V) of the data points. The convex hull 
 radius ($R^{h}$) is defined as the radius of a circle with an area equal to the $A$ area 
 of the convex hull of the data points.
\citet{Cartwright2004} define the elongation measure $\xi$ as follows:
\begin{equation}
\xi = \frac{R^{c}}{R^{h}}.
\end{equation}
Fig. \ref{sample_group} shows a sample group from the C3PO in the TAPO region with an elongation measure of $\xi=3.4$.

We calculated the elongation measure for all the 205 larger groups (\ie\ with more than 3 members) 
in the C3PO all-sky data. We found a mean elongation of $\sim$2.7 in the TAPO region and $\sim$2.6 in the all-sky, see Table~\ref{tab:groups}. 
The mean elongation of the filaments in the Monte-Carlo simulated samples (see Sect.~\ref{sec_groups}) 
does not differ from that in the C3PO in these regions, see Table~\ref{tab:groups}. 
We note here the very low value of $N_{G4}$ in the simulated samples.
We also investigated the mean elongation in the C3PO data for different cut-off lengths and found
that the averaged elongation of the groups is insensitive to the  cut-off length
and is always around $\sim$2.6,

\subsection{HI shells and superloops}

A first look at the all-sky distribution of the C3PO sources reveals
loop-like structures at medium and large scales.  Here we quantify
this impression and study the associations between these loops and
known shells and supershells, already stressed to be locations of star
formation. The link between the cold sub-structures detected by
{\Planck} and the star formation regions is also discussed within the
framework  of the triggering scenario \citep[e.g. ][]{Deharveng2005,
  Zavagno2010a, Zavagno2010b}.

\subsubsection{HI shells}

\begin{table}[t!]
\begin{center}
\begin{tabular}{llcc}
\hline
\hline
Region & & C3PO & MC \\
\hline
 & IN & 0.764 & $0.756 \pm  0.022$ \\
TAPO  & ON & 0.912 & $0.853 \pm 0.023$ \\
 & OFF & 0.792 & $1.058 \pm 0.054$ \\
 \hline
 & IN & 0.184 & $0.175 \pm  0.002$ \\
All-sky  & ON & 0.322 & $0.238 \pm 0.005$ \\
 & OFF & 0.056 & $0.097 \pm 0.002$ \\
 \hline
\end{tabular}
\caption{Surface density of C3PO sources and Monte-Carlo simulations for HI supershells in the 3 cases:  
IN shell, ON shell and OFF areas. Values for both TAPO region and all-sky are presented. 
For Monte-Carlo simulations, the mean value of the distribution is given with the 1-$\sigma$ discrepancy.}
\label{HI}
\end{center}
\end{table}

\begin{table}[t!]
\begin{center}
\begin{tabular}{llcc}
\hline
\hline
Region & & C3PO & MC \\
\hline
 & IN & 0.652 & $1.024 \pm  0.033$ \\
TAPO  & ON & 1.212 & $0.906 \pm 0.027$ \\
 & OFF & 0.557 & $0.600 \pm 0.023$ \\
 \hline
 & IN & 0.119 & $0.135 \pm  0.004$ \\
All-sky  & ON & 0.193 & $0.122 \pm 0.003$ \\
 & OFF & 0.144 & $0.173 \pm 0.002$ \\
 \hline
\end{tabular}
\caption{Same as Table~\ref{HI} for IRAS Loops}
\label{IRAS}
\end{center}
\end{table}

Initially we compared the distribution of C3PO sources and the HI supershells  from \citet{Heiles1984}.
This list of shells  contains HI shells, supershells 
and also shell-like objects and `worms', restricted to shells crossing $|b|=10$ or located 
entirely outside $|b|=10$. 
These objects were originally derived from two 21\,cm line surveys \citep{Weaver1973,Heiles1974}
exhibiting filamentary structure associated with high-velocity gas and radio continuum loops. 
\citet{Heiles1979} observed that HI shells do not seem to be significantly
  correlated with any other types of object, except perhaps young stellar clusters.
  
The comparison of HI shells and C3PO distribution gives the following results:
For the TAPO region,  682 (46$\%$) of the clumps are found on the shells, 635 (43$\%$) within them
and 153 (11$\%$) outside the shells. For the all-sky sample the corresponding
values are as follows: 1869 (33$\%$), 2985 (52$\%$), 873 (15$\%$). 

To assess the statistical reliability of these numbers, we performed an equivalent analysis on
 the 1000 Monte-Carlo simulations introduced in Sect.~\ref{sec_groups}. 
The simulated surface density of clumps on the shells is slightly lower ($~ 3 \sigma$) than in the real sample, see Table~\ref{HI}.

\subsubsection{IRAS loops}

\begin{figure*}[t!]
\center
\includegraphics[width=11.7cm, viewport=50 20 510 100]{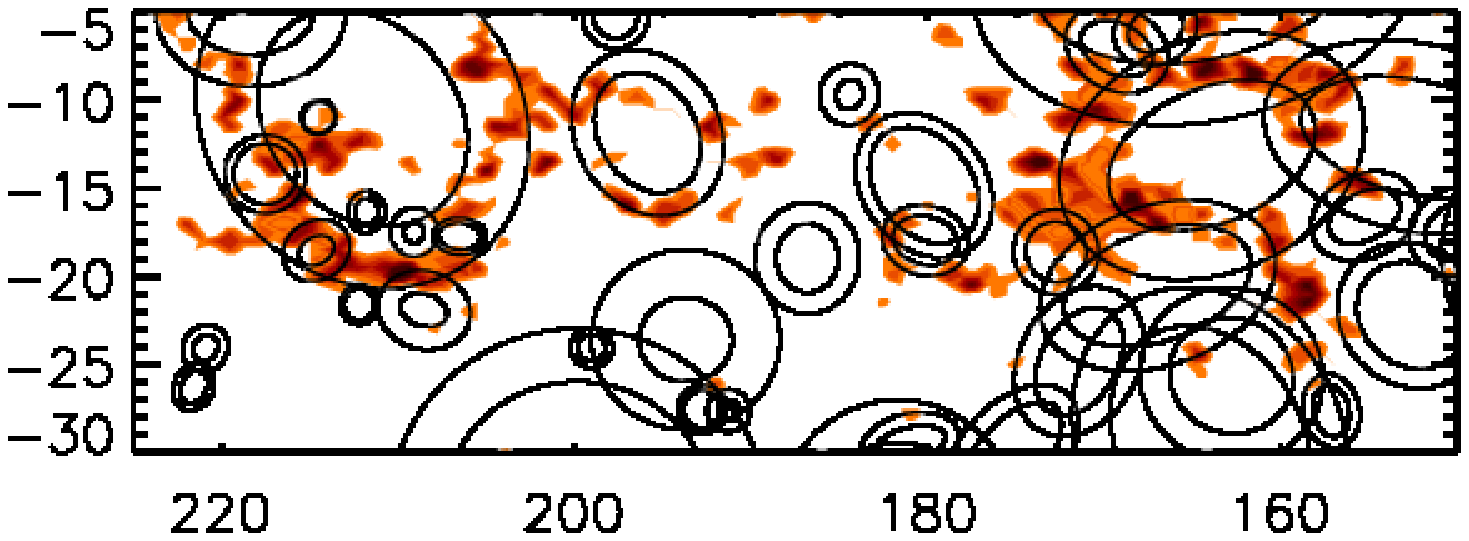}
\includegraphics[width=6.6cm, viewport=50 0 450 400]{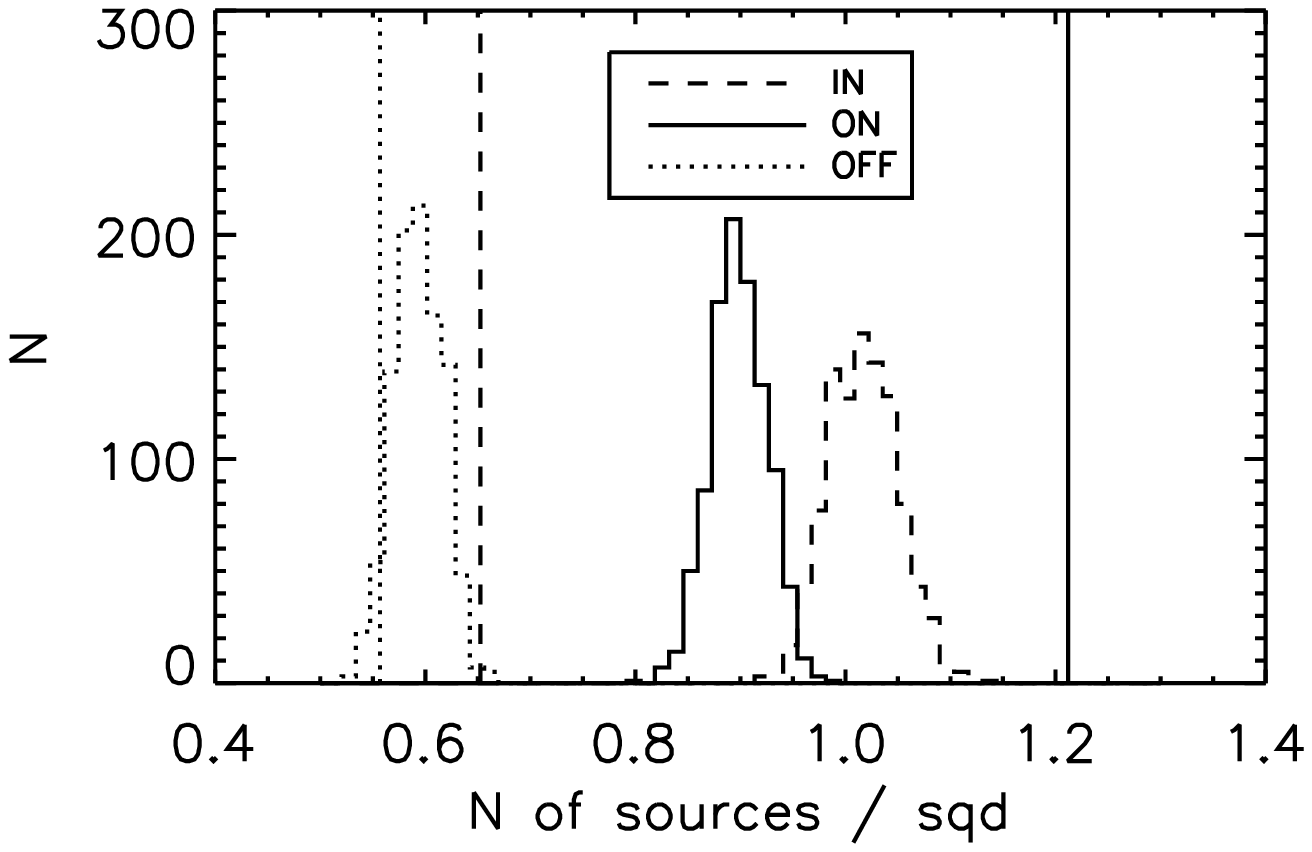}
\caption{ Left: Surface density grayscale map of the C3PO sources with the inner and outer boundaries of FIR loops \citep{Konyves2006} overlaid. 
Right: Histogram of the surface density of 1000 Monte-Carlo simulated sources for ON (solid line), IN (dashed line) and OFF (dotted line) areas. 
Vertical lines show the same for the C3PO clumps, respectively.}
\label{contour}
\end{figure*}

We also compared the distribution of {\Planck} cold clumps to the FIR
loops in the IRAS Galactic Infrared Loops of K\"onyves et al. (2006).
These IRAS loops were identified in an investigation of the large-scale
structure of the diffuse interstellar medium \citep{Kiss2004} based
on 60 and $100~\mu\rm{m}$ ISSA plates \citep[IRAS Sky Survey
Atlas][]{Wheelock1994}.  They have been identified in the Galaxy as
surfaces of high and low density ISM \citep{Toth2007}, and have been
proposed as locations for star formation \citep{Kiss2006}.  These
loops by definition must show an excess FIR intensity confined to an
arc-like feature extending to  at least 60\% of a complete ellipse-shaped
ring. Dust IR emission maps by \citet{Schlegel1998} were investigated
to derive the parameters describing the loop features
\citep{Kiss2004}.  Comparing the distribution of IRAS loops and the
C3PO clumps gives a hint of the correlation between star formation
sites and the cold sub-structures of the ISM.

Fig.~\ref{contour} shows the surface density map
of C3PO in TAPO with the FIR loop boundaries overlaid. We found 810 (55$\%$), 
312 (21$\%$) and 348 clumps (24$\%$) on the FIR loop shells, inside the loops and in the
area between the FIR loops, respectively. Values for the all-sky study are as follows: 
1928 (34$\%$), 877 (15$\%$), 2922 (51$\%$). 
We performed the same analysis as with the Heiles supershells using the  Monte-Carlo 
simulations as described in Sect.~\ref{sec_groups}, see Table~\ref{IRAS}.
We present the histogram of surface density distribution of the Monte-Carlo realizations for the IN, ON and OFF 
(between loops) areas in Fig.~\ref{contour}b. The vertical lines give the C3PO observations.
A significant ($> 30\%$) excess of C3PO clumps on the 
FIR loop shells in the TAPO region is found in  compared to the simulations.
The same behaviour is observed on the all-sky analysis (see Table~\ref{IRAS}).
 
This analysis shows that the all-sky distribution of the C3PO cold clumps
shows a significant correlation with known large scale loops,
identified to be star formation regions. It is even more reliable in
the case of IRAS loops for which this correlation is detected at
23\,$\sigma$ compared to Monte-Carlo simulations. It is detected at
15\,$\sigma$ for HI supershells on the all-sky data, but only at 2.6\,$\sigma$ in
the TAPO region.

\subsection{Triggering}

\begin{figure*}[p]
\center
\includegraphics[width=15cm]{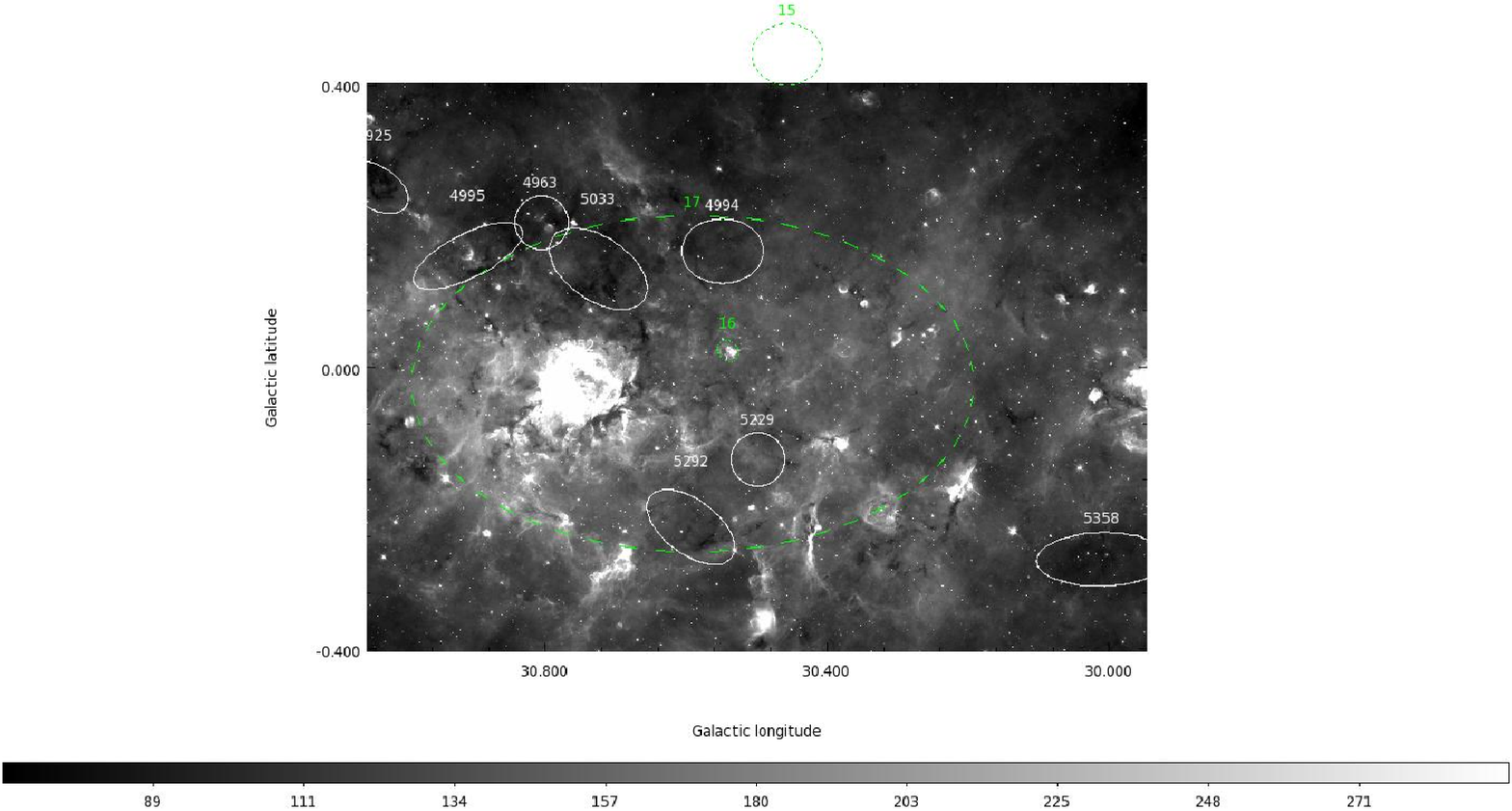}
\includegraphics[width=15cm]{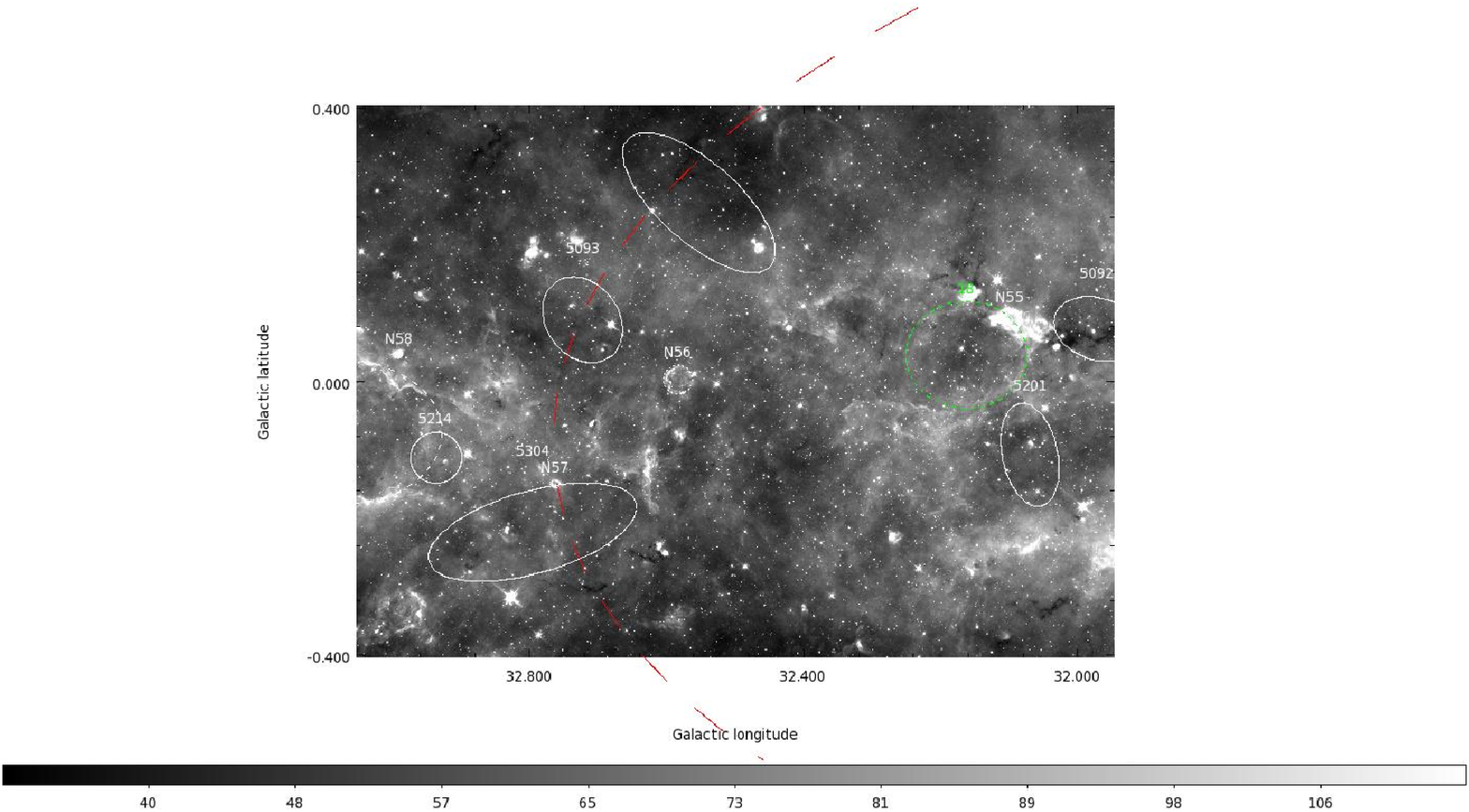}
\includegraphics[width=15cm]{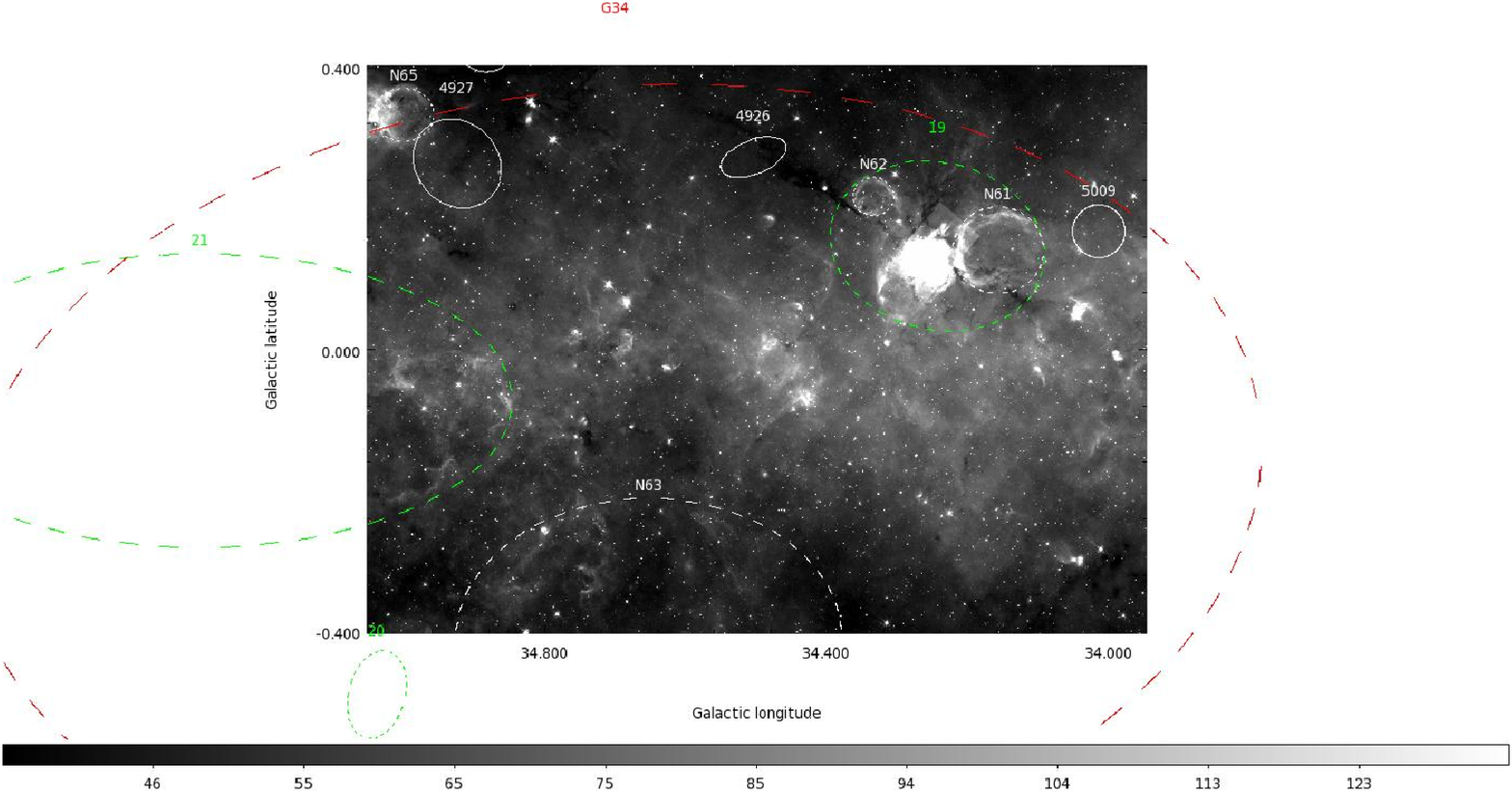}
\caption[]{Spatial location of: {\Planck} clumps (white solid line), Churchwell's bubbles (white dashed 
line), SFCs from \citet{Rahman2010} (green dashed line), WMAP sources from \citet{Murray2010}, red 
dashed line). Plain numbers correspond to entries in the C3PO catalogue, while N$\*$ are names of bubbles in Churchwell's 
compilation. Background data are from GLIMPSE 8$\mu$m.}
\label{fig:wmap_triggering}
\end{figure*}

\begin{figure*}
\center
\includegraphics[width=17cm, viewport=95 60 555 250]{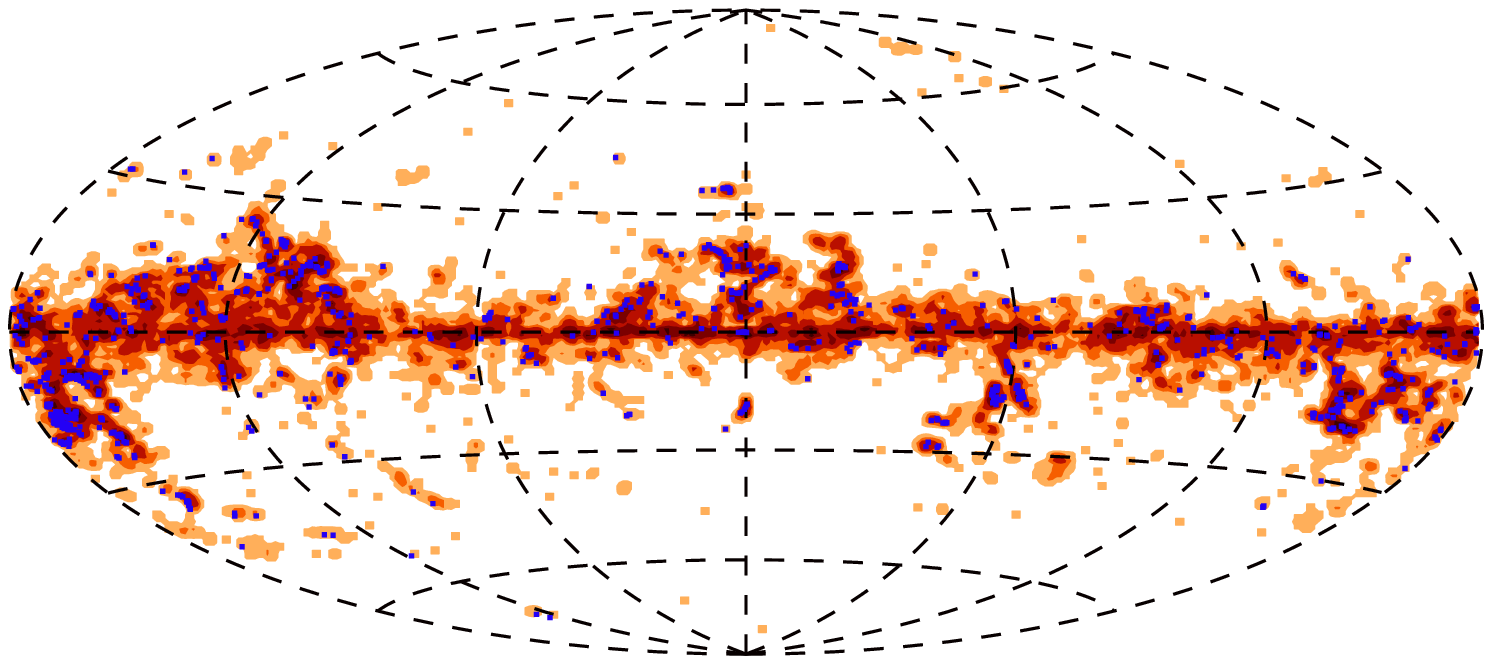}
\caption{Distribution of the ECC objects (blue squares) over-plotted on the C3PO density all-sky map.}
\label{fig:cc_ecc_spatial_distribution}
\end{figure*}

In this subsection, we explore, from a qualitative point of view, 
correlations between the spatial location of {\Planck} cold
clumps and that of candidate sources of star formation triggering. A
significant body of evidence \citep[e.g. ][]{Deharveng2005,
  Zavagno2010a, Zavagno2010b} has recently been found to support the
so-called {\em{collect and collapse}} scenario, in which the swept up
shell generated by the radiation pressure associated with  newly
born stars is the preferential location for the formation of a second generation of
stars.  Alternatively, triggered star formation may occur within
the swept up shell itself, as discussed by \citet{Elmegreen1998}.

In this study we consider, in addition to the {\Planck} cold clumps,
three classes of objects, namely: the 13 most luminous sources
recently identified by \citet{Murray2010} in the WMAP \citep[Wilkinson
Microwave Anisotropy Probe, ][]{Bennett2003a} free-free map and
thought to be responsible for the bulk of the ionizing luminosity of
the Galaxy; the 41 very massive star formation complexes (SFCs) found
by \citet{Rahman2010} to be associated with the WMAP sources; the
catalogue of bubbles by \citet{Churchwell2006}, the majority of which
are thought to be signposts of HII regions.  The ensemble of these
three classes of objects might be representative, on different angular
scales, of different stages of a triggering scenario, with the WMAP
sources being the {\em{super-giant}} bubbles (scale of multiple
supermassive star formation complexes), the Rahman $\&$ Murray's SFCs
acting as the {\em{super}} bubbles (scale of a supermassive star
formation complex), and the Churchwell et al.'s sources representing
the {\em{common}} bubbles (scale of a typical OB association).  Given
the preliminary nature of the analysis here described, and due to the
relatively low {\Planck} angular resolution, we do not take into
account smaller star formation complexes, such as compact and 
ultra-compact HII regions.

The 13 WMAP sources are all located along the Galactic Plane. As for
the SFCs of \citet{Rahman2010} and the bubbles in the catalogue by
\citet{Churchwell2006}, these have been mostly identified using the
data from the Spitzer GLIMPSE \citep[Galactic Legacy Infrared
Mid-Plane Survey Extraordinaire][]{Benjamin2003} survey, which only
covers the inner Galactic Plane (first and fourth quadrant). Taking
into account the size of the WMAP sources and the coverage of the
GLIMPSE survey, we restrict our analysis to the longitude range
10$^{\circ} <$ l $<$ 65$^{\circ}$ and 295$^{\circ} <$ l $<$
350$^{\circ}$, $|b| <$ 2$^{\circ}$.  Despite the limited area, this is
where the bulk of the Galactic star formation takes place. At the same
time, it is important to bear in mind that nearby molecular clouds
such as Perseus and Taurus, have not been included in this study.

In the restricted coordinate region defined about, there are 802
\Planck\ candidate prestellar clumps.  We compare the spatial location
of these objects using, for the C3PO clumps, the size of the major and
minor axis, and the position angles, derived from the fits, as
described in Sect.~\ref{sec_step1}. Likewise, we make use of the same
type of information, when provided by the authors, for the rest of the
sources. Since at this stage we are only interested in a qualitative,
rather than quantitative, analysis, to assess a spatial correlation we
simply overlay the positions of the {\Planck} sources with those of the other
classes of object and count the number of (partial or complete)
overlaps between them.  In total, we find that 147 {\Planck} sources
($\sim$18$\%$) appear to be spatially correlated with either the WMAP
sources or the SFCs (Fig.~\ref{fig:wmap_triggering}).  Of these, 50
($\sim$35$\%$) candidate cold clumps are located at the edges of the
{\em{super-giant}} or {\em{giant}} shells, while the remaining objects
lie within  the area delineated by the shells. The relatively low
correlation might appear surprising. However, three potentially important
caveats should be noted:  (1) the WMAP sources are visually identified
in the free-free map, hence the completeness of the sample is highly
uncertain; (2) likewise for the SFCs, given that they have been
extracted with the same technique. In particular, the completeness
will be non-uniform across the Galactic plane, with the more confused
lines of sight being the most affected by incompleteness; (3) the
thickness of the shells traced by the WMAP sources and the SFCs is not
provided by the authors. As a consequence, many C3PO clumps which are
currently near but not close enough to overlap with the other objects, might
actually be found to overlap when the thickness of the
{\em{super-giant}} and {\em{giant}} bubbles is taken into
account.

We also do not find a significant spatial correlation with Churchwell
et al.'s bubbles.  However, in this case, the lack of correlation can
be explained by the fact these bubbles and the C3PO sources
typically have comparable angular sizes. In fact, often Churchwell et
al.'s bubbles are smaller than the Planck candidate clumps, suggesting
that a triggering scenario in which the formation of a Planck clump is
induced by the expansion of a GLIMPSE bubble is  not very
likely.

We strongly emphasize that these results will have to be confirmed by
a more detailed and quantitative analysis.  In particular, effects due
to incompleteness, chance superposition and foreground/background
sources along the line of sight will be investigated using Monte Carlo
simulations, and information on the distances of all the
objects, where available.

\section{The Early Release Compact Source Catalogue subsample}
\label{sec_ecc}

\subsection{Selection Criteria}

\begin{figure*}
\center
\includegraphics[width=8cm, viewport=0 360 350 300, angle=90]{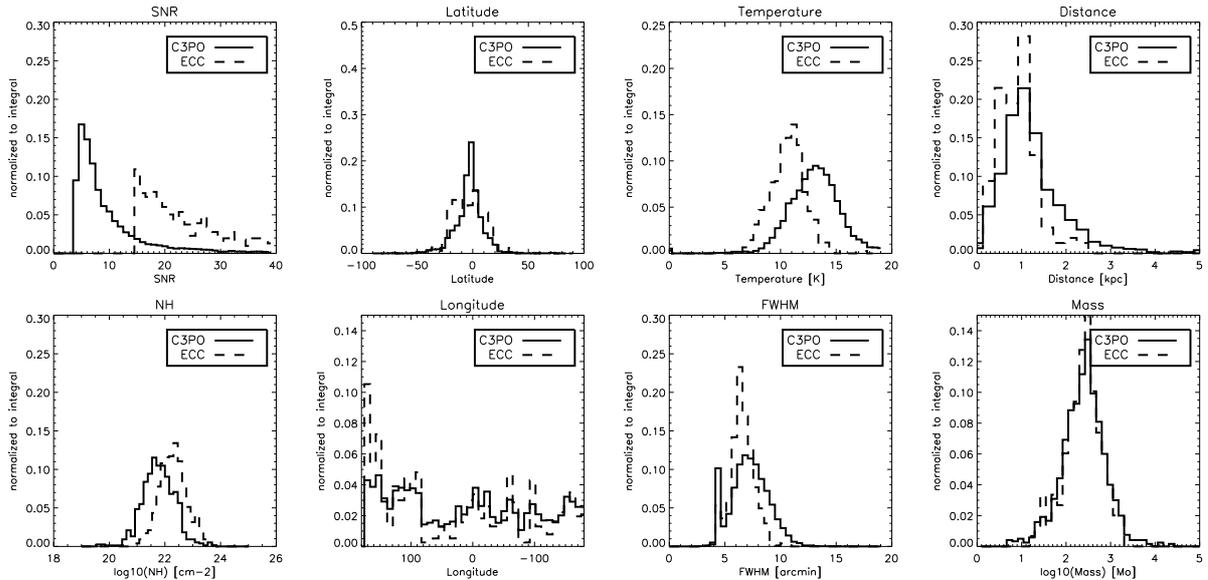}
\caption{Statistical comparison between ECC and C3PO.}
\label{fig:cc_ecc_comparison}
\end{figure*}

The Early Release Compact Source Catalogue \citep{planck2011-1.10} subsample, called the 
Early Cold Cores (ECC) catalogue, is a subsample of the full C3PO legacy catalogue. 
It is obtained by applying the following two criteria:
\begin{enumerate}
\item ${\rm SNR} > 15$,
\item $T_{\rm ECC} < 14 \,\rm{K}$,
\end{enumerate}
where ${\rm SNR}$ is the signal-to-noise ratio of the detection, and
$T_{\rm ECC}$ is the temperature derived from aperture photometry.
However, this temperature $T_{\rm ECC}$ is derived differently from the
C3PO. In the ECC, the photometry is done on the original {\Planck}
maps, \ie\ maps that still retain the warm component of the diffuse
dust emission, by placing an aperture of 5\arcmin\ radius on top of
the detection. The background is estimated from an annulus around the
aperture with an inner radius of 5\arcmin\ and an outer radius of
10$\arcmin$. Temperature is derived from a fit to all four bands used
in the detection. While the annulus is efficient in subtracting the
diffuse warm background, any warmer surrounding envelope associated
with the cold clump remains. This estimate of the temperature does not
take into account the real shape of the sources nor the
subtraction of the warm envelope linked to the clump. It  does, however, provide a
simple and straightforward estimate of the temperature that is used only for
the selection process.

The signal-to-noise cut at $\rm{SNR} = 15$ was chosen based on the Monte Carlo 
Quality Assessment (MCQA) results. a reliability of 90\% was required, where the reliability is defined as the fraction of sources 
that have a recovered flux within 30\% of that of the  injected sources.
The MCQA analysis has shown that the choice of $\rm{SNR}=15$ ensures a reliability of 90\% over the temperature 
range $T_{\rm ECC}<14\,\rm{K}$ in all three Planck wavelengths
\citep[cf Fig.~23 of ][]{planck2011-1.10sup}. Hence, the ECC is 90\% reliable, 
where the reliability is based on flux.

These selection criteria lead to a sub-sample 
of 915 objects over the sky and is distributed as shown in Fig.~\ref{fig:cc_ecc_spatial_distribution}.
Of these objects 118 have no Simbad entry and are  new detections. 
Thus at least 13\% are new detections, as discussed  in Sect.\ref{sec_xcheck_simbad}.

\subsection{Comparison with C3PO catalogue}

Fig.~\ref{fig:cc_ecc_comparison} shows the histograms of various properties of the C3PO and ECC catalogues, normalized 
to the integral of each histogram. The selection criteria (high SNR, low temperature) explain the differences
between the two catalogues.
Because of the SNR selection criteria, the ECC is largely the high SNR tail of C3PO. The clear suppression of objects in the 
Galactic plane as shown by the latitude histogram is explained by two characteristics of the detections in this
region: (1) the plane is warmer and hence CoCoCoDet is able to find less cold objects in the plane;
(2) the plane is a confused region, and the SNR tends to be lower than for more isolated sources. 
The temperature shown in Fig.~\ref{fig:cc_ecc_comparison} is not $T_{\rm ECC}$ used in the selection criteria, 
but the derived temperature 
of the clump itself $T_{\rm C}$. The ECC sources are clearly colder than 14 \,K as a rule, and even colder than the bulk of the C3PO. 

The ECC is clearly weighted towards smaller distances, where SNR is higher due to
 the flux of the cold clump, while more distant sources have lower SNR. Similar reasons 
 explain the differences in the mass histograms; while the peak is in the same place, the ECC 
 misses very low mass clumps, which have a poor SNR since they are faint, and misses very massive 
 clumps, which tend to be farther away and hence harder to detect. The histograms of hydrogen column density 
 show that ECC is weighted towards  sight-lines with higher column density, again 
 due to the temperature. The longitude histograms are complex, but the peak in the direction of the
anti-center is caused by the 
 Taurus-Perseus-Aurigae Complex, as also seen on Fig.~\ref{fig:cc_ecc_spatial_distribution}.

Finally, the extent of the clumps, measured here by the full width half maximum of the fitted 
Gaussian profile, shows that the ECC sources are shifted slightly towards smaller sizes. 
This can be partly explained by the method of photometry used, but also by the fact that 
the smaller clumps tend to be colder.

The ECC is not intended to be a complete catalogue, but a subset of high SNR, reliable detections 
of cold clumps. It is biased towards local, cold objects which have angular sizes comparable to
or smaller than the Planck beam.

\section{Discussion}

\subsection{Nature of the C3PO objects}
\label{sec_nature_clumps}

The C3PO sources have been selected by their colour properties as the relatively coldest Galactic sources over the sky. 
However their nature is still uncertain: are they cold clouds, clumps or cores? Are they really starless?
The low resolution of {\Planck} does not allow us to probe the inner structures of
these objects to answer these questions. Here we will use the statistical properties
discussed in the preceeding sections to constrain the  nature of the CP3O objects.

The analysis of the mass-luminosity relation (see Sect.~\ref{sec_luminosity}) has shown that 85\% 
of the catalogue display properties of an early stage population,
whereas the remaining 15\% could be associated with objects that have already started the accretion-powered phase. 
Thus C3PO sources are mainly pre-stellar objects, but not systematically pre-stellar cores. 
Table~\ref{tab:physical_properties} summarizes the properties of the C3PO population 
in term of temperature, mass, physical size, molecular column density and density.
Following the categorization of \citet{Williams2000} and \citet{Bergin2007} into three
populations (clouds, clumps and cores), the {\Planck} detections seem mostly associated to clumps and clouds, 
but not to cores that are defined to be smaller than 0.2\,pc, with a density $n> 10^4\rm{cm}^{-3}$ and mass $M<5\,M_{\sun}$.
Indeed a first study based on the prestellar core population 
revealed by Herschel in Aquila \citep{Konyves2010} and the {\Planck} detections in the same region
underlies the fact that {\Planck} cold clumps are often associated with at least two Herschel cores. 
The same question is addressed in more detail in Paper II \citep{planck2011-7.7a}, in which
a comparison between {\Planck} cold clumps maps and higher resolution maps obtained with Herschel follow-up has been performed.

\begin{figure}
\center
\includegraphics[width=8cm]{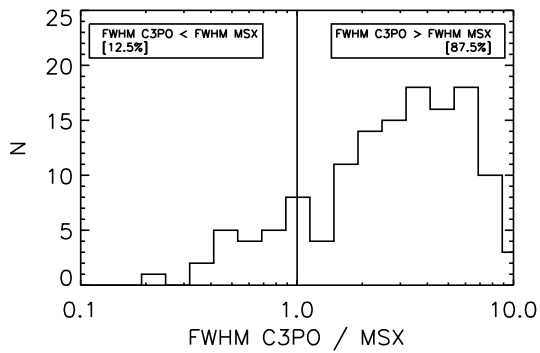}
\caption{Ratio of the angular extension of C3PO clumps and MSX IRDCs}
\label{fig:c3po_msx_fwhm}
\end{figure}

As {\Planck} cold clumps are tracing pre-stellar objects, a comparison
with IRDC properties is very instructive.  A cross-match between C3PO
sources and the MSX catalogue of 10931 IRDCs \citep{Simon2006a} leads
to 469 (136 inn the {\it photometric reliable} catalogue) {\Planck}
Cold Clumps associated with 1007 (296 respectively) MSX IRDCs.  Notice
also that less than 8\% of the {\Planck} clumps inside the MSX region
are not directly associated with an IRDC, or are located in the border
of the spatial domain covered by the MSX survey.  A comparison of the
angular size of these objects is presented on
Fig.~\ref{fig:c3po_msx_fwhm} and clearly shows that the C3PO objects
are more extended than the IRDCs, confirming their nature as clumps
and not cores. Moreover the fact that most of the IRDCs of \citet{Simon2006a}
are themselves divided into cores, gives further evidence of the nature of the {\Planck} detections that are only extended clumps.

\begin{table}
\center
\begin{tabular}{lccc}
\hline
\hline
Quantity &  min & $< >$ & max \\
\hline
$\rm{T}_{C}$ [K] & 7 & 13 & 17 \\
$N_{\rm{H}} [Hatom.cm^{-2}]$ &  $10^{20}$& $2\cdot10^{21}$ & $2 \cdot10^{23}$\\
Size [pc] & 0.2 & 1.2 & 18 \\
Ellipticity & 0.4 & 0.8 & 1\\
Mass [$\rm{M}_{\sun}$]  &   0.4 & 88 & 24000   \\
Mean density [cm$^{-3}$] & $10^2$ & $2 \cdot 10^3$ & $10^5$\\
\hline
\end{tabular}
\caption{Statistical physical properties of the {\Planck} C3PO catalogue of cold clumps.}
\label{tab:physical_properties}
\end{table}

\begin{figure}
\center
\includegraphics[width=8cm]{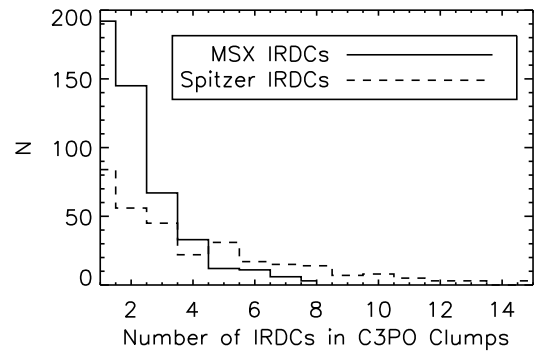}
\caption{Number of IRDCs found inside the {\Planck} Cold Clumps when a match has been obtained
with MSX catalogue \citep{Simon2006a} and Spitzer catalogue \citep{Peretto2009}.}
\label{fig:irdcs_in_c3po}
\end{figure}

We also perform a cross-match with the Spitzer catalogue of IRDCs \citep{Peretto2009}, leading to
321 (78 on the {\it photometric reliable} catalogue) {\Planck} Cold Clumps associated with 1382 (356 respectively) Spitzer IRDCs.
As shown on Fig.~\ref{fig:irdcs_in_c3po} C3PO clumps could contains up to 8 MSX IRDCs and 15 Spitzer IRDCs.
Moreover, as already stressed in Sect.~\ref{sec_physical_properties_extension}, 
the mean extension of the C3PO sources (about 1.4 times the {\Planck} beam)
is an indicator of the intrinsic scaling of $r^{-2}$ for  the profile of the extended envelope 
and an indicator of the physical link between these clumps and the larger scale  Galactic diatribution.
 All these observations  favour the fragmentation scenario \citep{Falgarone1985, Williams2000}, 
 and place the C3PO objects in the middle of this process, between clouds and cores, as cold sub-structures.

\begin{figure}
\center
\includegraphics[width=8cm]{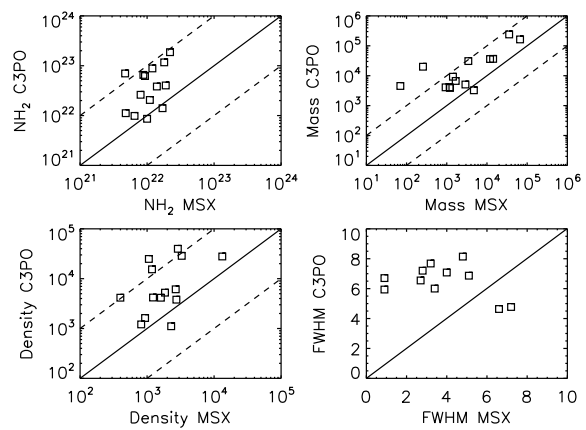}
\caption{Comparison of the physical properties between 14 sources in common in the C3PO catalogue and 
the MSX IRDCs, and for which a distance estimate is available.}
\label{fig:c3po_msx}
\end{figure}

 A detailed comparison has been performed on a smaller sample of 14
 objects for which a cross-match with MSX IRDCs has been obtained, and
 providing also kinematic distance, mass and density column estimates
 \citep{Simon2006b}.  Fig.~\ref{fig:c3po_msx} shows the comparison of
 the mass, column density, density and FWHM.  The solid line is the
 `1:1' relation and the dashed lines show the `1:10' and `1:0.1'
 relations. As already stressed above, the extension of the C3PO
 clumps is greater than for IRDCs. The C3PO clumps are 3 to 4 times
 more massive than the IRDCs, thus the mass of the cores represents
 only 25\% to 30\% of the total mass, as observed in the CygX cold
 core population \citep{Roy2010}.  More surprising is that the column
 density and the density of the clumps are about 2 to 3 times greater
 than those of the IRDCs.  This can be interpreted as a variation 
 of the dust opacity $\kappa_{\nu}$, as discussed in
 Sect.~\ref{sec_nh}. We adopted the value of \citet{Beckwith1990} whereas
 it has been shown that it can vary by a factor of 3 or more  from diffuse to dense, cold
 regions  \citep{Ossenkopf1994, Kruegel1994,
   Stepnik2003, Juvela2010}. This  could partially explain the discrepancy
 observed here.  As an example, the \citet{Ossenkopf1994} value of
 $\kappa_{\nu}$ is about twice the estimate of \citet{Beckwith1990}
 and would tend to lower the observed bias between the column
 densities of C3PO clumps and MSX IRDCs. However \citet{Simon2001}
 gave limitations on the column density estimates derived from CO
 lines that are known to be upper limits and to have an uncertainty
 of a few. This would then result in acceptable compatibility between C3PO and MSX estimates inside the error bars, 
 without exploring changes of the dust opacity values.

\subsection{T, $\beta$ relation}
\label{sec_t_beta}

The variability of the dust spectral index $\beta$ is still a
contentious issue.  Using Pronaos data \citet{Dupac2003} claimed to
find an anti-correlation of the dust spectral index with
temperature. Since then, many investigations have been carried out on
Archeops cold cores \citep{Desert2008}, on Boomerang data
\citep{Veneziani2010} and recently on Herschel data
\citep{Paradis2010}, in which the same behaviour was
found. Nevertheless, the issue is not straightforward because the
inferred  T-$\beta$ relation suffers from a strong
degeneracy between the two parameters.

\begin{figure}
\center\includegraphics[width=8cm]{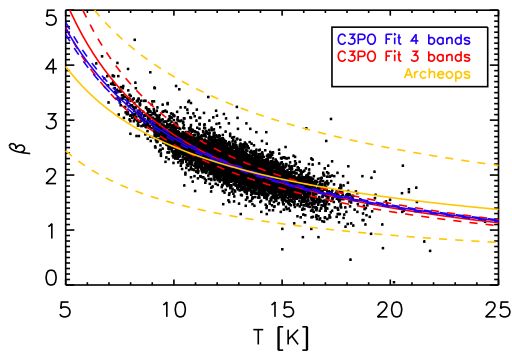}
\caption[]{The relation between the temperature and spectral index in
the catalogue and the fitted $\beta(T)$ relations. The estimates are
based on bias corrected flux values. The relation
estimated from IRAS and the three highest frequency HFI channels is shown in blue line (with 1$\sigma$ error in dashed line), 
when the red line is the relation obtained using only IRAS and two highest frequency HFI channels. 
The relation obtained on Archeops cold cores \citep{Desert2008} is shown in orange.}
\label{fig:TB_1}
\end{figure}

In the  analysis described in this paper,  flux values were first corrected for 
small biases estimated form  Monte Carlo simulation. The derived ($T$, $\beta$) values were
fitted with a formula 
\begin{equation}
\beta=(\delta + \omega T)^{-1}, 
\end{equation}
\citep[see][]{Dupac2003}. Using the IRAS 100\,$\mu$m data and the
three highest HFI frequency bands, the least squares fit gave
parameters $\delta=0.020$ and $\omega=0.035$ (see
Fig.~\ref{fig:TB_1}). When the analysis was repeated without the
353\,GHz band, the fit give parameters $\delta=0.053$ and
$\omega=0.032$. In particular, the presence of the 353\,GHz band has
only little effect on $\omega$, \ie\ on the magnitude of the temperature
dependence.

\begin{figure}
\center
\includegraphics[width=8cm]{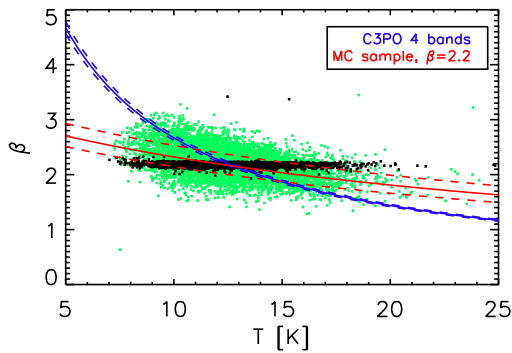}
\caption[]{
Monte Carlo analysis of the noise effects on the $\beta(T)$ relation.
The black dots correspond to a synthetic sample where the source
fluxes have been scaled to be consistent with a constant $\beta$ value.
The green points show the same sources after adding the estimated
noise in the flux values, and the red line is the fit to these points.
The observed relation (blue line) is significantly steeper.
}
\label{fig:TB_2}
\end{figure}

To estimate the uncertainty of the parameters, we carried out Monte
Carlo simulations starting with the source fluxes. The uncertainties
of the flux values were assumed to be 40\% for the IRAS 100\,$\mu$m
and 8\% for the three {\Planck} bands. In addition, for each Monte Carlo
realization of the catalogue, we included variations in the
calibration. The calibration factors  were generated from normal distribution and
were assumed to be 13.5\% for the IRAS observations, 7\% for the
545\,GHz and 857\,GHz and 2\% for the 353\,GHz
{\Planck} bands, the same scaling being always applied to the two highest frequency {\Planck} bands, 
as their calibration is based on the correlation with  the same FIRAS template 
(see \ref{sec_calib_uncertainty}). 
With the estimated statistical uncertainties, the result for the analysis of the four 
bands is $\delta=0.020 \pm 0.001$ and $\omega = 0.035 \pm 0.001$.

In addition, we used Monte
Carlo simulations to determine to what extent noise could produce an
apparent correlation between $T$ and $\beta$. We started with the
fluxes, temperatures, and spectral indices listed in the catalogue.
For each source, the fluxes were scaled with $\nu^{2.2-\beta}$ to
produce a synthetic sample in which  the spectral index is approximately
constant but where the deviations from the grey body curve are
still consistent with the observed sample. The resulting ($T$, $\beta$)
points are plotted in Fig.~\ref{fig:TB_2} as black dots. In the 
Monte-Carlo study, noise is added to the fluxes, and the observed relation is no longer flat (green
points and the red least squares line). The scatter of the points is
larger than for the observed points (Fig.~\ref{fig:TB_1}), suggesting that
the noise has  not been underestimated. 
Nevertheless, the fitted $\beta(T)$ curve is still flat compared to the relation observed in the catalogue (blue curve)
and the hypothesis of a flat $\beta(T)$ relation can be excluded
with more than 99\% confidence.

The recovered temperature dependence of the spectral index is much
stronger than the one found in PRONAOS data \citep{Dupac2003}. It is
also slightly steeper than the relation reported for Archeops cores
\citep[see Fig.~\ref{fig:TB_2} ;][]{Desert2008} although the Archeops relations
is clearly compatible with our results given the uncertainties of the
Archeops relation.

This kind of a $\beta(T)$ relation is expected for amorphous grains
with disordered structure \citep{Meny2007, Boudet2005}. Moreover
recent laboratory experiments on silica and silicates have shown a dependence of 
the spectral index on temperature and wavelength \citep{Coupeaud2011a,Coupeaud2011b}. 
As pointed out by \citet{Desert2008}, the spectral index is
expected to vary also with wavelength and this could have explained
part of the difference between PRONAOS at $200 - 500\mu$m and Archeops
at longer wavelengths, $500\mu$m--$2$\,mm. However, with data covering
wavelengths from 100\,$\mu$m to 850\,$\mu$m, we also find a very
steep $\beta(T)$ relation. When the spectral index is estimated
without our longest wavelength band (353\,GHz or 850\,$\mu$m) the
$\beta(T)$ dependence becomes marginally steeper (see Fig.~\ref{fig:TB_1}).
Although the effect is only slight, it does suggest some flattening of
the emission spectrum beyond 500\,$\mu$m.

The observed colour temperature $T_{\rm C}$ and spectral index are not
identical to the corresponding mass averaged quantities along the 
line of sight. In the presence of any observational errors, the $T$ and
$\beta$ parameters become anticorrelated 
\citep{Schwartz1982, Dupac2003}. \citet{Shetty2009a, Shetty2009b} studied
the effects of both observational noise and line of sight temperature
variations on these parameters. The colour temperature is biased
towards the warmest regions that emit more radiation. Therefore,
$T_{\rm C}$ overestimates the real dust temperature and the spectral
indices are correspondingly underestimated. With a simple model of two
dust layers \citet{Shetty2009b} showed that $T_{\rm C}$
may rise above the real dust temperature anywhere along the
line of sight. Similar results were obtained by
\citet{Malinen2010} in connection with more complex models that
combined magneto-hydrodynamic  simulations with radiative transfer modeling. 
 \citet{Malinen2010} concluded that the spectral indices become more biased close to
local radiation sources and this could affect the derived $\beta$--$T$
relation when the sample also includes star forming clouds. 
 Because of all
these effects, some caution is needed when interpreting the value and variation of the observed spectral
index.
However we see observationally the dust emission spectral index $\beta$ increasing towards the cold clumps and, because temperature variations 
tend to decrease $\beta$ especially towards the clump where the temperature variations are the largest, 
the actual variations in dust properties may be even more pronounced.

\section{Conclusion}

We have applied a dedicated source extraction algorithm, {\it
  CoCoCoDeT} \citep{Montier2010}, to the {\Planck} data combined with
the IRAS $100~\mu\rm{m}$, to build a robust catalogue of cold sources
over the whole sky.  This Cold Core Catalogue of {\Planck} Objects
(C3PO) is the first objectively selected all-sky catalogue of cold objects. 
It has been built using the local colour signature of relative colder objects
embedded in a warmer background. We stress that this method could lead to  
missing sources due to an already cold background or spurious warm detections due to
very hot backgrounds. 
The catalogue consists of  10783 objects, from which we have selected the 7608 sources
with  a complete set of robust physical characteristics such as
fluxes, temperature, angular extension, ellipticity and 
column density.  A second sub-sample has been constructed  with 2619 sources
with  distance estimates, physical sizes, masses, and
densities. In this paper we have performed a  statistical analysis of this
complete sample of cold objects and analysed their  physical properties.
 
A dedicated method has been applied to derive the photometry of the
clumps themselves, and to look at their local properties inside the warmer
envelopes embedded in the Galactic environment. The temperature of the
{\Planck} cold clumps spans from 7\,K to 17\,K and peaks at around 13\,K,
in agreement with previous studies.  The advantage of the C3PO
catalogue is that it provides a high number of very cold objects,
about 600 objects have a temperature $T_{\rm C}<10\,\rm{K}$. 
It has been shown that the data are not consistent with a constant value of the dust spectral index 
$\beta$ over the whole range of temperature. Several possible scenario are possible, such as the effect
of multiple temperature components folded into the measurements, and also $\beta(T)$.
The question
of the dependence of the dust spectral index $\beta$ with 
temperature has been discussed here and constrained, especially using this
low temperature sample.  The mean value of $\beta$ is around 2.1 and
$\beta(T)$ follows a function of the temperature:
$\beta=(\delta + \omega T)^{-1}$ with $\delta=0.020$ and
$\omega=0.035$.  Monte-Carlo simulations have demonstrated that such
an anti-correlation can not be explained by a fitting degeneracy
given our error bars.

 The mean density of these objects varies between $30\,\rm{cm}^{-3}$ and  $10^5\rm{cm}^{-3}$ 
 with an averaged value of  $2\times10^3\rm{cm}^{-3}$. Following the prescription of \citet{Williams2000}, 
such objects are classified as clumps, but not cores. The mass range of the catalogue objects varies
between  0.3 and $2.5\times10^4\,M_{\sun}$, the physical sizes span from 0.2 to 18\,pc.
These parameters of temperature, mass, density and size match  well the definition of clumps  given by \citet{Williams2000}.
 A cross-match between C3PO sources and IRDCs from MSX and Spitzer has shown that each C3PO cold clump could
 contain up to 15 IRDCs, and that the C3PO sources are statistically more extended.
 Thus {\Planck} cold detections appear to be mainly cold clumps, intermediate structures of the fragmentation scenario
 \citep{Falgarone1985}, between
  large clouds and very dense cold cores. Moreover,  the fact that they are significantly extended (by a factor of almost 1.4, 
 compared to the PSF of the {\Planck}-HFI instrument) indicates that these compact sources are linked to larger envelopes 
 following a power-law profile with index in the range  $-2$ to $-1$.
 These aspects are investigated in further detail in \citet{planck2011-7.7a}, which analyses Herschel observations of a few
 \Planck\ sources.

 The {\Planck} cold clump population is strongly associated with
 Galactic structures, especially the molecular component, and is
 mainly distributed within the Galactic plane. Nevertheless, as the
 detection of cold sources inside the thin Galactic plane is difficult
 due to the high confusion level, most of the  detections are located
 in the Solar neighbourhood, within a distance of 7\,kpc.  By studying
 the correlation between the cold clumps and the large scale
 structures such as the IRAS loops \citep{Konyves2007} or the WMAP
 triggering features \citep{Murray2010}, it has been shown in this
 paper that the cold clump population is preferably distributed on the
 borders of these large shells, where star formation is very active.
 Another interesting aspect of the C3PO is the filamentary
 structure of the clumps. This has been statistically 
 quantified over the whole sky and is also apparent in the high
 degree of ellipticity of the sources.  The large variety of objects
 covered by this all-sky catalogue gives the opportunity to perform a
 tentative classification using the physical properties,
 environment, and the evolution stage of these objects.  In that
 perspective, the Herschel follow-up key program Galactic Cold Cores is a
 unique tool to select a representative sample of cold clumps inside
 this large C3PO catalogue, and then obtain high resolution
 information with Herschel-SPIRE/PACS. Such an analysis has already
 been carried out by \citet{Juvela2010} on Herschel Science Demonstration Phase
 Data, and is  continued in \citet{planck2011-7.7a}.

 Finally, we stress that a robust sub-sample of this large C3PO
 catalogue has been delivered to the community: the Early Cold Core Catalogue
 (ECC).  It provides a list of 915 objects over the whole sky, with
 similar statistics at high fluxes to the C3PO, and provides opportunities for
 ground-based or Herschel follow-up of cold clumps. The final legacy
 catalogue C3PO will benefit from two further {\Planck} sky surveys.
 It will increase its robustness, via increased redundancy over the
 sky, and also increase the number of detections because of the
 reduction of the noise level.  The polarization information will be
 also available in this final version of the catalogue. A special
 effort will be made to improve the distance estimates and so increase
 the volume of Galaxy explored, and also to improve the completeness
 of the statistical analysis.

\appendix

\section{Photometry Monte-Carlo Quality Assessment}

\begin{figure*}[!]
\center
\includegraphics[width=15cm]{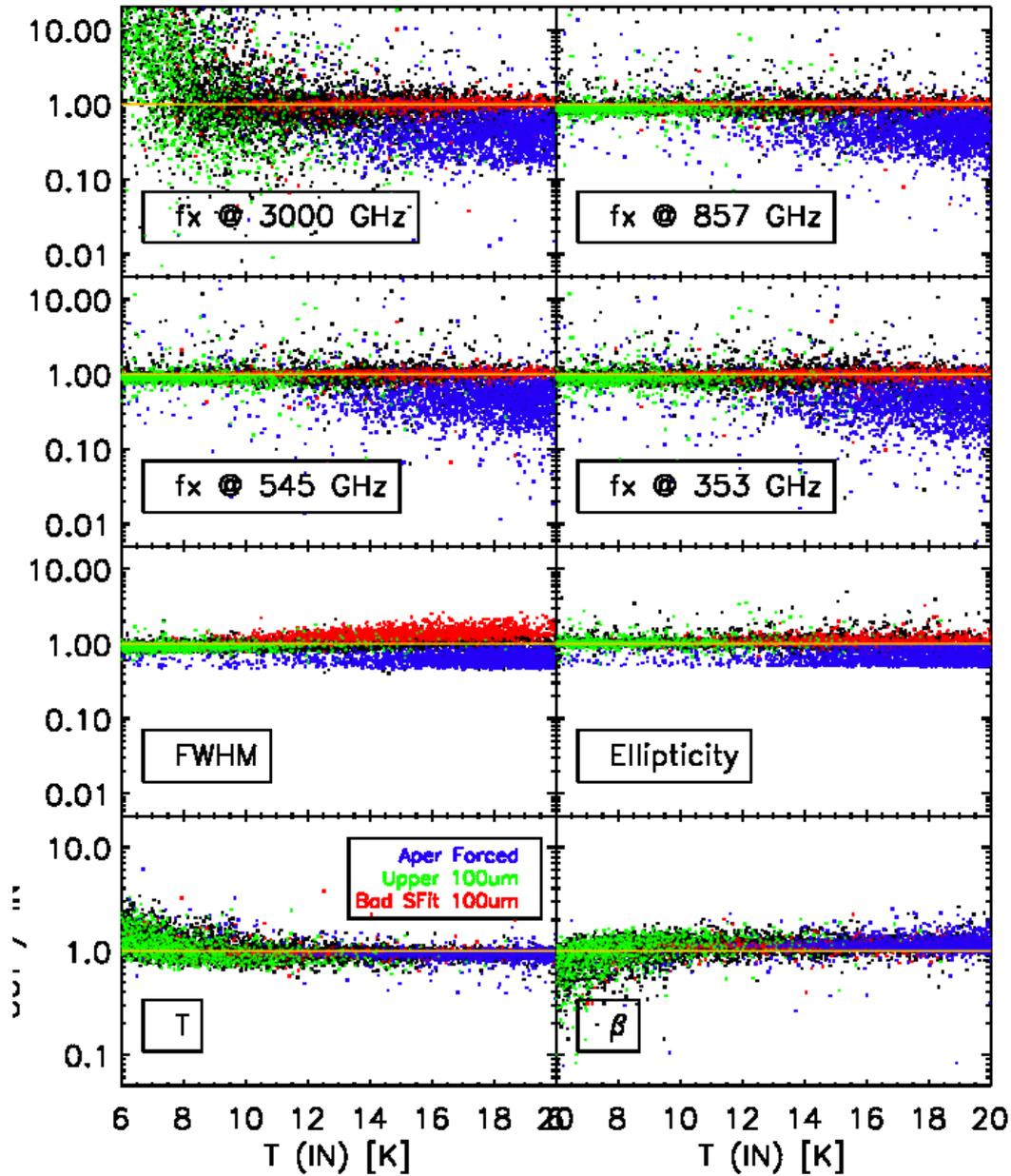} 
\caption{Comparison of the output and injected values of the Monte-Carlo simulations performed 
to assess the quality of the photometry algorithm.
Black dots refer to the {\it nominal} case, when red, blue and green dots stand for various flags
of the photometry algorithm: {\it Bad Sfit 100$\mu$m}, {\it Aper Forced} and {\it Upper 100$\mu$m} respectively.}
\label{fig:mc_analysis}
\end{figure*}

As presented in Sect.~\ref{sec_mcqa}, we have performed a Monte-carlo Analysis to assess the robustness of the photometry algorithm 
described in Sect.~\ref{sec_photometry}. A set of 10000 sources have been injected in {\Planck} all-sky maps following 
a distribution of temperature, dust spectral index and flux in agreement with the observations.
Fig.~\ref{fig:mc_analysis} compiles the relative errors between output 
and input quantities (fluxes, FWHM, Ellipticity, temperature and spectral index) 
for the complete set of simulated sources as a function of the flags raised during the photometry estimate.
Three flags have been introduced: {\it Aper Forced} (blue), {\it Bad Sfit 100\,$\mu $m} (red) and {\it Upper 100\,$\mu$m} (green).

\section{Impact of Calibration Uncertainty}
\label{sec_calib_uncertainty}

\begin{figure*}
\center
\includegraphics[width=15cm, viewport=50 0 850 600]{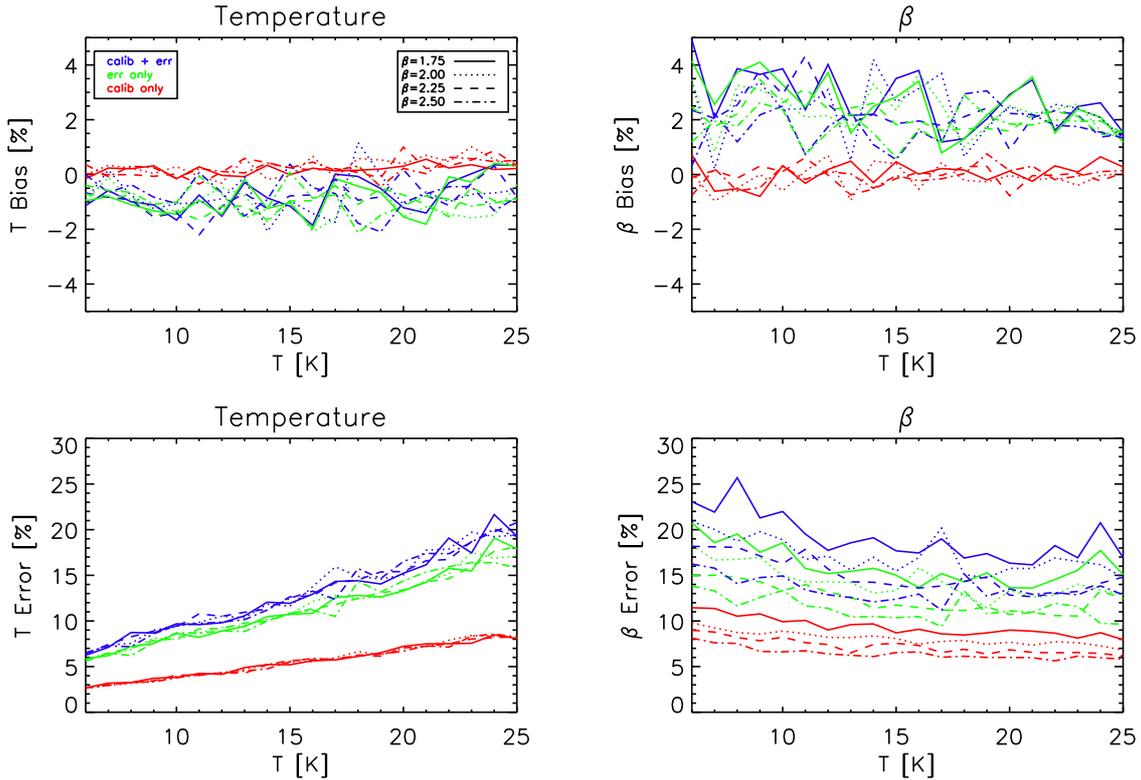}
\caption{Bias and 1$\sigma$ discrepancy of the temperature $T$ and the dust spectral index $\beta$ estimates derived from
a Monte-Carlo analysis dedicated to the study of the impact of the calibration error. three cases have been studied : calibration 
uncertainty only (red lines),  local variance (green) and calibration plus local variance (blue).}
\label{fig:mc_calib_uncertainty}
\end{figure*}

The calibration uncertainty is known to be 13.5\% for the IRIS $100~\mu\rm{m}$ \citep{Miville2005}, and
about 7\% for the {\Planck} high frequency bands  at 857\,GHz and 545\,GHz, and 2\% at 353\,GHz
\citep{planck2011-1.5}. We investigate here what is the impact of this error on 
temperature and spectral index measurements, considering that it is not independent from one band to the others. 
The principle is to constrain the 2 top {\Planck} bands (545 and 857\,GHz) to follow the same relative calibration error, when the $100~\mu\rm{m}$
 and the 353\,GHz remain independent.
For a given temperature $T$ and a dust spectral index $\beta$, SEDs are simulated using a modified black-body modeling, 
including colour correction. A 13.5\% relative noise is added on the $100~\mu\rm{m}$ fluxes, when another 
 7\% noise realization is added
at 545\,Ghz and 857\,GHz simultaneously, and a 2\% noise realization is added on the 353\,GHz fluxes. 
We also study the combination of a calibration error
plus an independent error band per band, set to 
20\% and 10\% for $100~\mu\rm{m}$ and {\Planck} bands respectively. A third set of data is defined as our reference case
and does not include any calibration error, but only an independent error per band. 
The T,$\beta$ fitting algorithm described in Sect.~\ref{sec_physical_properties_temperature} is applied on these 3 sets of SEDs, 
leading to an estimate of the bias and 1$\sigma$ error of
the output distribution, 
for T ranging from 6\,K to 25\,K and 
$\beta$ spanning from 1.75 to 2.5. 

The results are shown in Fig.~\ref{fig:mc_calib_uncertainty}. 
The impact of the calibration uncertainty (in red) is very low compared to the impact of the measurement error only (in green).
In all cases, the bias on temperature and $\beta$ is negligible or significantly lower than 1\%. 
In both cases the bias level remains independent of T and $\beta$.
On the other hand  the 1$\sigma$ error on the temperature goes from 3\% to 8\% at respectively 6\,K and 25\.K 
due to calibration uncertainty only, and remains independent of $\beta$.
The 1-$\sigma$ discrepancy on $\beta$ is about 10\% to 7\% in the temperature range and it decreases of 3\% when  
$\beta$ goes from 1.75 to 2.5.
This error is larger and much more sensitive to $\beta$ in the case of measurement uncertainty (green curves).
Thus the impact of the calibration uncertainty seems very small compared to the impact of the measurement error of the fluxes. 
Moreover in the case of a combination of calibration and measurement uncertainties (in blue), 
both resulting uncertainties on T and $\beta$ seem to add quadratically.

For this work, we propose to consider separately the impact of the calibration uncertainty, 
and just add the quadratic errors at the end of the processing.
 The values of the calibration uncertainties for T and $\beta$ are provided in Annex A 
 Table \ref{tab:calib_uncertainty_t} \& \ref{tab:calib_uncertainty_beta} for a given set of T,$\beta$ parameters.

\begin{table*}
\begin{center}

\newdimen\digitwidth 
\setbox0=\hbox{\rm 0} 
\digitwidth=\wd0 
\catcode`*=\active 
\def*{\kern\digitwidth} 

\newdimen\signwidth 
\setbox0=\hbox{+} 
\signwidth=\wd0 
\catcode`!=\active 
\def!{\kern\signwidth} 

\newdimen\signwidth 
\setbox0=\hbox{.} 
\signwidth=\wd0 
\catcode`?=\active 
\def?{\kern\signwidth} 

\begin{tabular}{l|cc|cc|cc|cc}
\hline
\hline
  & \multicolumn{2}{|c}{$\beta =  1.75$} & \multicolumn{2}{|c}{$\beta =  2.00$} & \multicolumn{2}{|c}{$\beta =  2.25$} & \multicolumn{2}{|c}{$\beta =  2.50$} \\
T [K] & Bias [\%] & $1-\sigma$ [\%] & Bias [\%] & $1-\sigma$ [\%] & Bias [\%] & $1-\sigma$ [\%] & Bias [\%] & $1-\sigma$ [\%] \\
\hline
 *6 &  -0.15 &   2.68 &  -0.01 &   2.59 &   !0.03 &   2.70 &   !0.36 &   2.68 \\
 *7 &   !0.23 &   3.19 &   !0.34 &   2.94 &  -0.01 &   3.04 &  -0.08 &   2.95 \\
 *8 &   !0.21 &   3.25 &   !0.24 &   3.16 &  -0.00 &   3.25 &   !0.31 &   3.29 \\
 *9 &   !0.32 &   3.72 &   !0.04 &   3.41 &  -0.07 &   3.44 &  !0.25 &   3.33 \\
10 &  -0.15 &   3.81 &   !0.29 &   3.92 &   !0.02 &   4.08 &  -0.15 &   3.78 \\
11 &   !0.28 &   4.25 &  -0.06 &   4.17 &  -0.35 &   4.04 &   !0.22 &   4.09 \\
12 &  -0.03 &   4.15 &  -0.31 &   4.32 &   !0.09 &   4.32 &   !0.11 &   4.27 \\
13 &  -0.08 &   4.89 &   !0.58 &   4.69 &   !0.58 &   4.28 &   !0.06 &   4.52 \\
14 &   !0.34 &   5.22 &  -0.18 &   5.26 &   !0.43 &   5.20 &   !0.41 &   4.76 \\
15 &  -0.01 &   5.20 &  -0.06 &   5.46 &   !0.25 &   5.66 &   !0.24 &   5.54 \\
16 &   !0.23 &   5.64 &   !0.66 &   5.54 &   !0.16 &   5.75 &   !0.45 &   5.92 \\
17 &   !0.13 &   5.73 &  -0.12 &   5.97 &   !0.21 &   5.83 &   !0.15 &   5.74 \\
18 &   !0.16 &   6.12 &   !0.58 &   6.64 &   !0.08 &   6.24 &   !0.31 &   6.24 \\
19 &   !0.20 &   6.52 &   !0.53 &   6.30 &  -0.12 &   6.14 &  -0.45 &   6.35 \\
20 &   !0.31 &   7.22 &   !0.40 &   7.04 &   !1.00 &   7.05 &   !0.22 &   6.69 \\
21 &   !0.56 &   7.49 &   !0.70 &   7.20 &   !0.22 &   7.22 &   !0.03 &   7.28 \\
22 &   !0.24 &   7.51 &   !0.47 &   7.89 &   !0.31 &   7.49 &   !0.63 &   7.13 \\
23 &   !0.36 &   7.56 &   !1.01 &   8.41 &   !0.53 &   8.02 &   !0.58 &   8.33 \\
24 &   !0.18 &   8.48 &   !0.55 &   8.30 &   !0.87 &   8.17 &   !0.47 &   8.56 \\
25 &   !0.22 &   8.08 &   !0.29 &   8.21 &   !0.33 &   8.27 &   !0.51 &   8.26 \\
\hline
\hline
\end{tabular}

\caption{Bias and $1-\sigma$ error on the temperature due to the calibration uncertainty 
on fluxes in the IRIS $100 \mu m$ and {\Planck} 857, 545 and 353\,GHz bands, and 
estimated for each couple of (T,$\beta$).}
\label{tab:calib_uncertainty_t}
\end{center}
\end{table*}

\begin{table*}
\begin{center}

\newdimen\digitwidth 
\setbox0=\hbox{\rm 0} 
\digitwidth=\wd0 
\catcode`*=\active 
\def*{\kern\digitwidth} 

\newdimen\signwidth 
\setbox0=\hbox{+} 
\signwidth=\wd0 
\catcode`!=\active 
\def!{\kern\signwidth} 

\newdimen\signwidth 
\setbox0=\hbox{.} 
\signwidth=\wd0 
\catcode`?=\active 
\def?{\kern\signwidth} 

\begin{tabular}{l|cc|cc|cc|cc}
\hline
\hline
  & \multicolumn{2}{|c}{$\beta =  1.75$} & \multicolumn{2}{|c}{$\beta =  2.00$} & \multicolumn{2}{|c}{$\beta =  2.25$} & \multicolumn{2}{|c}{$\beta =  2.50$} \\
T [K] & Bias [\%] & $1-\sigma$ [\%] & Bias [\%] & $1-\sigma$ [\%] & Bias [\%] & $1-\sigma$ [\%] & Bias [\%] & $1-\sigma$ [\%] \\
\hline
 *6 &   !0.72 &  11.45 &   !0.07 &   9.79 &  -0.11 &   9.00 &  -0.77 &   8.12 \\
 *7 &  -0.61 &  11.37 &  -0.97 &   9.24 &   !0.15 &   8.77 &   !0.21 &   7.61 \\
 *8 &  -0.52 &  10.52 &  -0.50 &   8.73 &  -0.14 &   8.19 &  -0.60 &   7.53 \\
 *9 &  -0.80 &  10.77 &   !0.00 &   8.56 &   !0.31 &   7.83 &  -0.48 &   6.66 \\
10 &   0.32 &   9.91 &  -0.59 &   8.89 &   !0.11 &   8.23 &   !0.29 &   6.62 \\
11 &  -0.32 &  10.06 &   !0.12 &   8.59 &   !0.61 &   7.62 &  -0.18 &   6.73 \\
12 &   !0.18 &   9.01 &   !0.70 &   8.20 &  -0.02 &   7.42 &   !0.10 &   6.39 \\
13 &   !0.49 &   9.60 &  -0.95 &   8.19 &  -0.74 &   6.51 &   !0.02 &   6.24 \\
14 &  -0.30 &   9.67 &   !0.49 &   8.36 &  -0.53 &   7.42 &  -0.50 &   6.10 \\
15 &   !0.47 &   8.69 &   !0.41 &   8.07 &  -0.15 &   7.55 &  -0.04 &   6.54 \\
16 &   !0.01 &   9.09 &  -0.53 &   7.46 &   !0.08 &   7.31 &  -0.23 &   6.61 \\
17 &   !0.22 &   8.58 &   !0.43 &   7.76 &  -0.04 &   6.53 &  -0.16 &   6.03 \\
18 &   !0.16 &   8.46 &  -0.21 &   7.86 &   !0.29 &   6.90 &   !0.03 &   6.13 \\
19 &  -0.20 &   8.70 &  -0.27 &   7.45 &   !0.30 &   6.36 &   !0.75 &   6.01 \\
20 &   !0.12 &   9.00 &  -0.04 &   7.68 &  -0.78 &   6.84 &   !0.14 &   6.01 \\
21 &  -0.25 &   8.88 &  -0.21 &   7.34 &  -0.04 &   6.55 &   !0.30 &   5.98 \\
22 &   !0.32 &   8.66 &  -0.12 &   7.73 &  -0.02 &   6.51 &  -0.19 &   5.63 \\
23 &   !0.09 &   8.12 &  -0.39 &   7.62 &   !0.05 &   6.55 &  -0.06 &   6.14 \\
24 &   !0.65 &   8.71 &  -0.03 &   7.27 &  -0.23 &   6.44 &   !0.11 &   5.97 \\
25 &   !0.28 &   7.95 &   !0.03 &   6.80 &   !0.06 &   6.17 &   !0.14 &   5.81 \\
\hline
\hline
\end{tabular}

\caption{Bias and $1-\sigma$ error on the spectral index $\beta$ due to the calibration uncertainty 
on fluxes in the IRIS $100 \mu m$ and {\Planck} 857, 545 and 353\,GHz bands, and 
estimated for each couple of (T,$\beta$).}
\label{tab:calib_uncertainty_beta}
\end{center}
\end{table*}

\begin{acknowledgements}
A description of the Planck Collaboration and a list of its members can
be found at
http://www.rssd.esa.int/index.php?project=\ PLANCK\&page=Planck\_Collaboration
\end{acknowledgements}

\allearlypapers 
\bibliographystyle{aa}
\bibliography{biblio_v1.1,planck_bib_v4.0}

\end{document}

%% file: Planck.tex
\def\setsymbol#1#2{\expandafter\def\csname #1\endcsname{#2}}
\def\getsymbol#1{\csname #1\endcsname}

\def\Planck{{\it Planck\/}}

\def\allearlypapers{\nocite{planck2011-1.1, planck2011-1.3, planck2011-1.4, planck2011-1.5, planck2011-1.6, planck2011-1.7, planck2011-1.10, planck2011-1.10sup, planck2011-5.1a, planck2011-5.1b, planck2011-5.2a, planck2011-5.2b, planck2011-5.2c, planck2011-6.1, planck2011-6.2, planck2011-6.3a, planck2011-6.4a, planck2011-6.4b, planck2011-6.6, planck2011-7.0, planck2011-7.2, planck2011-7.3, planck2011-7.7a, planck2011-7.7b, planck2011-7.12, planck2011-7.13}}

\newbox\tablebox    \newdimen\tablewidth
\def\leaderfil{\leaders\hbox to 5pt{\hss.\hss}\hfil}
\def\tablenote#1 #2\par{\begingroup \parindent=0.8em
    \abovedisplayshortskip=0pt\belowdisplayshortskip=0pt
    \noindent
    $$\hss\vbox{\hsize\tablewidth \hangindent=\parindent \hangafter=1 \noindent
    \hbox to \parindent{\sup{\rm #1}\hss}\strut#2\strut\par}\hss$$
    \endgroup}

\def\L2{\ifmmode L_2\else $L_2$\fi}

\def\DeltaT{\ifmmode \Delta T\else $\Delta T$\fi}
\def\deltat{\ifmmode \Delta t\else $\Delta t$\fi}
\def\fknee{\ifmmode f_{\rm knee}\else $f_{\rm knee}$\fi}
\def\Fmax{\ifmmode F_{\rm max}\else $F_{\rm max}$\fi}
\def\solar{\ifmmode{\rm M}_{\mathord\odot}\else${\rm M}_{\mathord\odot}$\fi}

\def\inv{\ifmmode^{-1}\else$^{-1}$\fi}
\def\mo{\ifmmode^{-1}\else$^{-1}$\fi}
\def\sup#1{\ifmmode ^{\rm #1}\else $^{\rm #1}$\fi}
\def\expo#1{\ifmmode \times 10^{#1}\else $\times 10^{#1}$\fi}
\def\,{\thinspace}
\def\lsim{\mathrel{\raise .4ex\hbox{\rlap{$<$}\lower 1.2ex\hbox{$\sim$}}}}
\def\gsim{\mathrel{\raise .4ex\hbox{\rlap{$>$}\lower 1.2ex\hbox{$\sim$}}}}

\def\simprop{\mathrel{\raise .4ex\hbox{\rlap{$\propto$}\lower 1.2ex\hbox{$\sim$}}}}
\def\deg{\ifmmode^\circ\else$^\circ$\fi}
\def\pdeg{\ifmmode $\setbox0=\hbox{$^{\circ}$}\rlap{\hskip.11\wd0 .}$^{\circ}
          \else \setbox0=\hbox{$^{\circ}$}\rlap{\hskip.11\wd0 .}$^{\circ}$\fi}
\def\arcs{\ifmmode {^{\scriptstyle\prime\prime}}
          \else $^{\scriptstyle\prime\prime}$\fi}
\def\arcm{\ifmmode {^{\scriptstyle\prime}}
          \else $^{\scriptstyle\prime}$\fi}
\newdimen\sa  \newdimen\sb
\def\parcs{\sa=.07em \sb=.03em
     \ifmmode \hbox{\rlap{.}}^{\scriptstyle\prime\kern -\sb\prime}\hbox{\kern -\sa}
     \else \rlap{.}$^{\scriptstyle\prime\kern -\sb\prime}$\kern -\sa\fi}
\def\parcm{\sa=.08em \sb=.03em
     \ifmmode \hbox{\rlap{.}\kern\sa}^{\scriptstyle\prime}\hbox{\kern-\sb}
     \else \rlap{.}\kern\sa$^{\scriptstyle\prime}$\kern-\sb\fi}
\def\ra[#1 #2 #3.#4]{#1\sup{h}#2\sup{m}#3\sup{s}\llap.#4}
\def\dec[#1 #2 #3.#4]{#1\deg#2\arcm#3\arcs\llap.#4}
\def\deco[#1 #2 #3]{#1\deg#2\arcm#3\arcs}
\def\rra[#1 #2]{#1\sup{h}#2\sup{m}}

\def\dots{\relax\ifmmode \ldots\else $\ldots$\fi}
\def\WHzsr{\ifmmode $W\,Hz\mo\,sr\mo$\else W\,Hz\mo\,sr\mo\fi}
\def\mHz{\ifmmode $\,mHz$\else \,mHz\fi}
\def\GHz{\ifmmode $\,GHz$\else \,GHz\fi}
\def\mKs{\ifmmode $\,mK\,s$^{1/2}\else \,mK\,s$^{1/2}$\fi}
\def\muKs{\ifmmode \,\mu$K\,s$^{1/2}\else \,$\mu$K\,s$^{1/2}$\fi}
\def\muKRJs{\ifmmode \,\mu$K$_{\rm RJ}$\,s$^{1/2}\else \,$\mu$K$_{\rm RJ}$\,s$^{1/2}$\fi}
\def\muKHz{\ifmmode \,\mu$K\,Hz$^{-1/2}\else \,$\mu$K\,Hz$^{-1/2}$\fi}
\def\MJysr{\ifmmode \,$MJy\,sr\mo$\else \,MJy\,sr\mo\fi}
\def\MJysrmK{\ifmmode \,$MJy\,sr\mo$\,mK$_{\rm CMB}\mo\else \,MJy\,sr\mo\,mK$_{\rm CMB}\mo$\fi}
\def\microns{\ifmmode \,\mu$m$\else \,$\mu$m\fi}

\def\muK{\ifmmode \,\mu$K$\else \,$\mu$\hbox{K}\fi}
\def\microK{\ifmmode \,\mu$K$\else \,$\mu$\hbox{K}\fi}
\def\muW{\ifmmode \,\mu$W$\else \,$\mu$\hbox{W}\fi}
\def\kms{\ifmmode $\,km\,s$^{-1}\else \,km\,s$^{-1}$\fi}
\def\kmsMpc{\ifmmode $\,\kms\,Mpc\mo$\else \,\kms\,Mpc\mo\fi}
\setsymbol{LFI:center:frequency:70GHz:units}{70.3\,GHz}
\setsymbol{LFI:center:frequency:44GHz:units}{44.1\,GHz}
\setsymbol{LFI:center:frequency:30GHz:units}{28.5\,GHz}

\setsymbol{LFI:center:frequency:70GHz}{70.3}
\setsymbol{LFI:center:frequency:44GHz}{44.1}
\setsymbol{LFI:center:frequency:30GHz}{28.5}

\setsymbol{LFI:center:frequency:LFI18:Rad:M:units}{71.7\GHz}
\setsymbol{LFI:center:frequency:LFI19:Rad:M:units}{67.5\GHz}
\setsymbol{LFI:center:frequency:LFI20:Rad:M:units}{69.2\GHz}
\setsymbol{LFI:center:frequency:LFI21:Rad:M:units}{70.4\GHz}
\setsymbol{LFI:center:frequency:LFI22:Rad:M:units}{71.5\GHz}
\setsymbol{LFI:center:frequency:LFI23:Rad:M:units}{70.8\GHz}
\setsymbol{LFI:center:frequency:LFI24:Rad:M:units}{44.4\GHz}
\setsymbol{LFI:center:frequency:LFI25:Rad:M:units}{44.0\GHz}
\setsymbol{LFI:center:frequency:LFI26:Rad:M:units}{43.9\GHz}
\setsymbol{LFI:center:frequency:LFI27:Rad:M:units}{28.3\GHz}
\setsymbol{LFI:center:frequency:LFI28:Rad:M:units}{28.8\GHz}
\setsymbol{LFI:center:frequency:LFI18:Rad:S:units}{70.1\GHz}
\setsymbol{LFI:center:frequency:LFI19:Rad:S:units}{69.6\GHz}
\setsymbol{LFI:center:frequency:LFI20:Rad:S:units}{69.5\GHz}
\setsymbol{LFI:center:frequency:LFI21:Rad:S:units}{69.5\GHz}
\setsymbol{LFI:center:frequency:LFI22:Rad:S:units}{72.8\GHz}
\setsymbol{LFI:center:frequency:LFI23:Rad:S:units}{71.3\GHz}
\setsymbol{LFI:center:frequency:LFI24:Rad:S:units}{44.1\GHz}
\setsymbol{LFI:center:frequency:LFI25:Rad:S:units}{44.1\GHz}
\setsymbol{LFI:center:frequency:LFI26:Rad:S:units}{44.1\GHz}
\setsymbol{LFI:center:frequency:LFI27:Rad:S:units}{28.5\GHz}
\setsymbol{LFI:center:frequency:LFI28:Rad:S:units}{28.2\GHz}

\setsymbol{LFI:center:frequency:LFI18:Rad:M}{71.7}
\setsymbol{LFI:center:frequency:LFI19:Rad:M}{67.5}
\setsymbol{LFI:center:frequency:LFI20:Rad:M}{69.2}
\setsymbol{LFI:center:frequency:LFI21:Rad:M}{70.4}
\setsymbol{LFI:center:frequency:LFI22:Rad:M}{71.5}
\setsymbol{LFI:center:frequency:LFI23:Rad:M}{70.8}
\setsymbol{LFI:center:frequency:LFI24:Rad:M}{44.4}
\setsymbol{LFI:center:frequency:LFI25:Rad:M}{44.0}
\setsymbol{LFI:center:frequency:LFI26:Rad:M}{43.9}
\setsymbol{LFI:center:frequency:LFI27:Rad:M}{28.3}
\setsymbol{LFI:center:frequency:LFI28:Rad:M}{28.8}
\setsymbol{LFI:center:frequency:LFI18:Rad:S}{70.1}
\setsymbol{LFI:center:frequency:LFI19:Rad:S}{69.6}
\setsymbol{LFI:center:frequency:LFI20:Rad:S}{69.5}
\setsymbol{LFI:center:frequency:LFI21:Rad:S}{69.5}
\setsymbol{LFI:center:frequency:LFI22:Rad:S}{72.8}
\setsymbol{LFI:center:frequency:LFI23:Rad:S}{71.3}
\setsymbol{LFI:center:frequency:LFI24:Rad:S}{44.1}
\setsymbol{LFI:center:frequency:LFI25:Rad:S}{44.1}
\setsymbol{LFI:center:frequency:LFI26:Rad:S}{44.1}
\setsymbol{LFI:center:frequency:LFI27:Rad:S}{28.5}
\setsymbol{LFI:center:frequency:LFI28:Rad:S}{28.2}

\setsymbol{LFI:white:noise:sensitivity:70GHz:units}{134.7\muKs}
\setsymbol{LFI:white:noise:sensitivity:44GHz:units}{164.7\muKs}
\setsymbol{LFI:white:noise:sensitivity:30GHz:units}{143.4\muKs}

\setsymbol{LFI:white:noise:sensitivity:70GHz}{134.7}
\setsymbol{LFI:white:noise:sensitivity:44GHz}{164.7}
\setsymbol{LFI:white:noise:sensitivity:30GHz}{143.4}

\setsymbol{LFI:white:noise:sensitivity:LFI18:Rad:M:units}{512.0\muKs}
\setsymbol{LFI:white:noise:sensitivity:LFI19:Rad:M:units}{581.4\muKs}
\setsymbol{LFI:white:noise:sensitivity:LFI20:Rad:M:units}{590.8\muKs}
\setsymbol{LFI:white:noise:sensitivity:LFI21:Rad:M:units}{455.2\muKs}
\setsymbol{LFI:white:noise:sensitivity:LFI22:Rad:M:units}{492.0\muKs}
\setsymbol{LFI:white:noise:sensitivity:LFI23:Rad:M:units}{507.7\muKs}
\setsymbol{LFI:white:noise:sensitivity:LFI24:Rad:M:units}{462.2\muKs}
\setsymbol{LFI:white:noise:sensitivity:LFI25:Rad:M:units}{413.6\muKs}
\setsymbol{LFI:white:noise:sensitivity:LFI26:Rad:M:units}{478.6\muKs}
\setsymbol{LFI:white:noise:sensitivity:LFI27:Rad:M:units}{277.7\muKs}
\setsymbol{LFI:white:noise:sensitivity:LFI28:Rad:M:units}{312.3\muKs}
\setsymbol{LFI:white:noise:sensitivity:LFI18:Rad:S:units}{465.7\muKs}
\setsymbol{LFI:white:noise:sensitivity:LFI19:Rad:S:units}{555.6\muKs}
\setsymbol{LFI:white:noise:sensitivity:LFI20:Rad:S:units}{623.2\muKs}
\setsymbol{LFI:white:noise:sensitivity:LFI21:Rad:S:units}{564.1\muKs}
\setsymbol{LFI:white:noise:sensitivity:LFI22:Rad:S:units}{534.4\muKs}
\setsymbol{LFI:white:noise:sensitivity:LFI23:Rad:S:units}{542.4\muKs}
\setsymbol{LFI:white:noise:sensitivity:LFI24:Rad:S:units}{399.2\muKs}
\setsymbol{LFI:white:noise:sensitivity:LFI25:Rad:S:units}{392.6\muKs}
\setsymbol{LFI:white:noise:sensitivity:LFI26:Rad:S:units}{418.6\muKs}
\setsymbol{LFI:white:noise:sensitivity:LFI27:Rad:S:units}{302.9\muKs}
\setsymbol{LFI:white:noise:sensitivity:LFI28:Rad:S:units}{285.3\muKs}

\setsymbol{LFI:white:noise:sensitivity:LFI18:Rad:M}{512.0}
\setsymbol{LFI:white:noise:sensitivity:LFI19:Rad:M}{581.4}
\setsymbol{LFI:white:noise:sensitivity:LFI20:Rad:M}{590.8}
\setsymbol{LFI:white:noise:sensitivity:LFI21:Rad:M}{455.2}
\setsymbol{LFI:white:noise:sensitivity:LFI22:Rad:M}{492.0}
\setsymbol{LFI:white:noise:sensitivity:LFI23:Rad:M}{507.7}
\setsymbol{LFI:white:noise:sensitivity:LFI24:Rad:M}{462.2}
\setsymbol{LFI:white:noise:sensitivity:LFI25:Rad:M}{413.6}
\setsymbol{LFI:white:noise:sensitivity:LFI26:Rad:M}{478.6}
\setsymbol{LFI:white:noise:sensitivity:LFI27:Rad:M}{277.7}
\setsymbol{LFI:white:noise:sensitivity:LFI28:Rad:M}{312.3}
\setsymbol{LFI:white:noise:sensitivity:LFI18:Rad:S}{465.7}
\setsymbol{LFI:white:noise:sensitivity:LFI19:Rad:S}{555.6}
\setsymbol{LFI:white:noise:sensitivity:LFI20:Rad:S}{623.2}
\setsymbol{LFI:white:noise:sensitivity:LFI21:Rad:S}{564.1}
\setsymbol{LFI:white:noise:sensitivity:LFI22:Rad:S}{534.4}
\setsymbol{LFI:white:noise:sensitivity:LFI23:Rad:S}{542.4}
\setsymbol{LFI:white:noise:sensitivity:LFI24:Rad:S}{399.2}
\setsymbol{LFI:white:noise:sensitivity:LFI25:Rad:S}{392.6}
\setsymbol{LFI:white:noise:sensitivity:LFI26:Rad:S}{418.6}
\setsymbol{LFI:white:noise:sensitivity:LFI27:Rad:S}{302.9}
\setsymbol{LFI:white:noise:sensitivity:LFI28:Rad:S}{285.3}

\setsymbol{LFI:knee:frequency:70GHz:units}{29.5\mHz}
\setsymbol{LFI:knee:frequency:44GHz:units}{56.2\mHz}
\setsymbol{LFI:knee:frequency:30GHz:units}{113.7\mHz}

\setsymbol{LFI:knee:frequency:70GHz}{29.5}
\setsymbol{LFI:knee:frequency:44GHz}{56.2}
\setsymbol{LFI:knee:frequency:30GHz}{113.7}

\setsymbol{LFI:knee:frequency:LFI18:Rad:M:units}{16.3\mHz}
\setsymbol{LFI:knee:frequency:LFI19:Rad:M:units}{15.1\mHz}
\setsymbol{LFI:knee:frequency:LFI20:Rad:M:units}{18.7\mHz}
\setsymbol{LFI:knee:frequency:LFI21:Rad:M:units}{37.2\mHz}
\setsymbol{LFI:knee:frequency:LFI22:Rad:M:units}{12.7\mHz}
\setsymbol{LFI:knee:frequency:LFI23:Rad:M:units}{34.6\mHz}
\setsymbol{LFI:knee:frequency:LFI24:Rad:M:units}{46.2\mHz}
\setsymbol{LFI:knee:frequency:LFI25:Rad:M:units}{24.9\mHz}
\setsymbol{LFI:knee:frequency:LFI26:Rad:M:units}{67.6\mHz}
\setsymbol{LFI:knee:frequency:LFI27:Rad:M:units}{187.4\mHz}
\setsymbol{LFI:knee:frequency:LFI28:Rad:M:units}{122.2\mHz}
\setsymbol{LFI:knee:frequency:LFI18:Rad:S:units}{17.7\mHz}
\setsymbol{LFI:knee:frequency:LFI19:Rad:S:units}{22.0\mHz}
\setsymbol{LFI:knee:frequency:LFI20:Rad:S:units}{8.7\mHz}
\setsymbol{LFI:knee:frequency:LFI21:Rad:S:units}{25.9\mHz}
\setsymbol{LFI:knee:frequency:LFI22:Rad:S:units}{15.8\mHz}
\setsymbol{LFI:knee:frequency:LFI23:Rad:S:units}{129.8\mHz}
\setsymbol{LFI:knee:frequency:LFI24:Rad:S:units}{100.9\mHz}
\setsymbol{LFI:knee:frequency:LFI25:Rad:S:units}{38.9\mHz}
\setsymbol{LFI:knee:frequency:LFI26:Rad:S:units}{58.9\mHz}
\setsymbol{LFI:knee:frequency:LFI27:Rad:S:units}{104.4\mHz}
\setsymbol{LFI:knee:frequency:LFI28:Rad:S:units}{40.7\mHz}

\setsymbol{LFI:knee:frequency:LFI18:Rad:M}{16.3}
\setsymbol{LFI:knee:frequency:LFI19:Rad:M}{15.1}
\setsymbol{LFI:knee:frequency:LFI20:Rad:M}{18.7}
\setsymbol{LFI:knee:frequency:LFI21:Rad:M}{37.2}
\setsymbol{LFI:knee:frequency:LFI22:Rad:M}{12.7}
\setsymbol{LFI:knee:frequency:LFI23:Rad:M}{34.6}
\setsymbol{LFI:knee:frequency:LFI24:Rad:M}{46.2}
\setsymbol{LFI:knee:frequency:LFI25:Rad:M}{24.9}
\setsymbol{LFI:knee:frequency:LFI26:Rad:M}{67.6}
\setsymbol{LFI:knee:frequency:LFI27:Rad:M}{187.4}
\setsymbol{LFI:knee:frequency:LFI28:Rad:M}{122.2}
\setsymbol{LFI:knee:frequency:LFI18:Rad:S}{17.7}
\setsymbol{LFI:knee:frequency:LFI19:Rad:S}{22.0}
\setsymbol{LFI:knee:frequency:LFI20:Rad:S}{8.7}
\setsymbol{LFI:knee:frequency:LFI21:Rad:S}{25.9}
\setsymbol{LFI:knee:frequency:LFI22:Rad:S}{15.8}
\setsymbol{LFI:knee:frequency:LFI23:Rad:S}{129.8}
\setsymbol{LFI:knee:frequency:LFI24:Rad:S}{100.9}
\setsymbol{LFI:knee:frequency:LFI25:Rad:S}{38.9}
\setsymbol{LFI:knee:frequency:LFI26:Rad:S}{58.9}
\setsymbol{LFI:knee:frequency:LFI27:Rad:S}{104.4}
\setsymbol{LFI:knee:frequency:LFI28:Rad:S}{40.7}

\setsymbol{LFI:slope:70GHz:units}{$-1.03$\mHz}
\setsymbol{LFI:slope:44GHz:units}{$-0.89$\mHz}
\setsymbol{LFI:slope:30GHz:units}{$-0.87$\mHz}

\setsymbol{LFI:slope:70GHz}{$-1.03$}
\setsymbol{LFI:slope:44GHz}{$-0.89$}
\setsymbol{LFI:slope:30GHz}{$-0.87$}

\setsymbol{LFI:slope:LFI18:Rad:M:units}{$-1.04$\mHz}
\setsymbol{LFI:slope:LFI19:Rad:M:units}{$-1.09$\mHz}
\setsymbol{LFI:slope:LFI20:Rad:M:units}{$-0.69$\mHz}
\setsymbol{LFI:slope:LFI21:Rad:M:units}{$-1.56$\mHz}
\setsymbol{LFI:slope:LFI22:Rad:M:units}{$-1.01$\mHz}
\setsymbol{LFI:slope:LFI23:Rad:M:units}{$-0.96$\mHz}
\setsymbol{LFI:slope:LFI24:Rad:M:units}{$-0.83$\mHz}
\setsymbol{LFI:slope:LFI25:Rad:M:units}{$-0.91$\mHz}
\setsymbol{LFI:slope:LFI26:Rad:M:units}{$-0.95$\mHz}
\setsymbol{LFI:slope:LFI27:Rad:M:units}{$-0.87$\mHz}
\setsymbol{LFI:slope:LFI28:Rad:M:units}{$-0.88$\mHz}
\setsymbol{LFI:slope:LFI18:Rad:S:units}{$-1.15$\mHz}
\setsymbol{LFI:slope:LFI19:Rad:S:units}{$-1.00$\mHz}
\setsymbol{LFI:slope:LFI20:Rad:S:units}{$-0.95$\mHz}
\setsymbol{LFI:slope:LFI21:Rad:S:units}{$-0.92$\mHz}
\setsymbol{LFI:slope:LFI22:Rad:S:units}{$-1.01$\mHz}
\setsymbol{LFI:slope:LFI23:Rad:S:units}{$-0.95$\mHz}
\setsymbol{LFI:slope:LFI24:Rad:S:units}{$-0.73$\mHz}
\setsymbol{LFI:slope:LFI25:Rad:S:units}{$-1.16$\mHz}
\setsymbol{LFI:slope:LFI26:Rad:S:units}{$-0.79$\mHz}
\setsymbol{LFI:slope:LFI27:Rad:S:units}{$-0.82$\mHz}
\setsymbol{LFI:slope:LFI28:Rad:S:units}{$-0.91$\mHz}

\setsymbol{LFI:slope:LFI18:Rad:M}{$-1.04$}
\setsymbol{LFI:slope:LFI19:Rad:M}{$-1.09$}
\setsymbol{LFI:slope:LFI20:Rad:M}{$-0.69$}
\setsymbol{LFI:slope:LFI21:Rad:M}{$-1.56$}
\setsymbol{LFI:slope:LFI22:Rad:M}{$-1.01$}
\setsymbol{LFI:slope:LFI23:Rad:M}{$-0.96$}
\setsymbol{LFI:slope:LFI24:Rad:M}{$-0.83$}
\setsymbol{LFI:slope:LFI25:Rad:M}{$-0.91$}
\setsymbol{LFI:slope:LFI26:Rad:M}{$-0.95$}
\setsymbol{LFI:slope:LFI27:Rad:M}{$-0.87$}
\setsymbol{LFI:slope:LFI28:Rad:M}{$-0.88$}
\setsymbol{LFI:slope:LFI18:Rad:S}{$-1.15$}
\setsymbol{LFI:slope:LFI19:Rad:S}{$-1.00$}
\setsymbol{LFI:slope:LFI20:Rad:S}{$-0.95$}
\setsymbol{LFI:slope:LFI21:Rad:S}{$-0.92$}
\setsymbol{LFI:slope:LFI22:Rad:S}{$-1.01$}
\setsymbol{LFI:slope:LFI23:Rad:S}{$-0.95$}
\setsymbol{LFI:slope:LFI24:Rad:S}{$-0.73$}
\setsymbol{LFI:slope:LFI25:Rad:S}{$-1.16$}
\setsymbol{LFI:slope:LFI26:Rad:S}{$-0.79$}
\setsymbol{LFI:slope:LFI27:Rad:S}{$-0.82$}
\setsymbol{LFI:slope:LFI28:Rad:S}{$-0.91$}

\setsymbol{LFI:FWHM:70GHz:units}{13\parcm01}
\setsymbol{LFI:FWHM:44GHz:units}{27\parcm92}
\setsymbol{LFI:FWHM:30GHz:units}{32\parcm65}

\setsymbol{LFI:FWHM:70GHz}{13.01}
\setsymbol{LFI:FWHM:44GHz}{27.92}
\setsymbol{LFI:FWHM:30GHz}{32.65}

\setsymbol{LFI:FWHM:LFI18:units}{13\parcm39}
\setsymbol{LFI:FWHM:LFI19:units}{13\parcm01}
\setsymbol{LFI:FWHM:LFI20:units}{12\parcm75}
\setsymbol{LFI:FWHM:LFI21:units}{12\parcm74}
\setsymbol{LFI:FWHM:LFI22:units}{12\parcm87}
\setsymbol{LFI:FWHM:LFI23:units}{13\parcm27}
\setsymbol{LFI:FWHM:LFI24:units}{22\parcm98}
\setsymbol{LFI:FWHM:LFI25:units}{30\parcm46}
\setsymbol{LFI:FWHM:LFI26:units}{30\parcm31}
\setsymbol{LFI:FWHM:LFI27:units}{32\parcm65}
\setsymbol{LFI:FWHM:LFI28:units}{32\parcm66}

\setsymbol{LFI:FWHM:LFI18}{13.39}
\setsymbol{LFI:FWHM:LFI19}{13.01}
\setsymbol{LFI:FWHM:LFI20}{12.75}
\setsymbol{LFI:FWHM:LFI21}{12.74}
\setsymbol{LFI:FWHM:LFI22}{12.87}
\setsymbol{LFI:FWHM:LFI23}{13.27}
\setsymbol{LFI:FWHM:LFI24}{22.98}
\setsymbol{LFI:FWHM:LFI25}{30.46}
\setsymbol{LFI:FWHM:LFI26}{30.31}
\setsymbol{LFI:FWHM:LFI27}{32.65}
\setsymbol{LFI:FWHM:LFI28}{32.66}

\setsymbol{LFI:FWHM:uncertainty:LFI18:units}{0.170\arcm}
\setsymbol{LFI:FWHM:uncertainty:LFI19:units}{0.174\arcm}
\setsymbol{LFI:FWHM:uncertainty:LFI20:units}{0.170\arcm}
\setsymbol{LFI:FWHM:uncertainty:LFI21:units}{0.156\arcm}
\setsymbol{LFI:FWHM:uncertainty:LFI22:units}{0.164\arcm}
\setsymbol{LFI:FWHM:uncertainty:LFI23:units}{0.171\arcm}
\setsymbol{LFI:FWHM:uncertainty:LFI24:units}{0.652\arcm}
\setsymbol{LFI:FWHM:uncertainty:LFI25:units}{1.075\arcm}
\setsymbol{LFI:FWHM:uncertainty:LFI26:units}{1.131\arcm}
\setsymbol{LFI:FWHM:uncertainty:LFI27:units}{1.266\arcm}
\setsymbol{LFI:FWHM:uncertainty:LFI28:units}{1.287\arcm}

\setsymbol{LFI:FWHM:uncertainty:LFI18}{0.170}
\setsymbol{LFI:FWHM:uncertainty:LFI19}{0.174}
\setsymbol{LFI:FWHM:uncertainty:LFI20}{0.170}
\setsymbol{LFI:FWHM:uncertainty:LFI21}{0.156}
\setsymbol{LFI:FWHM:uncertainty:LFI22}{0.164}
\setsymbol{LFI:FWHM:uncertainty:LFI23}{0.171}
\setsymbol{LFI:FWHM:uncertainty:LFI24}{0.652}
\setsymbol{LFI:FWHM:uncertainty:LFI25}{1.075}
\setsymbol{LFI:FWHM:uncertainty:LFI26}{1.131}
\setsymbol{LFI:FWHM:uncertainty:LFI27}{1.266}
\setsymbol{LFI:FWHM:uncertainty:LFI28}{1.287}

\setsymbol{HFI:center:frequency:100GHz:units}{100\,GHz}
\setsymbol{HFI:center:frequency:143GHz:units}{143\,GHz}
\setsymbol{HFI:center:frequency:217GHz:units}{217\,GHz}
\setsymbol{HFI:center:frequency:353GHz:units}{353\,GHz}
\setsymbol{HFI:center:frequency:545GHz:units}{545\,GHz}
\setsymbol{HFI:center:frequency:857GHz:units}{857\,GHz}

\setsymbol{HFI:center:frequency:100GHz}{100}
\setsymbol{HFI:center:frequency:143GHz}{143}
\setsymbol{HFI:center:frequency:217GHz}{217}
\setsymbol{HFI:center:frequency:353GHz}{353}
\setsymbol{HFI:center:frequency:545GHz}{545}
\setsymbol{HFI:center:frequency:857GHz}{857}

\setsymbol{HFI:Ndetectors:100GHz}{8}
\setsymbol{HFI:Ndetectors:143GHz}{11}
\setsymbol{HFI:Ndetectors:217GHz}{12}
\setsymbol{HFI:Ndetectors:353GHz}{12}
\setsymbol{HFI:Ndetectors:545GHz}{3}
\setsymbol{HFI:Ndetectors:857GHz}{4}

\setsymbol{HFI:FWHM:Maps:100GHz:units}{9\parcm88}
\setsymbol{HFI:FWHM:Maps:143GHz:units}{7\parcm18}
\setsymbol{HFI:FWHM:Maps:217GHz:units}{4\parcm87}
\setsymbol{HFI:FWHM:Maps:353GHz:units}{4\parcm65}
\setsymbol{HFI:FWHM:Maps:545GHz:units}{4\parcm72}
\setsymbol{HFI:FWHM:Maps:857GHz:units}{4\parcm39}
\setsymbol{HFI:FWHM:Maps:100GHz}{9.88}
\setsymbol{HFI:FWHM:Maps:143GHz}{7.18}
\setsymbol{HFI:FWHM:Maps:217GHz}{4.87}
\setsymbol{HFI:FWHM:Maps:353GHz}{4.65}
\setsymbol{HFI:FWHM:Maps:545GHz}{4.72}
\setsymbol{HFI:FWHM:Maps:857GHz}{4.39}

\setsymbol{HFI:beam:ellipticity:Maps:100GHz}{1.15}
\setsymbol{HFI:beam:ellipticity:Maps:143GHz}{1.01}
\setsymbol{HFI:beam:ellipticity:Maps:217GHz}{1.06}
\setsymbol{HFI:beam:ellipticity:Maps:353GHz}{1.05}
\setsymbol{HFI:beam:ellipticity:Maps:545GHz}{1.14}
\setsymbol{HFI:beam:ellipticity:Maps:857GHz}{1.19}

\setsymbol{HFI:FWHM:Mars:100GHz:units}{9\parcm37}
\setsymbol{HFI:FWHM:Mars:143GHz:units}{7\parcm04}
\setsymbol{HFI:FWHM:Mars:217GHz:units}{4\parcm68}
\setsymbol{HFI:FWHM:Mars:353GHz:units}{4\parcm43}
\setsymbol{HFI:FWHM:Mars:545GHz:units}{3\parcm80}
\setsymbol{HFI:FWHM:Mars:857GHz:units}{3\parcm67}

\setsymbol{HFI:FWHM:Mars:100GHz}{9.37}
\setsymbol{HFI:FWHM:Mars:143GHz}{7.04}
\setsymbol{HFI:FWHM:Mars:217GHz}{4.68}
\setsymbol{HFI:FWHM:Mars:353GHz}{4.43}
\setsymbol{HFI:FWHM:Mars:545GHz}{3.80}
\setsymbol{HFI:FWHM:Mars:857GHz}{3.67}

\setsymbol{HFI:beam:ellipticity:Mars:100GHz}{1.18}
\setsymbol{HFI:beam:ellipticity:Mars:143GHz}{1.03}
\setsymbol{HFI:beam:ellipticity:Mars:217GHz}{1.14}
\setsymbol{HFI:beam:ellipticity:Mars:353GHz}{1.09}
\setsymbol{HFI:beam:ellipticity:Mars:545GHz}{1.25}
\setsymbol{HFI:beam:ellipticity:Mars:857GHz}{1.03}

\setsymbol{HFI:CMB:relative:calibration:100GHz}{$\lsim 1\%$}
\setsymbol{HFI:CMB:relative:calibration:143GHz}{$\lsim 1\%$}
\setsymbol{HFI:CMB:relative:calibration:217GHz}{$\lsim 1\%$}
\setsymbol{HFI:CMB:relative:calibration:353GHz}{$\lsim 1\%$}
\setsymbol{HFI:CMB:relative:calibration:545GHz}{}
\setsymbol{HFI:CMB:relative:calibration:857GHz}{}

\setsymbol{HFI:CMB:absolute:calibration:100GHz}{$\lsim 2\%$}
\setsymbol{HFI:CMB:absolute:calibration:143GHz}{$\lsim 2\%$}
\setsymbol{HFI:CMB:absolute:calibration:217GHz}{$\lsim 2\%$}
\setsymbol{HFI:CMB:absolute:calibration:353GHz}{$\lsim 2\%$}
\setsymbol{HFI:CMB:absolute:calibration:545GHz}{}
\setsymbol{HFI:CMB:absolute:calibration:857GHz}{}

\setsymbol{HFI:FIRAS:gain:calibration:accuracy:statistical:100GHz}{}
\setsymbol{HFI:FIRAS:gain:calibration:accuracy:statistical:143GHz}{}
\setsymbol{HFI:FIRAS:gain:calibration:accuracy:statistical:217GHz}{}
\setsymbol{HFI:FIRAS:gain:calibration:accuracy:statistical:353GHz}{2.5\%}
\setsymbol{HFI:FIRAS:gain:calibration:accuracy:statistical:545GHz}{1\%}
\setsymbol{HFI:FIRAS:gain:calibration:accuracy:statistical:857GHz}{0.5\%}

\setsymbol{HFI:FIRAS:gain:calibration:accuracy:systematic:100GHz}{}
\setsymbol{HFI:FIRAS:gain:calibration:accuracy:systematic:143GHz}{}
\setsymbol{HFI:FIRAS:gain:calibration:accuracy:systematic:217GHz}{}
\setsymbol{HFI:FIRAS:gain:calibration:accuracy:systematic:353GHz}{}
\setsymbol{HFI:FIRAS:gain:calibration:accuracy:systematic:545GHz}{7\%}
\setsymbol{HFI:FIRAS:gain:calibration:accuracy:systematic:857GHz}{7\%}

\setsymbol{HFI:FIRAS:zero:point:accuracy:100GHz:units}{0.8\MJysr}
\setsymbol{HFI:FIRAS:zero:point:accuracy:143GHz:units}{}
\setsymbol{HFI:FIRAS:zero:point:accuracy:217GHz:units}{}
\setsymbol{HFI:FIRAS:zero:point:accuracy:353GHz:units}{1.4\MJysr}
\setsymbol{HFI:FIRAS:zero:point:accuracy:545GHz:units}{2.2\MJysr}
\setsymbol{HFI:FIRAS:zero:point:accuracy:857GHz:units}{1.7\MJysr}

\setsymbol{HFI:FIRAS:zero:point:accuracy:100GHz}{0.8}
\setsymbol{HFI:FIRAS:zero:point:accuracy:143GHz}{}
\setsymbol{HFI:FIRAS:zero:point:accuracy:217GHz}{}
\setsymbol{HFI:FIRAS:zero:point:accuracy:353GHz}{1.4}
\setsymbol{HFI:FIRAS:zero:point:accuracy:545GHz}{2.2}
\setsymbol{HFI:FIRAS:zero:point:accuracy:857GHz}{1.7}

\setsymbol{HFI:unit:conversion:100GHz:units}{0.2415\MJysrmK}
\setsymbol{HFI:unit:conversion:143GHz:units}{0.3694\MJysrmK}
\setsymbol{HFI:unit:conversion:217GHz:units}{0.4811\MJysrmK}
\setsymbol{HFI:unit:conversion:353GHz:units}{0.2883\MJysrmK}
\setsymbol{HFI:unit:conversion:545GHz:units}{0.05826\MJysrmK}
\setsymbol{HFI:unit:conversion:857GHz:units}{0.002238\MJysrmK}

\setsymbol{HFI:unit:conversion:100GHz}{0.2415}
\setsymbol{HFI:unit:conversion:143GHz}{0.3694}
\setsymbol{HFI:unit:conversion:217GHz}{0.4811}
\setsymbol{HFI:unit:conversion:353GHz}{0.2883}
\setsymbol{HFI:unit:conversion:545GHz}{0.05826}
\setsymbol{HFI:unit:conversion:857GHz}{0.002238}

\setsymbol{HFI:colour:correction:alpha=-2:V1.01:100GHz}{0.9893}
\setsymbol{HFI:colour:correction:alpha=-2:V1.01:143GHz}{0.9759}
\setsymbol{HFI:colour:correction:alpha=-2:V1.01:217GHz}{1.0007}
\setsymbol{HFI:colour:correction:alpha=-2:V1.01:353GHz}{1.0028}
\setsymbol{HFI:colour:correction:alpha=-2:V1.01:545GHz}{1.0019}
\setsymbol{HFI:colour:correction:alpha=-2:V1.01:857GHz}{0.9889}

\setsymbol{HFI:colour:correction:alpha=0:V1.01:100GHz}{1.0008}
\setsymbol{HFI:colour:correction:alpha=0:V1.01:143GHz}{1.0148}
\setsymbol{HFI:colour:correction:alpha=0:V1.01:217GHz}{0.9909}
\setsymbol{HFI:colour:correction:alpha=0:V1.01:353GHz}{0.9888}
\setsymbol{HFI:colour:correction:alpha=0:V1.01:545GHz}{0.9878}
\setsymbol{HFI:colour:correction:alpha=0:V1.01:857GHz}{1.0014}

%% file: Proj_Ref_7_7b_authors_and_institutes.tex
\author{\small
Planck Collaboration:
P.~A.~R.~Ade\inst{68}
\and
N.~Aghanim\inst{45}
\and
M.~Arnaud\inst{55}
\and
M.~Ashdown\inst{53, 74}
\and
J.~Aumont\inst{45}
\and
C.~Baccigalupi\inst{66}
\and
A.~Balbi\inst{27}
\and
A.~J.~Banday\inst{72, 6, 60}
\and
R.~B.~Barreiro\inst{50}
\and
J.~G.~Bartlett\inst{3, 51}
% ...
}
\institute{\small
Aalto University Mets\"{a}hovi Radio Observatory, Mets\"{a}hovintie 114, FIN-02540 Kylm\"{a}l\"{a}, Finland\\
\and
Agenzia Spaziale Italiana Science Data Center, c/o ESRIN, via Galileo Galilei, Frascati, Italy\\
\and
Astroparticule et Cosmologie, CNRS (UMR7164), Universit\'{e} Denis Diderot Paris 7, B\^{a}timent Condorcet, 10 rue A. Domon et L\'{e}onie Duquet, Paris, France\\
\and
Atacama Large Millimeter/submillimeter Array, ALMA Santiago Central Offices Alonso de Cordova 3107, Vitacura, Casilla 763 0355, Santiago, Chile\\
\and
CITA, University of Toronto, 60 St. George St., Toronto, ON M5S 3H8, Canada\\
\and
CNRS, IRAP, 9 Av. colonel Roche, BP 44346, F-31028 Toulouse cedex 4, France\\
\and
California Institute of Technology, Pasadena, California, U.S.A.\\
\and
DAMTP, Centre for Mathematical Sciences, Wilberforce Road, Cambridge CB3 0WA, U.K.\\
\and
DSM/Irfu/SPP, CEA-Saclay, F-91191 Gif-sur-Yvette Cedex, France\\
\and
DTU Space, National Space Institute, Juliane Mariesvej 30, Copenhagen, Denmark\\
\and
D\'{e}partement de physique, de g\'{e}nie physique et d'optique, Universit\'{e} Laval, Qu\'{e}bec, Canada\\
\and
Departamento de F\'{\i}sica, Universidad de Oviedo, Avda. Calvo Sotelo s/n, Oviedo, Spain\\
\and
Department of Astronomy and Astrophysics, University of Toronto, 50 Saint George Street, Toronto, Ontario, Canada\\
\and
Department of Astronomy and Earth Sciences, Tokyo Gakugei University, Koganei, Tokyo 184-8501, Japan\\
\and
Department of Physics \& Astronomy, University of British Columbia, 6224 Agricultural Road, Vancouver, British Columbia, Canada\\
\and
Department of Physics, Gustaf H\"{a}llstr\"{o}min katu 2a, University of Helsinki, Helsinki, Finland\\
\and
Department of Physics, Princeton University, Princeton, New Jersey, U.S.A.\\
\and
Department of Physics, Purdue University, 525 Northwestern Avenue, West Lafayette, Indiana, U.S.A.\\
\and
Department of Physics, University of California, Berkeley, California, U.S.A.\\
\and
Department of Physics, University of California, One Shields Avenue, Davis, California, U.S.A.\\
\and
Department of Physics, University of California, Santa Barbara, California, U.S.A.\\
\and
Department of Physics, University of Illinois at Urbana-Champaign, 1110 West Green Street, Urbana, Illinois, U.S.A.\\
\and
Dipartimento di Fisica G. Galilei, Universit\`{a} degli Studi di Padova, via Marzolo 8, 35131 Padova, Italy\\
\and
Dipartimento di Fisica, Universit\`{a} La Sapienza, P. le A. Moro 2, Roma, Italy\\
\and
Dipartimento di Fisica, Universit\`{a} degli Studi di Milano, Via Celoria, 16, Milano, Italy\\
\and
Dipartimento di Fisica, Universit\`{a} degli Studi di Trieste, via A. Valerio 2, Trieste, Italy\\
\and
Dipartimento di Fisica, Universit\`{a} di Roma Tor Vergata, Via della Ricerca Scientifica, 1, Roma, Italy\\
\and
Discovery Center, Niels Bohr Institute, Blegdamsvej 17, Copenhagen, Denmark\\
\and
Dpto. Astrof\'{i}sica, Universidad de La Laguna (ULL), E-38206 La Laguna, Tenerife, Spain\\
\and
E\"{o}tv\"{o}s Lor\'{a}nd University, Department of Astronomy, P\'{a}zm\'{a}ny P\'{e}ter s\'{e}t\'{a}ny 1/A, 1117 Budapest, Hungary\\
\and
European Southern Observatory, ESO Vitacura, Alonso de Cordova 3107, Vitacura, Casilla 19001, Santiago, Chile\\
\and
European Space Agency, ESAC, Planck Science Office, Camino bajo del Castillo, s/n, Urbanizaci\'{o}n Villafranca del Castillo, Villanueva de la Ca\~{n}ada, Madrid, Spain\\
\and
European Space Agency, ESTEC, Keplerlaan 1, 2201 AZ Noordwijk, The Netherlands\\
\and
Helsinki Institute of Physics, Gustaf H\"{a}llstr\"{o}min katu 2, University of Helsinki, Helsinki, Finland\\
\and
INAF - Osservatorio Astrofisico di Catania, Via S. Sofia 78, Catania, Italy\\
\and
INAF - Osservatorio Astronomico di Padova, Vicolo dell'Osservatorio 5, Padova, Italy\\
\and
INAF - Osservatorio Astronomico di Roma, via di Frascati 33, Monte Porzio Catone, Italy\\
\and
INAF - Osservatorio Astronomico di Trieste, Via G.B. Tiepolo 11, Trieste, Italy\\
\and
INAF/IASF Bologna, Via Gobetti 101, Bologna, Italy\\
\and
INAF/IASF Milano, Via E. Bassini 15, Milano, Italy\\
\and
INRIA, Laboratoire de Recherche en Informatique, Universit\'{e} Paris-Sud 11, B\^{a}timent 490, 91405 Orsay Cedex, France\\
\and
IPAG: Institut de Plan\'{e}tologie et d'Astrophysique de Grenoble, Universit\'{e} Joseph Fourier, Grenoble 1 / CNRS-INSU, UMR 5274, Grenoble, F-38041, France\\
\and
Imperial College London, Astrophysics group, Blackett Laboratory, Prince Consort Road, London, SW7 2AZ, U.K.\\
\and
Infrared Processing and Analysis Center, California Institute of Technology, Pasadena, CA 91125, U.S.A.\\
\and
Institut d'Astrophysique Spatiale, CNRS (UMR8617) Universit\'{e} Paris-Sud 11, B\^{a}timent 121, Orsay, France\\
\and
Institut d'Astrophysique de Paris, CNRS UMR7095, Universit\'{e} Pierre \& Marie Curie, 98 bis boulevard Arago, Paris, France\\
\and
Institute of Astronomy and Astrophysics, Academia Sinica, Taipei, Taiwan\\
\and
Institute of Theoretical Astrophysics, University of Oslo, Blindern, Oslo, Norway\\
\and
Instituto de Astrof\'{\i}sica de Canarias, C/V\'{\i}a L\'{a}ctea s/n, La Laguna, Tenerife, Spain\\
\and
Instituto de F\'{\i}sica de Cantabria (CSIC-Universidad de Cantabria), Avda. de los Castros s/n, Santander, Spain\\
\and
Jet Propulsion Laboratory, California Institute of Technology, 4800 Oak Grove Drive, Pasadena, California, U.S.A.\\
\and
Jodrell Bank Centre for Astrophysics, Alan Turing Building, School of Physics and Astronomy, The University of Manchester, Oxford Road, Manchester, M13 9PL, U.K.\\
\and
Kavli Institute for Cosmology Cambridge, Madingley Road, Cambridge, CB3 0HA, U.K.\\
\and
LERMA, CNRS, Observatoire de Paris, 61 Avenue de l'Observatoire, Paris, France\\
\and
Laboratoire AIM, IRFU/Service d'Astrophysique - CEA/DSM - CNRS - Universit\'{e} Paris Diderot, B\^{a}t. 709, CEA-Saclay, F-91191 Gif-sur-Yvette Cedex, France\\
\and
Laboratoire Traitement et Communication de l'Information, CNRS (UMR 5141) and T\'{e}l\'{e}com ParisTech, 46 rue Barrault F-75634 Paris Cedex 13, France\\
\and
Laboratoire de Physique Subatomique et de Cosmologie, CNRS, Universit\'{e} Joseph Fourier Grenoble I, 53 rue des Martyrs, Grenoble, France\\
\and
Laboratoire de l'Acc\'{e}l\'{e}rateur Lin\'{e}aire, Universit\'{e} Paris-Sud 11, CNRS/IN2P3, Orsay, France\\
\and
Lawrence Berkeley National Laboratory, Berkeley, California, U.S.A.\\
\and
Max-Planck-Institut f\"{u}r Astrophysik, Karl-Schwarzschild-Str. 1, 85741 Garching, Germany\\
\and
MilliLab, VTT Technical Research Centre of Finland, Tietotie 3, Espoo, Finland\\
\and
National University of Ireland, Department of Experimental Physics, Maynooth, Co. Kildare, Ireland\\
\and
Niels Bohr Institute, Blegdamsvej 17, Copenhagen, Denmark\\
\and
Observational Cosmology, Mail Stop 367-17, California Institute of Technology, Pasadena, CA, 91125, U.S.A.\\
\and
Optical Science Laboratory, University College London, Gower Street, London, U.K.\\
\and
SISSA, Astrophysics Sector, via Bonomea 265, 34136, Trieste, Italy\\
\and
SUPA, Institute for Astronomy, University of Edinburgh, Royal Observatory, Blackford Hill, Edinburgh EH9 3HJ, U.K.\\
\and
School of Physics and Astronomy, Cardiff University, Queens Buildings, The Parade, Cardiff, CF24 3AA, U.K.\\
\and
Space Sciences Laboratory, University of California, Berkeley, California, U.S.A.\\
\and
Spitzer Science Center, 1200 E. California Blvd., Pasadena, California, U.S.A.\\
\and
Stanford University, Dept of Physics, Varian Physics Bldg, 382 Via Pueblo Mall, Stanford, California, U.S.A.\\
\and
Universit\'{e} de Toulouse, UPS-OMP, IRAP, F-31028 Toulouse cedex 4, France\\
\and
Universities Space Research Association, Stratospheric Observatory for Infrared Astronomy, MS 211-3, Moffett Field, CA 94035, U.S.A.\\
\and
University of Cambridge, Cavendish Laboratory, Astrophysics group, J J Thomson Avenue, Cambridge, U.K.\\
% ...
}